\documentclass[aps,prd,10pt,twocolumn,superscriptaddress,showpacs,longbibliography]{revtex4-1}

\pdfoutput=1
\usepackage{hyperref}
\usepackage{graphicx}
\usepackage{amsfonts,amsmath,amssymb,bm,bbm}
\usepackage{color}
\usepackage{enumitem}
\usepackage{url}
\usepackage{multirow}
\usepackage[version=4]{mhchem}
\usepackage{slashed}

\hypersetup{
    pdfnewwindow=true,
    colorlinks=true,  
    linkcolor=blue,   
    citecolor=blue,   
    filecolor=blue,   
    urlcolor=blue     
}

\def\bi{\begin{itemize}[noitemsep,leftmargin=*]
\setlength\itemsep{1em}
        }
\def\ei{\end{itemize}}

\begin{document}

\title{Neutrino-nucleus cross sections for $W$-boson and trident production}

\author{Bei Zhou}
\email{zhou.1877@osu.edu}
\thanks{\scriptsize \!\!  \href{https://orcid.org/0000-0003-1600-8835}{orcid.org/0000-0003-1600-8835}}
\affiliation{Center for Cosmology and AstroParticle Physics (CCAPP), Ohio State University, Columbus, OH 43210}
\affiliation{Department of Physics, Ohio State University, Columbus, OH 43210}

\author{John F. Beacom}
\email{beacom.7@osu.edu}
\thanks{\scriptsize \!\!  \href{http://orcid.org/0000-0002-0005-2631}{orcid.org/0000-0002-0005-2631}}
\affiliation{Center for Cosmology and AstroParticle Physics (CCAPP), Ohio State University, Columbus, OH 43210}
\affiliation{Department of Physics, Ohio State University, Columbus, OH 43210}
\affiliation{Department of Astronomy, Ohio State University, Columbus, OH 43210}

\date{February 18, 2020} 

\begin{abstract}
    The physics of neutrino-nucleus cross sections is a critical probe of the Standard Model and beyond. A precise understanding is also needed to accurately deduce astrophysical neutrino spectra. At energies above $\sim 5$ GeV, the cross section is dominated by deep inelastic scattering, mediated by weak bosons. In addition, there are subdominant processes where the hadronic coupling is through virtual photons, $\gamma^\ast$: (on-shell) $W$-boson production (e.g., where the underlying interaction is $\nu_\ell + \gamma^\ast \rightarrow \ell^- + W^+$) and trident production (e.g., where it is $\nu + \gamma^\ast \rightarrow \nu + \ell_1^- + \ell_2^+$). These processes become increasingly relevant at TeV--PeV energies. We undertake the first systematic approach to these processes (and those with hadronic couplings through virtual $W$ and $Z$ bosons), treating them together, avoiding common approximations, considering all neutrino flavors and final states, and covering the energy range $10\,$--$10^8$ GeV.  In particular, we present the first complete calculation of $W$-boson production and the first calculation of trident production at TeV--PeV energies. When we use the same assumptions as in prior work, we recover all of their major results. In a companion paper~\cite{Beacom:2019pzs}, we show that these processes should be taken into account for IceCube-Gen2.
\end{abstract}

\maketitle

\section{Introduction}
\label{sec_introduction}

The interactions of neutrinos with quarks, nucleons, and nuclei are a cornerstone of the Standard Model.  These test neutrino couplings to hadrons and probe the internal structure of hadronic states~\cite{LlewellynSmith:1971uhs, Hayato:2002sd, Andreopoulos:2009rq, Buss:2011mx, Lalakulich:2011eh, Golan:2012rfa, Formaggio:2013kya}.  Increasingly precise measurements of cross sections allow increasingly precise tests of neutrino mixing and beyond the Standard Model physics~\cite{Cornet:2001gy, AlvarezMuniz:2002ga, Altmannshofer:2014pba, Coloma:2017ppo, Bertuzzo:2018itn, Capozzi:2018dat, Machado:2019oxb}. Understanding the cross section is also crucial to neutrino astrophysics~\cite{Gaisser:1994yf, Gandhi:1995tf, Bahcall:1997eg, Gandhi:1998ri, Langanke:2004vx, Cocco:2007za, Yoshida:2008zb, CooperSarkar:2011pa, Connolly:2011vc, Laha:2013eev, Chen:2013dza}.  In the laboratory, neutrino scattering has been well measured up to $E_\nu \sim 10^2$ GeV~\cite{Seligman:1997fe, Tzanov:2005kr, Tanabashi:2018oca}.  Above $\sim 5$ GeV, the dominant interaction is deep inelastic scattering (DIS), where neutrinos couple via weak bosons to the quark degrees of freedom, with the nucleon and nuclear structure being less important but still relevant.

New scientific opportunities have arisen with IceCube, as atmospheric and astrophysical neutrinos have been detected up to $E_\nu \sim 10^7$ GeV~\cite{Aartsen:2014gkd, Aartsen:2015knd}. Even though the spectra are not known a priori, and the statistics are low, important progress can be made.  For example, the neutrino cross section can be determined by comparing the event spectra due to neutrinos that have propagated through substantial Earth matter or not~\cite{Hooper:2002yq, Borriello:2007cs, Connolly:2011vc, Klein:2013xoa, Aartsen:2017kpd, Bustamante:2017xuy}.  And to the extent that the cross section is understood --- e.g., the claimed theoretical precision (from the parton-distribution functions) at $10^7$ GeV is $\simeq 2\%$~\cite{CooperSarkar:2011pa} or $\simeq 1.5\%$~\cite{Connolly:2011vc} --- the measured event spectra can be used to accurately deduce neutrino spectra and flavor ratios, allowing tests of both astrophysical emission models and neutrino properties (e.g., Refs.~\cite{Padovani:2014bha, Tamborra:2014xia, Murase:2014foa, Ng:2014pca, Ioka:2014kca, Rott:2014kfa, Gauld:2015kvh, Murase:2015xka, Murase:2015gea, Bustamante:2015waa, Bhattacharya:2015jpa, Bechtol:2015uqb, Bustamante:2016ciw, Kadler:2016ygj}). As IceCube accumulates statistics, and larger detectors are under consideration~\cite{Blaufuss:2015muc, Adrian-Martinez:2016fdl}, the opportunities --- and the need for a better theoretical understanding of neutrino-nucleus scattering --- increase.

There are neutrino-nucleus interactions in which the hadronic coupling is via a virtual photon, $\gamma^\ast$, and the diagrams are more complex than in ordinary DIS.  Although these photon interactions are subdominant, their importance grows rapidly with energy, becoming relevant in the TeV--PeV range.  In (on-shell) $W$-boson production (Fig.~\ref{fig_real_W_diagrams}), the neutrino interacts with a virtual photon from the nucleus to produce a $W$ boson and a charged lepton.  The cross section for this process has been claimed to reach $\sim 10\%$ of the DIS cross section at $\sim 10^5$--$10^7$ GeV~\cite{Seckel:1997kk, Alikhanov:2014uja, Alikhanov:2015kla}. (To set a scale, in the past 7.5 years, IceCube has detected 60 starting events with reconstructed energies above 60 TeV~\cite{Stachurska:2019srh, Schneider:2019ayi}.)  {\it More careful calculations are needed}.  In trident production (Fig.~\ref{fig_trident_FF_diagrams}), the neutrino interacts with a virtual photon from the nucleus to produce a neutrino, a charged lepton, and a charged lepton of opposite sign~\cite{Altmannshofer:2014pba, Magill:2016hgc, Ge:2017poy, Ballett:2018uuc, Altmannshofer:2019zhy, Gauld:2019pgt}.  The cross section for this process has never been calculated at TeV--PeV energies.  {\it A first calculation is needed. }

In this paper, we provide the first full calculations of both processes.  We treat them in a unified way, avoiding common approximations, considering all neutrino flavors and final states, and covering the energy range 10--$10^8$ GeV.  We recover all previous major results when we adopt their inputs.  In our companion paper~\cite{Beacom:2019pzs}, we detail the implications for IceCube-Gen2 measurements of neutrino spectra and flavor ratios, tests of neutrino properties, and tests of new physics.

In Sec.~\ref{sec_review}, we review the $W$-boson and trident production processes, identifying the shortcomings of previous work. In Sec.~\ref{sec_sgmnugamma}, we calculate the neutrino-real photon cross sections for both processes. The more complicated neutrino-nucleus cross sections in different regimes are calculated in Secs.~\ref{sec_elastic} and~\ref{sec_inelastic}, then added up (Fig.~\ref{fig_sgmnuA_tot_O16}) and discussed in Sec.~\ref{sec_total}. We conclude in Sec.~\ref{sec_concl}.

\section{Review of $W$-boson and trident production}
\label{sec_review}

\begin{figure}[t]
\includegraphics[width=\columnwidth]{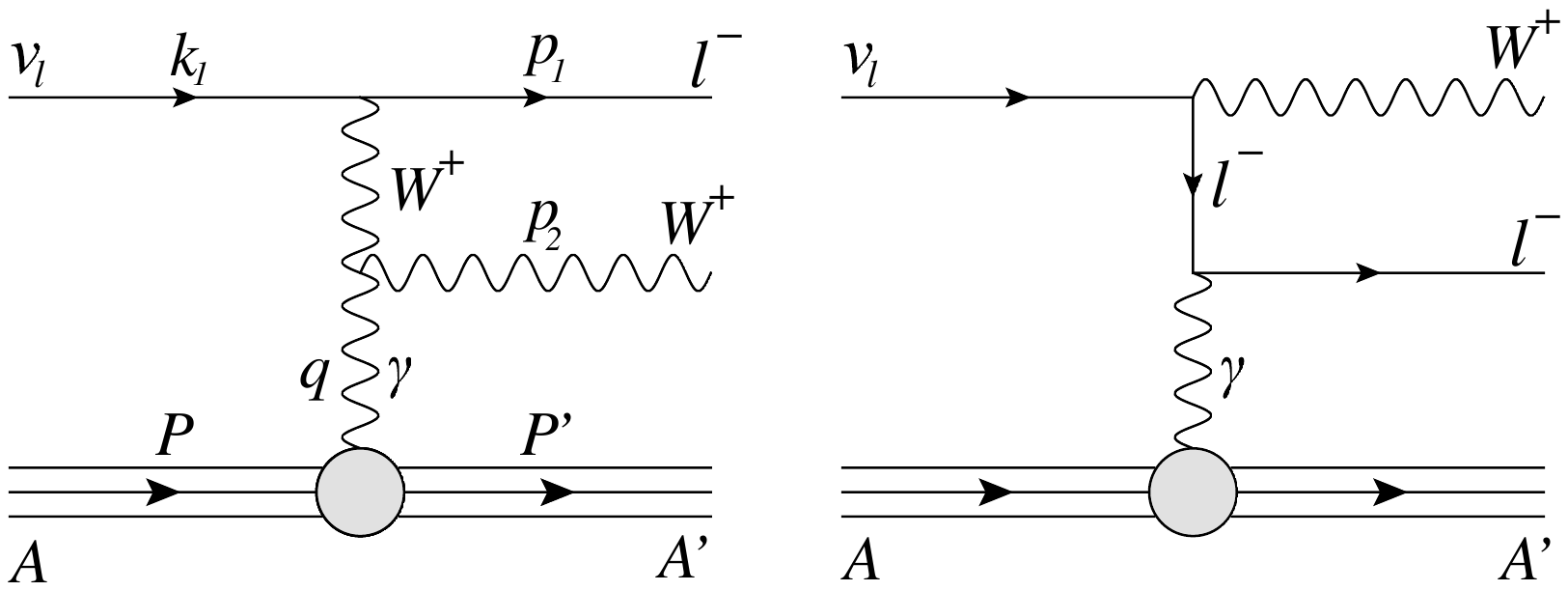}
    \vspace{0.001cm}
    \caption{Diagrams for (on-shell) $W$-boson production via photon exchange. $A$ and $A'$ are the initial- and final-state nuclei. (See Fig.~\ref{fig_trident_diagrams} and Sec.~\ref{sec_sgmnugamma_trident} for the connection with trident production.) For antineutrinos, take the CP transformation of the elementary particles. 
}
\label{fig_real_W_diagrams}
\end{figure}

The (on-shell) $W$-boson and trident production processes are, respectively, 
\begin{equation}
    \nu_\ell + A \rightarrow \ell^- + W^+ + A' \, ,
\label{eq_realW}
\end{equation}
\begin{equation}
    \nu + A \rightarrow \nu + \ell_1^- + \ell_2^+ + A' \, ,
\label{eq_trident}
\end{equation}
where $A$ and $A'$ are the initial and final-state nuclei and $\ell$ is a charged lepton. For trident production, for now we simplify the flavor information (for details, see Eq.~(\ref{eq_trident_details})). For antineutrinos, take the CP transformation of the elementary particles.

\begin{figure}[t!]
\includegraphics[width=\columnwidth]{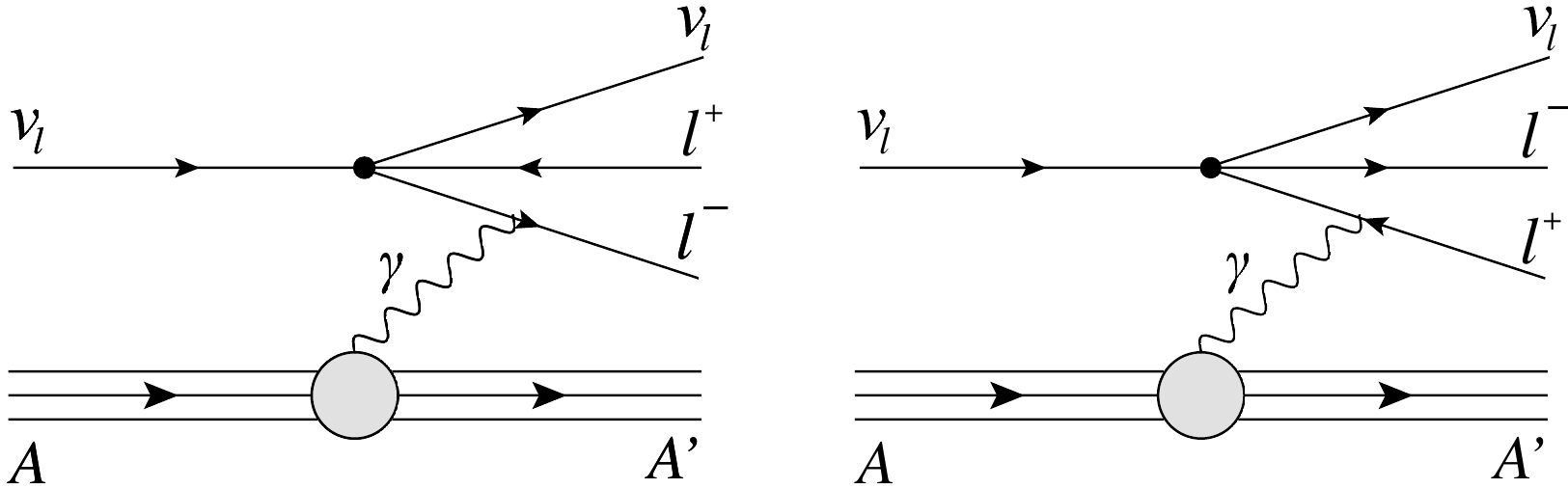}
    \vspace{0.001cm}
    \caption{Diagrams for trident production via photon exchange in the four-Fermi theory (see Fig.~\ref{fig_trident_diagrams} for the full Standard Model). For antineutrinos, take the CP transformation of the elementary particles. 
}
\label{fig_trident_FF_diagrams}
\end{figure}

Figures~\ref{fig_real_W_diagrams} and~\ref{fig_trident_FF_diagrams} show the diagrams for $W$-boson and trident production processes, respectively. {\it We also calculate diagrams, not shown, with $W$ and $Z$ boson couplings to the hadronic side; this is discussed in Sec.~\ref{sec_inelastic}.} For trident production, (for Fig.~\ref{fig_trident_FF_diagrams} only) we use the four-Fermi theory for simplicity, the diagrams of which nicely show the ``trident'' feature though hiding the connection to $W$-boson production (see Fig.~\ref{fig_trident_diagrams} for the full Standard-Model diagrams, on which our calculation is based). In both processes, a neutrino splits into charged particles (leptonic part) that couple to the photon from the nucleus (hadronic part). {\it The leptonic part is straightforward but depends on the process, while the hadronic part is complicated but independent of the process. }

In the rest of this section, we review the hadronic part (Sec.~\ref{sec_review_hadronic}), which also sets the foundation, then discuss the two processes respectively (Secs.~\ref{sec_review_realW} and~\ref{sec_review_trident}).

\subsection{Hadronic part} 
\label{sec_review_hadronic}

At most energies, the hadronic part is connected by a virtual photon from the nucleus. Above $\gtrsim 10^8$ GeV, the contributions of virtual weak bosons from the nucleus and mixing with the photon are not negligible (see Sec.~\ref{sec_inelastic}). 

The hadronic coupling can be in different regimes, including coherent ($\sigma \propto Z^2$), diffractive ($\sigma \propto Z$), and inelastic ($\sigma \propto Z$), in which the virtual photon couples to the whole nucleus, nucleon, and a single quark, respectively. (These three regimes are analogous to the coherent elastic neutrino-nucleus scattering, quasi-elastic scattering, and deep-inelastic scattering, respectively, for the usual neutrino-nucleus interaction, in which the hadronic coupling is through a $W/Z$ boson.) Adding the cross sections in different regimes gives the total cross section.

The coherent ($A' = A$) and diffractive ($A' \neq A$) regimes are both elastic, on the nucleus and nucleon, respectively. The former is usually described by a nuclear form factor and the latter by a nucleon form factor.

Two different calculational frameworks have been used in previous work, i.e., using or not using the equivalent photon approximation (EPA, or Weizs\"acker--Williams Approximation)~\cite{Fermi:1924tc, vonWeizsacker:1934nji, Williams:1934ad}. The EPA assumes the photon mediator (to the nucleus) to be on shell, i.e., $q^2=0$. This is motivated by the fact that the photon is usually very soft when the beam particle is very energetic (e.g., high-energy electron scattering on nuclei). Using real photons significantly simplifies the calculation, because then one does not need to take into consideration the photon virtuality or longitudinal polarization. However, Refs.~\cite{Kozhushner1961, Shabalin1963, Czyz:1964zz, Ballett:2018uuc} pointed out that EPA is not valid for most cases, especially for electron final states, leading to an overestimation of the cross section that, in some cases, is by more than 200\%. {\it The simplest reason is that, though the beam neutrino is very energetic, the charged particle that directly couples to the photon may not be.}

Inelastic scattering ($A' \neq A$) could also happen, with nucleon breakup. The hadronic part is usually described by parton-distribution functions (PDFs) for photon, quarks, etc.  The inelastic regime has two subprocesses, photon-initiated (related to the photon PDF) and quark-initiated (related to the quark PDFs)~\cite{Schmidt:2015zda}. See Sec.~\ref{sec_inelastic} for details.

\subsection{$W$-boson production}
\label{sec_review_realW}

The $W$-boson production process (Fig.~\ref{fig_real_W_diagrams}) initially raised interest in the 1960s and 1970s. The hypothetical (at that time) $W$ boson could be directly produced by a beam of $\nu_\mu$ scattering off the Coulomb field of a nucleus (Fig.~\ref{fig_real_W_diagrams}; e.g., Refs.~\cite{Lee:1960qv, Lee:1961jj, Bell:1996ms, Bell:1996mr, Brown:1971qr, Brown:1971xk}). If $W$ bosons were not detected, a lower bound on their mass could be set. Later, the discovery of the $W$ boson at a proton-antiproton collider~\cite{Arnison:1983rp}, and especially its large mass, significantly reduced the motivation to search for this process at fixed-target neutrino experiments. 
    
The interest in this process came back due to high-energy astrophysical neutrino detectors~\cite{Aartsen:2014gkd, Collaboration:2011nsa, Adrian-Martinez:2016fdl}, and was studied by Seckel~\cite{Seckel:1997kk} and Alikhanov~\cite{Alikhanov:2014uja, Alikhanov:2015kla}. In Ref.~\cite{Seckel:1997kk}, for the neutrino-nucleus cross sections, only the ratio of $\nu_e \rightarrow e^- W^+$ to charged-current (CC) DIS on $\ce{^{16}O}$ and $\ce{^{56}Fe}$ were shown, and only the coherent regime was considered (see Table~\ref{tab_summary_calc}). In Refs.~\cite{Alikhanov:2014uja, Alikhanov:2015kla}, all three flavors were considered and shown, and all three scattering regimes were considered. However, all three regimes used EPA, and nuclear effects (mainly Pauli blocking) were not included. Moreover, for the inelastic regime, only the photon-initiated subprocess was calculated (see Table~\ref{tab_summary_calc}).

\begin{figure}[t]
\includegraphics[width=\columnwidth]{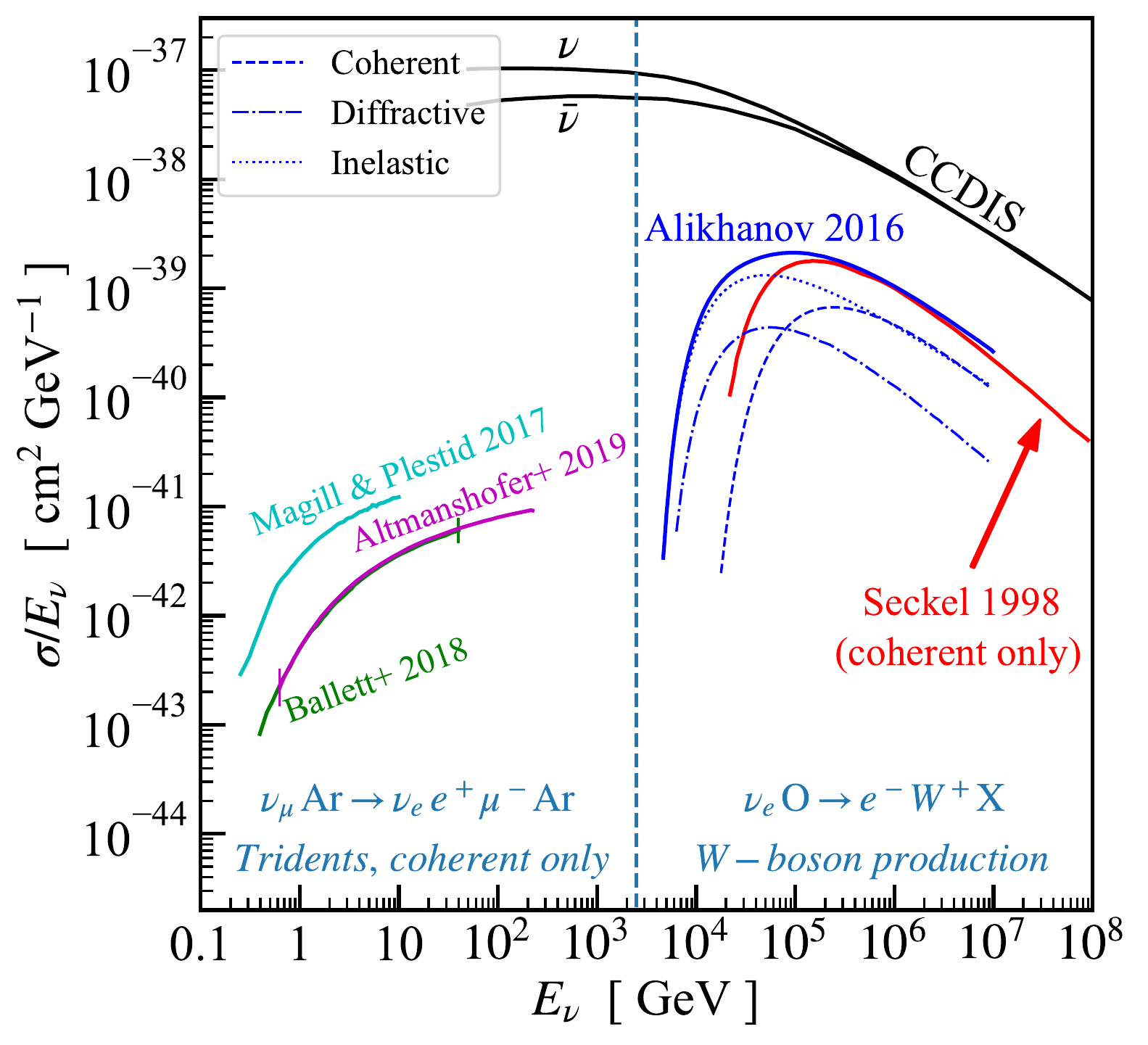}
    \caption{Summary of cross sections for $W$-boson and trident production from previous work, with the two processes separated as labeled. To simplify the figure, for {\it $W$-boson production}, we show only $\nu_e \rightarrow e^- W^+$ on $\ce{^{16}O}$ (by Seckel~\cite{Seckel:1997kk} and Alikhanov~\cite{Alikhanov:2014uja, Alikhanov:2015kla}), and for {\it trident production}, only the coherent regime (the dominant part) of $\nu_\mu \rightarrow \nu_e e^- \mu^+$ on $\ce{^{40}Ar}$ (by Magill \& Plestid~\cite{Magill:2016hgc}, Ballett {\it et al.}~\cite{Ballett:2018uuc}, and Altmannshofer {\it et al.}~\cite{Altmannshofer:2019zhy}). Also shown, for comparison, is the cross section of charged-current deep inelastic scattering (CCDIS)~\cite{CooperSarkar:2011pa}.
}
\label{fig_previous_xsecs}
\end{figure}

Figure~\ref{fig_previous_xsecs} shows their results. All three scattering regimes are important. The high threshold is set by the $W$-boson mass and the hadronic structure functions. The diffractive regime has a lower threshold than the coherent regime because larger $Q^2$ ($\equiv -q^2$; virtuality of the photon) can be probed by the nucleon form factor than the nuclear form factor. Above threshold, the coherent cross section ($\propto Z^2$) is larger than the diffractive cross section ($\propto Z$). 

The coherent cross section of Seckel~\cite{Seckel:1997kk} is about two times that of Alikhanov~\cite{Alikhanov:2014uja, Alikhanov:2015kla}, possibly due to their treating the nuclear form factor differently (as pointed out by Ref.~\cite{Alikhanov:2015kla}). (The origin of the factor of two between them could not be traced, as the details of the calculations are not given in Ref.~\cite{Seckel:1997kk}.)

Importantly, this cross section is claimed to be $\sim 10\%$ of the charged-current deep inelastic scattering (CCDIS) cross section~\cite{CooperSarkar:2011pa}, indicating this process is detectable by high-energy neutrino detectors like IceCube~\cite{Aartsen:2015knd}, KM3NeT~\cite{Adrian-Martinez:2016fdl}, and especially the forthcoming IceCube-Gen2~\cite{Blaufuss:2015muc}. With 60 starting events with energies above 60 TeV~\cite{Stachurska:2019srh, Schneider:2019ayi}, IceCube already has a nominal precision scale of 13\%, and IceCube-Gen2 would be 10 times larger. {\it However, on the theory side, due to the limitations above, more complete and careful calculations are needed (see Table~\ref{tab_summary_calc}).}

\begin{table*}[t]
    \caption{ \label{tab_summary_calc}  Summary of the features of previous calculations and of this work. ``+'' and ``$-$'' means ``considered'' and ``not considered'' in the calculation respectively. ``Full SM'' means using full Standard Model, instead of four-Fermi theory.
}  
\medskip
\renewcommand{\arraystretch}{1.1} \centering 
\begin{tabular*}{\textwidth}{c||c||c||c|c|c|c||c|c}
\hline 
                    &    &Full SM&Coherent &Diffractive&Beyond EPA & Pauli blocking & Inel., photon & Inel., quark \\

\hline  \hline 
    $W$-boson &  Seckel~\cite{Seckel:1997kk}          &   +  &      +    &      $-$    &  $-$   &$-$&       $-$        &    $-$      \\
    production   & Alikhanov~\cite{Alikhanov:2014uja, Alikhanov:2015kla}       &   +  &    +    &      +    &   $-$    &$-$&        +        &    $-$      \\

\hline \hline
    \multirow{5}{*}{  \begin{tabular}[c]{@{}c@{}}Trident\\production\end{tabular} }    & Altmannshofer {\it et al.}~\cite{Altmannshofer:2014pba}         &   $-$    &    +    &      $-$    &   $-$    &$-$&        $-$        &    $-$      \\
        & Magill \& Plestid~\cite{Magill:2016hgc} &   $-$    &    +    &      +    &   $-$    &$-$&        $-$        &    +      \\
    & Ge {\it et al.}~\cite{Ge:2017poy}              &   $-$    &    +    &      $-$   &   $-$   &$-$&        $-$        &    $-$      \\
    & Ballett {\it et al.}~\cite{Ballett:2018uuc}         &   $-$    &    +    &      +    &   +    &+&        $-$        &    $-$      \\
    & Altmannshofer {\it et al.}~\cite{Altmannshofer:2019zhy}   &   $-$    &    +    &      +    &   +    &+&        $-$        &    $-$      \\
\hline \hline
    \bf Both, unified    & \bf This work               &  \bf  +    &    \bf +    &      \bf +    &   \bf +    &\bf +&        \bf +        &    \bf +      \\

\hline 
\end{tabular*}
\end{table*}

\subsection{Trident production}
\label{sec_review_trident}

The trident processes (Fig.~\ref{fig_trident_FF_diagrams}) raised interest at a similar time to $W$-boson production, also as a process to probe the then-hypothetical $W / Z$ propagators in the weak interactions. Even if the weak bosons were not produced directly due to, e.g., their large masses, their existence could make the trident production rate different from that of the pure V-A theory (e.g., Refs.~\cite{Kozhushner1961, Shabalin1963, Czyz:1964zz, Lovseth:1971vv, Fujikawa:1971nx, Koike:1971tu, Koike:1971vg, Brown:1973ih, Belusevic:1987cw})

So far, only the $\nu_\mu \rightarrow \nu_\mu \mu^- \mu^+$ process has been observed, by the Charm-II~\cite{Geiregat:1990gz} and CCFR~\cite{Mishra:1991bv} experiments; NuTeV~\cite{Adams:1999mn} set an upper limit. These results are consistent with SM predictions.

The trident processes have been popular again in recent years, due to currently running and upcoming accelerator neutrino experiments (e.g., Refs.~\cite{Anelli:2015pba, Antonello:2015lea, Acciarri:2015uup, Soler:2015ada}) as well as Ref.~\cite{Altmannshofer:2014pba} showing first trident constraints on new physics such as $Z'$ models. Table~\ref{tab_summary_calc} summarizes the calculations of trident cross sections by Refs.~\cite{Altmannshofer:2014pba, Magill:2016hgc, Ge:2017poy} using EPA and by Refs.~\cite{Ballett:2018uuc, Altmannshofer:2019zhy} using an improved calculation. Usually, only the electron and muon flavors are considered, as the tau flavor is rare for accelerator neutrinos. The inelastic regime is very small, so not considered (except in Ref.~\cite{Magill:2016hgc}).

All previous work used the four-Fermi theory, instead of the full Standard Model (see Table~\ref{tab_summary_calc}). One reason is that these papers focused on accelerator neutrinos below $\sim 100$~GeV. Another reason is that the hadronic part (Sec.~\ref{sec_review_hadronic}) complicates the calculation a lot, so using the four-Fermi theory for leptonic part is significantly simpler.

Figure~\ref{fig_previous_xsecs} summarizes previous calculations. The threshold is set by the final-state lepton masses and hadronic structure functions. The difference between Magill \& Plestid~\cite{Magill:2016hgc} and Ballett {\it et al.}~\cite{Ballett:2018uuc}, Altmannshofer {\it et al.}~\cite{Altmannshofer:2019zhy} is due to the former using EPA, while the latter two not. 

Though at GeV energies the cross sections are $\sim 10^{-5}$ of CCDIS~\cite{CooperSarkar:2011pa}, they increase quickly. Therefore, it is interesting and important to know the cross sections at TeV--PeV energies. To this end, the full Standard Model is needed instead of the four-Fermi theory. In addition, our calculations fix several other shortcomings (see Table~\ref{tab_summary_calc}).


\section{Cross sections between neutrinos and real photons}
\label{sec_sgmnugamma}

In this section, we calculate the cross sections of $W$-boson and trident production between a neutrino and a real photon. This shows the underlying physics and the basic behavior of the cross sections. The connection between the two processes is also clearly revealed.

For the cross sections between elementary particles, we calculate the matrix elements and phase space integrals ourselves, and check the results using the public tools {\tt MadGraph (v2.6.4)}~\cite{Alwall:2014hca} and {\tt CalcHEP (v3.7.1)}~\cite{Belyaev:2012qa}. The calculational procedures set the basis for the off-shell cross sections in Sec.~\ref{sec_elastic}.

\subsection{$W$-boson production}
\label{sec_sgmnugamma_realW}

The leptonic part of $W$-boson production is (Fig.~\ref{fig_real_W_diagrams}),
\begin{equation}
    \nu_\ell\, +\, \gamma  \rightarrow \ell^- + W^+ \, .
\label{eq_realW_nugm}
\end{equation}

The cross section can be calculated using
\begin{equation}
    \sigma_{\nu \gamma}(s_{\nu \gamma}) =
    \frac{1}{2 s_{\nu \gamma}} \int \, \frac{1}{2} \sum_{\rm spins} |\mathcal{M}^{\rm WBP}|^2\, d{\rm PS_2}  \, ,
\label{eq_sgmnugm_realW}
\end{equation}
where $1/2s_{\nu \gamma}$ is the Lorentz-invariant flux factor, $s_{\nu \gamma} \equiv (k_1+q)^2$, 
$\frac{1}{2} \sum_{\rm spins} |\mathcal{M}^{\rm WBP}|^2$ the photon-spin averaged matrix element (Appendix~\ref{appdx_realW_amplitudes}), and $d{\rm PS_2} (= \frac{p_{\rm  CM}}{\sqrt{s_{\nu \gamma}}} \frac{d\cos\theta}{8 \pi}$) is the two-body phase space, of which $p_{\rm CM}$ is the momentum of the outgoing particle in the CM frame, with angle $\theta$ respect to the incoming particle. This process has been calculated by Refs.~\cite{Seckel:1997kk, Alikhanov:2014uja}. Our calculation gives the same results.

The diagrams for Eq.~(\ref{eq_realW_nugm}) are similar to those in Fig.~\ref{fig_real_W_diagrams}, but replacing the photon from the nucleus with a free (real) photon. Both diagrams, with a relative minus sign, need to be included to assure gauge invariance. Numerically, the first diagram dominates at small $s_{\nu \gamma}$, while the second dominates at large $s_{\nu \gamma}$. Neutrinos and antineutrinos have the same total and differential cross sections, as the matrix element is invariant under CP transformation.

\begin{figure*}[t!]
\includegraphics[width=\textwidth]{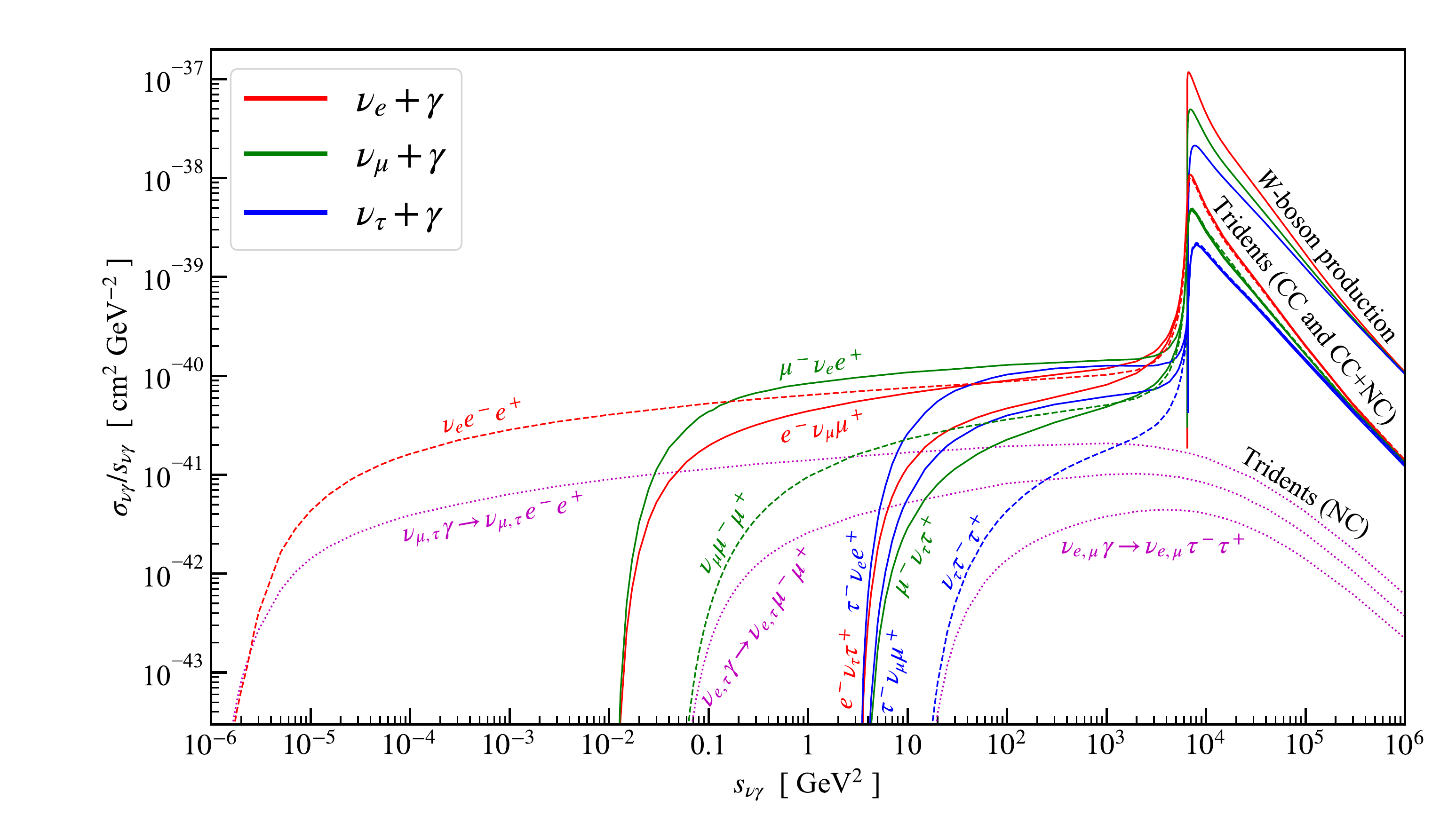}
    \caption{Our cross sections (actually $\sigma_{\nu \gamma}(s_{\nu \gamma})/s_{\nu \gamma}$) for $W$-boson and trident production, between a neutrino and a real photon as a function of their CM energy. {\bf Red}, {\bf green}, and {\bf blue} lines are $\nu_e$-, $\nu_\mu$-, and $\nu_\tau$-induced channels, respectively. {\bf Solid} lines are trident CC channels, and {\bf dashed} lines are trident CC+NC channels (we label only the final states for both). {\bf Magenta dotted} lines are trident NC channels, which depend on only the final-state charged leptons (we label both the initial and final states). The trident CC, NC, and CC+NC channels correspond to diagrams (1)--(3), (4)--(5), and (1)--(5) of Fig.~\ref{fig_trident_diagrams}. The corresponding antineutrino cross sections (i.e., obtained by CP-transforming the processes shown) are the same. See text for details.
}
\label{fig_sigma_nugm_all}
\end{figure*}

Figure~\ref{fig_sigma_nugm_all} shows $\sigma_{\nu \gamma}(s_{\nu \gamma})/s_{\nu \gamma}$ for $W$-boson production. We divide out $s_{\nu \gamma}$, the dominant trend, to highlight the deviations over the wide range of the x axis. The threshold is set by $s_{\nu \gamma} = (m_W+m_\ell)^2$. Just above threshold, the lepton propagator in the first diagram (Fig.~\ref{fig_real_W_diagrams}) gives a logarithmic term, $\sim \log[(...)/m_\ell^2]$ , which leads to $\sigma_{\nu_e \gamma} > \sigma_{\nu_\mu \gamma} > \sigma_{\nu_\tau \gamma}$~\cite{Seckel:1997kk}. For $s_{\nu \gamma} > 10^6$~GeV$^2$, the cross sections become constant and different flavors converge, with $\sigma_{\nu \gamma} \simeq 2\sqrt{2} \alpha G_F \simeq 10^{-34}$~cm$^{-2}$.

\subsection{Trident production}
\label{sec_sgmnugamma_trident}

\begin{figure}[t!]
\includegraphics[width=\columnwidth]{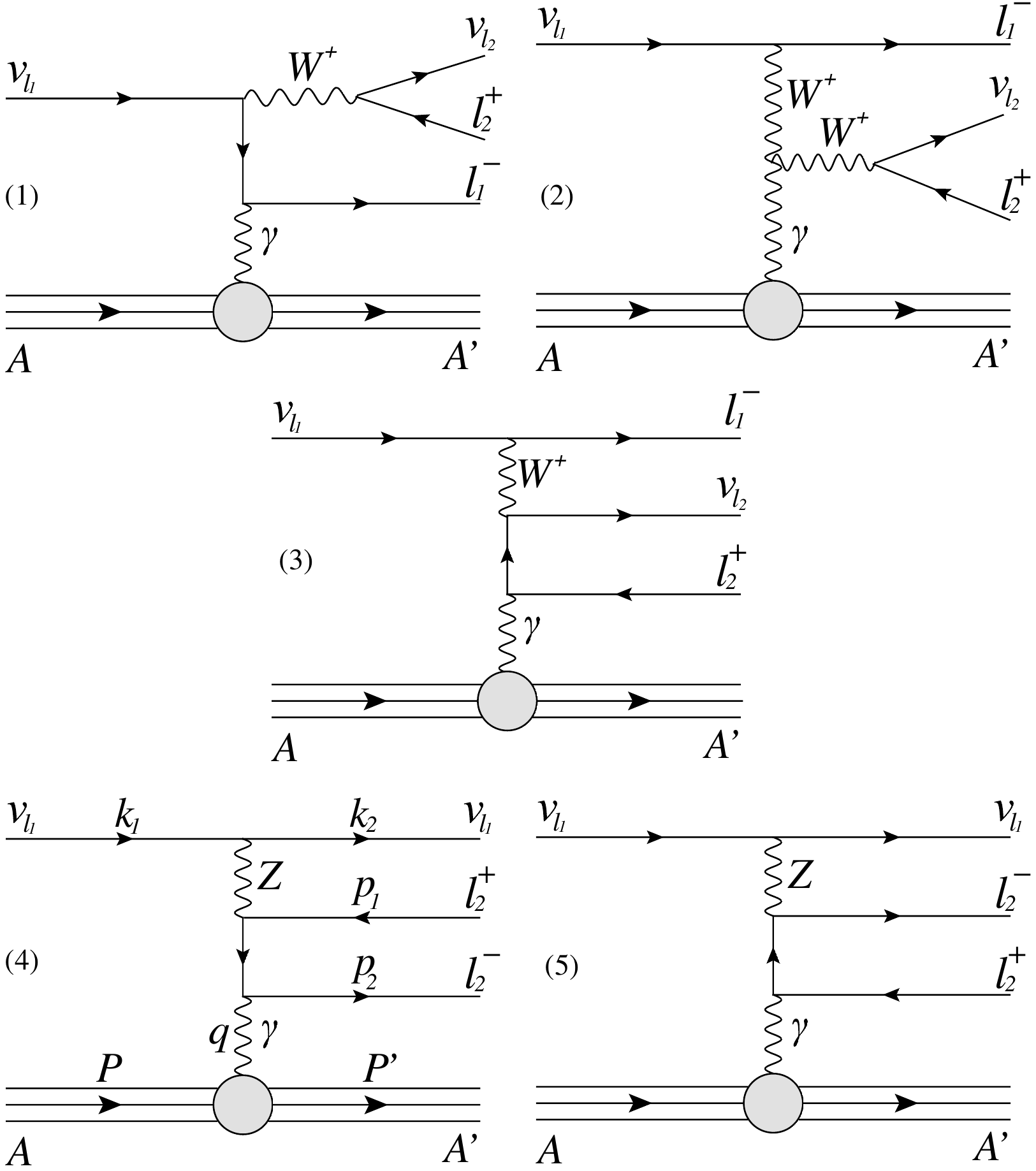}
    \vspace{0.001cm}
    \caption{Diagrams for trident production via photon exchange in the Standard Model, with the order, $\mathcal{M}^{\rm Tri}_1$--$\mathcal{M}^{\rm Tri}_5$, labeled in parentheses, and with the momenta labeled on the fourth diagram. The trident CC, NC, and CC+NC channels correspond to diagrams (1)--(3), (4)--(5), and (1)--(5). For antineutrinos, take the CP transformation of the elementary particles. The first and second diagrams are connected to $W$-boson production (Fig.~\ref{fig_real_W_diagrams}; also see Sec.~\ref{sec_sgmnugamma_trident} for details of the connection). \\
}
\label{fig_trident_diagrams}
\end{figure}

The leptonic part of trident production, for each incoming neutrino flavor, is (Fig.~\ref{fig_trident_diagrams})
\begin{subequations}
\begin{align}
    \nu_{\ell_1} + \gamma & \rightarrow \ell_1^- + \nu_{\ell_2}  + \ell_2^+ \ \  \text{ (CC)}  , \\
    \nu_{\ell_1} + \gamma & \rightarrow \nu_{\ell_1} + \ell_2^- + \ell_2^+  \ \  \text{ (NC)}  ,  \\
    \nu_\ell + \gamma     & \rightarrow \nu_\ell + \ell^- + \ell^+          \, \text{ (CC+NC)}  ,  
\end{align}
\label{eq_trident_details}
\end{subequations}
where $\ell_1$, $\ell_2$ = $e$, $\mu$ or $\tau$ and $\ell_1 \neq \ell_2$. So there are two, two and one CC, NC and CC+NC channels, respectively; details below. For antineutrinos, take the CP transformation of the elementary particles; details below.

The cross section can be calculated using~\cite{Vysotsky:2002ix, Altmannshofer:2014pba, Magill:2016hgc}
\begin{equation}
    \sigma_{\nu \gamma}(s_{\nu \gamma}) = \frac{1}{2 s_{\nu \gamma}} \int \frac{1}{2} \sum_{\rm spins} |\mathcal{M}^{\rm Tri}|^2\, d{\rm PS_3} \, ,
\label{eq_sgmnugm_trident}
\end{equation}
where $1/2 s_{\nu \gamma}$ is the Lorentz-invariant flux factor, $\frac{1}{2} \sum_{\rm spins} |\mathcal{M}^{\rm Tri}|^2$ the photon-spin averaged matrix element, and $d{\rm PS_3}$ is the three-body phase space (see below). Different from previous calculations, here we need to use the full Standard Model, instead of the four-Fermi theory, as we are also interested in TeV--PeV energies. 

Figure~\ref{fig_trident_diagrams} shows the five possible diagrams of trident production in the Standard Model. Note that the diagram involving a $W W \gamma$ vertex is not included by the four-Fermi theory, though it is suppressed at low energies. When $\ell_1 \neq \ell_2$, the top three diagrams (exclusively mediated by $W$) lead to CC channels, and the bottom two (exclusively mediated by $Z$) lead to NC channels.  When $\ell_1 = \ell_2 = \ell$, all five diagrams give the same final states, which lead to the CC+NC channels.

We work in the unitarity gauge, which is simpler for the tree level. The amplitudes for each diagram, $\mathcal{M}^{\rm Tri}_{1}$---$\mathcal{M}^{\rm Tri}_5$, can be found in Appendix~\ref{appdx_trident_amplitudes}, and relative signs between these diagrams are
\begin{equation}
    \mathcal{M}^{\rm Tri} = (\mathcal{M}^{\rm Tri}_1 - \mathcal{M}^{\rm Tri}_2 + \mathcal{M}^{\rm Tri}_3) - (\mathcal{M}^{\rm Tri}_4 + \mathcal{M}^{\rm Tri}_5) \, .
\label{eq_trident_relsign}
\end{equation}
The matrix element is calculated using {\tt FeynCalc}~\cite{Mertig:1990an, Shtabovenko:2016sxi}.

For antineutrinos, the total cross sections are the same as neutrinos, due to CP invariance~\cite{Czyz:1964zz,Ge:2017poy}. For the differential cross sections, they are only the same for the NC channels due to interchange symmetry of two charged leptons, which is not the case for CC and CC+NC channels~\cite{Ge:2017poy}. Therefore, for the following discussion, we take neutrinos only.

The three-body phase space in the case of real photon is~\cite{Vysotsky:2002ix, Altmannshofer:2014pba, Magill:2016hgc}
\begin{equation}
    d {\rm PS_3} = \frac{1}{2} \frac{1}{(4 \pi)^2} \frac{d t}{2 s_{\nu \gamma}} 
    \overline{\beta}(l)
    \frac{d l}{2 \pi} 
    \frac{d\Omega''}{4 \pi} \, ,
\label{eq_PS3}
\end{equation}
where $t \equiv 2 q \cdot (k_1-k_2)$, $l \equiv (p_1+p_2)^2$, $\Omega''$ the solid angle with respect to $q$ in the rest frame of $p_1 + p_2$, and
\begin{equation}
    \overline{\beta}(l) =  \sqrt{ 1 - \frac{2(m_1^2+m_2^2)}{l} + \frac{(m_1^2-m_2^2)^2}{l^2} } \, ,
\label{eq_PS3_beta}
\end{equation}
where $m_{1,\,2}$ is the mass of $p_{1,\,2}$. The integration over $l$ is done from $(m_1+m_2)^2$ to $s_{\nu\gamma}$, and $t$ from $l$ to $s_{\nu\gamma}$. Using these variables, we find that the numerical integration converges reasonably fast for both the four-Fermi theory case and the Standard-Model case.

Figure~\ref{fig_sigma_nugm_all} shows $\sigma_{\nu \gamma}(s_{\nu \gamma})/s_{\nu \gamma}$ for all 15 trident channels. The thresholds are set by the masses of final states, i.e., $s_{\nu \gamma} = (m_1 + m_2)^2$. The cross sections increase from threshold until $\sim 10^6$~GeV$^2$. For $s_{\nu\gamma} > 10^6$~GeV$^2$, same as for $W$-boson production, the cross sections become constant. 

For the CC and CC+NC channels, very interestingly, just above $s_{\nu\gamma} = m_W^2 \simeq 6.5 \times 10^3$~GeV$^2$, there is a sharp increase. This is due to the $s$-channel like part of the first and second diagrams in Fig.~\ref{fig_trident_diagrams}, which are mediated by $W$ bosons. For $s_{\nu\gamma} > m_W^2$, $W$-boson production is turned on. In the view of trident production, this is a $W$-boson resonance followed by decay to a neutrino and a charged lepton. Therefore the CC and CC+NC trident cross sections are enhanced by the $W$-boson production cross section (of same incoming neutrino flavor) times the corresponding decay branching ratio, $\Gamma_{W \rightarrow \nu_\ell \ell^-}/\Gamma_W$ ($\simeq 11\%$, with slight deviation for specific flavors~\cite{Tanabashi:2018oca}). The $W$-resonance contribution keeps dominating for $s_{\nu\gamma} > m_W^2$. This is different from usual resonance features, like the Glashow resonance $\bar{\nu}_e + e^- \rightarrow W^-$ or $e^- e^+ \rightarrow Z$. The reason is that the charged lepton, $\ell^-$, could take away additional 4-momentum, keeping the $s$-channel $W$ propagator on shell ($q_W^2 = m_W^2$). Contributions from the non-resonant part and from the other three non-resonant diagrams are negligible. So for $s_{\nu\gamma} > m_W^2$, the six CC and three CC+NC channels basically form three groups due three neutrino flavors. For $s_{\nu\gamma} > 10^6$~GeV$^2$, they all converge and become constant, i.e., $\sigma \simeq 2\sqrt{2} \alpha G_F \times \Gamma_{W \rightarrow \nu_\ell \ell^-}/\Gamma_W$, which $\simeq 10^{-35}$~cm$^{-2}$.

For the NC channels, the cross sections are much smaller, as there is no resonance. The channels that have same charged lepton final states have same cross sections, independent of incoming neutrino flavor, due to the same couplings and lepton propagators.


\section{Neutrino-nucleus cross sections: coherent and diffractive regimes}
\label{sec_elastic}

In this section, we calculate the photon-mediated neutrino-nucleus cross sections, $\sigma_{\nu A}$, for $W$-boson and trident production in the coherent and diffractive regimes, which are elastic on the nucleus and nucleon, respectively. We focus on the hadronic coupling through virtual photons, as the contribution through weak bosons is highly suppressed due to their large masses. We first describe the framework (Sec.~\ref{sec_elastic_framework}), which is independent of the leptonic part. Then we calculate the $W$-boson (Sec.~\ref{sec_elastic_realW}) and trident (Sec.~\ref{sec_elastic_trident}) production processes.

The framework we use, which is from Ballett {\it et al.}~\cite{Ballett:2018uuc}, is a complete treatment of the hadronic part instead of using EPA, which is known to be not a good approximation for trident production (see Fig.~\ref{fig_previous_xsecs} and Sec.~\ref{sec_review}). Moreover, the major nuclear effect, Pauli blocking, is included (see Refs.~\cite{Czyz:1964zz, Ballett:2018uuc} for details). {\it In this work, for the first time, we show that the EPA also does not work well for $W$-boson production. } Moreover, for trident production, we calculate all 15 possible channels, including for the $\tau$ flavor, and go to TeV--PeV energies, using the full Standard Model instead of the four-Fermi theory.

\subsection{Framework}
\label{sec_elastic_framework}

Both the coherent and diffractive cross sections can be calculated using~\cite{Ballett:2018uuc}
\begin{multline}
    \frac{d^{2} \sigma_{\nu \rm X}}{d Q^{2} d \hat{s}} =
    \frac{1}{32 \pi^{2}} \frac{1}{\hat{s} Q^{2}} 
    [ h_{\rm X}^{\rm T}\left(Q^{2}, \hat{s}\right) \sigma_{\nu \gamma}^{\rm T}\left(Q^{2}, \hat{s}\right)  \\
    +  h_{\rm X}^{\rm L}\left(Q^{2}, \hat{s}\right) \sigma_{\nu \gamma}^{\rm L}\left(Q^{2}, \hat{s}\right) ] \,, 
\label{eq_sgmnuA_elastic}
\end{multline}
where X is to distinguish coherent (X = c) and diffractive (X = d) regimes, $Q^2 \equiv -q^2$ the photon virtuality, and $\hat s \equiv 2(p_1 \cdot q) = s_{\nu \gamma} + Q^2 $. Note that Eq.~(\ref{eq_sgmnuA_elastic}) decomposes the $\sigma_{\nu \rm X}$ into $2 \times 2$ parts: transverse (``T'') and longitudinal (``L''), leptonic ($\sigma_{\nu \gamma}^{\rm T/L}$) and hadronic ($h_{\rm X}^{\rm T/L}$).

The leptonic parts, $\sigma_{\nu \gamma}^{\rm T/L}(Q^2, \hat{s})$, may be viewed as the cross sections between a neutrino and an off-shell photon, and it can be calculated as
\begin{subequations}
\begin{align}
    \sigma_{\mathrm{T}} = &\, \frac{1}{2 \hat{s}} \int \frac{1}{2} \sum_{\rm spins} \left(-g^{\mu\nu} + \frac{4 Q^2}{\hat{s}^2} k_1^\mu k_1^\nu \right) \mathrm{L}_\mu \mathrm{L}^\ast_\nu \, d{\rm PS_n} \, , \\
    \sigma_{\mathrm{L}} = &\, \frac{1}{\hat{s}} \int \sum_{\rm spins} \frac{4 Q^2}{\hat{s}^2} k_1^\mu k_1^\nu \mathrm{L}_\mu \mathrm{L}^\ast_\nu \, d{\rm PS_n} \, ,
\end{align}
\label{eq_sgmnugmTL}
\end{subequations}
where $L_\mu$ is the leptonic amplitudes, details in Appendices~\ref{appdx_realW_amplitudes} and~\ref{appdx_trident_amplitudes},
and $d {\rm PS_n}$ is the phase space of the leptonic part, with $n = 2, 3$ for the $W$-boson and trident production processes, respectively . 

A factor of 1/2 appears in the first equation because a virtual photon has two transverse polarizations. The $Q^2$ dependence should also be included in both the leptonic matrix element and phase space, which are process dependent. In the limit $Q^2 = 0$, the transverse cross section is the same as the real-photon case (Eqs.~(\ref{eq_sgmnugm_realW}) and~(\ref{eq_sgmnugm_trident})), and the longitudinal cross section vanishes. 

\begin{figure}[t]
\includegraphics[width=\columnwidth]{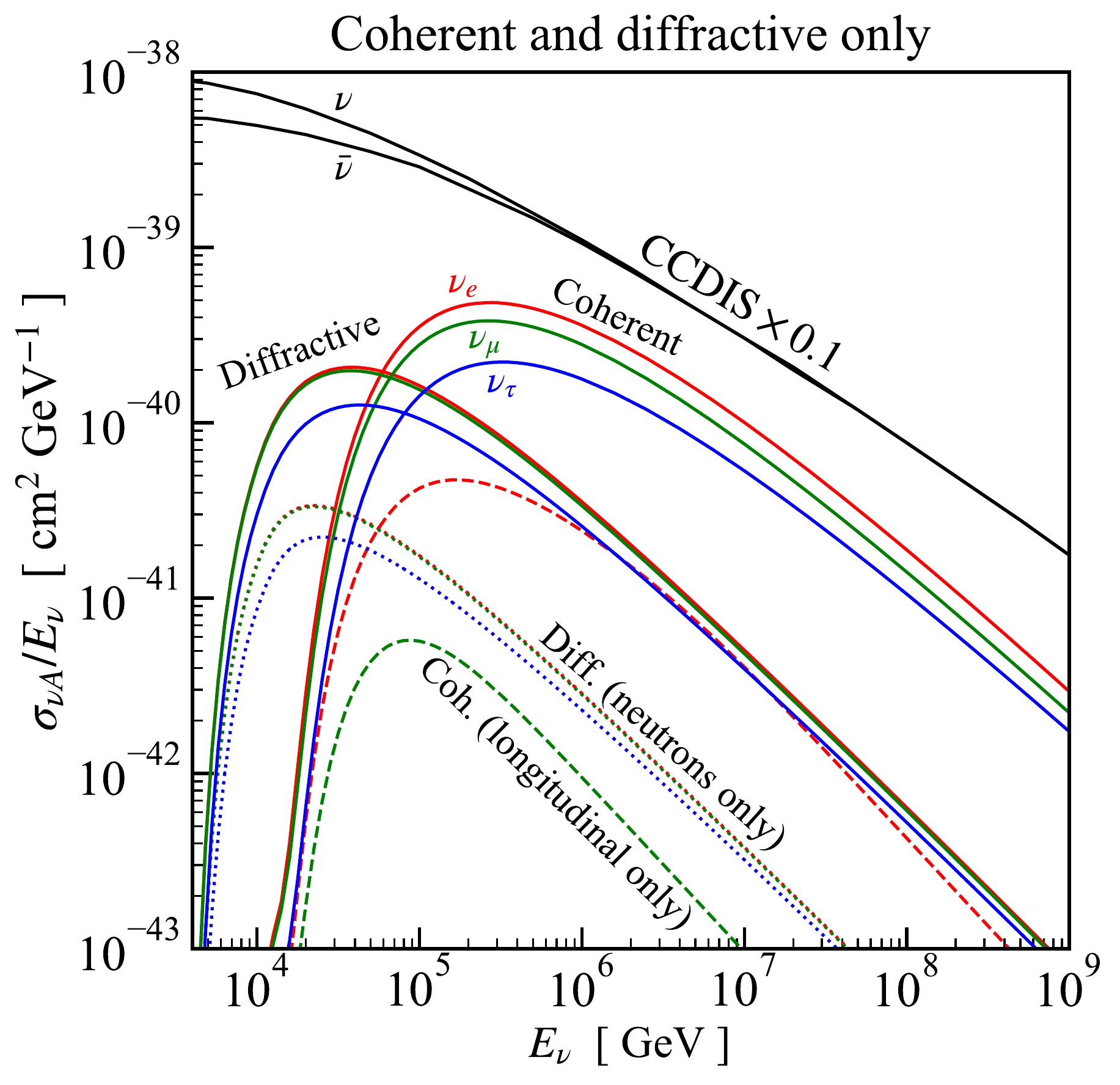}
    \caption{
        Our coherent and diffractive components of $W$-boson production cross sections, $\nu_\ell \rightarrow \ell^- + W^+$, on $\ce{^{16}O}$. {\bf Red}, {\bf green}, and {\bf blue} lines are $\nu_e$-, $\nu_\mu$-, and $\nu_\tau$-induced channels, respectively. {\bf Solid}: coherent (right bump) and diffractive (left bump) components. {\bf Dashed}: longitudinal contribution to the coherent regime, which is small, even for the largest case ($\nu_e$). The $\nu_\tau$ line is not shown due to being below the bound of the y axis. {\bf Dotted}: contribution from neutrons to the diffractive regime, which is small. The corresponding antineutrino cross sections are the same.
    }
\label{fig_sgmnuA_realW_Coh_Diff}
\end{figure}

\begin{figure}[t]
\vspace{0cm}
\includegraphics[width=\columnwidth]{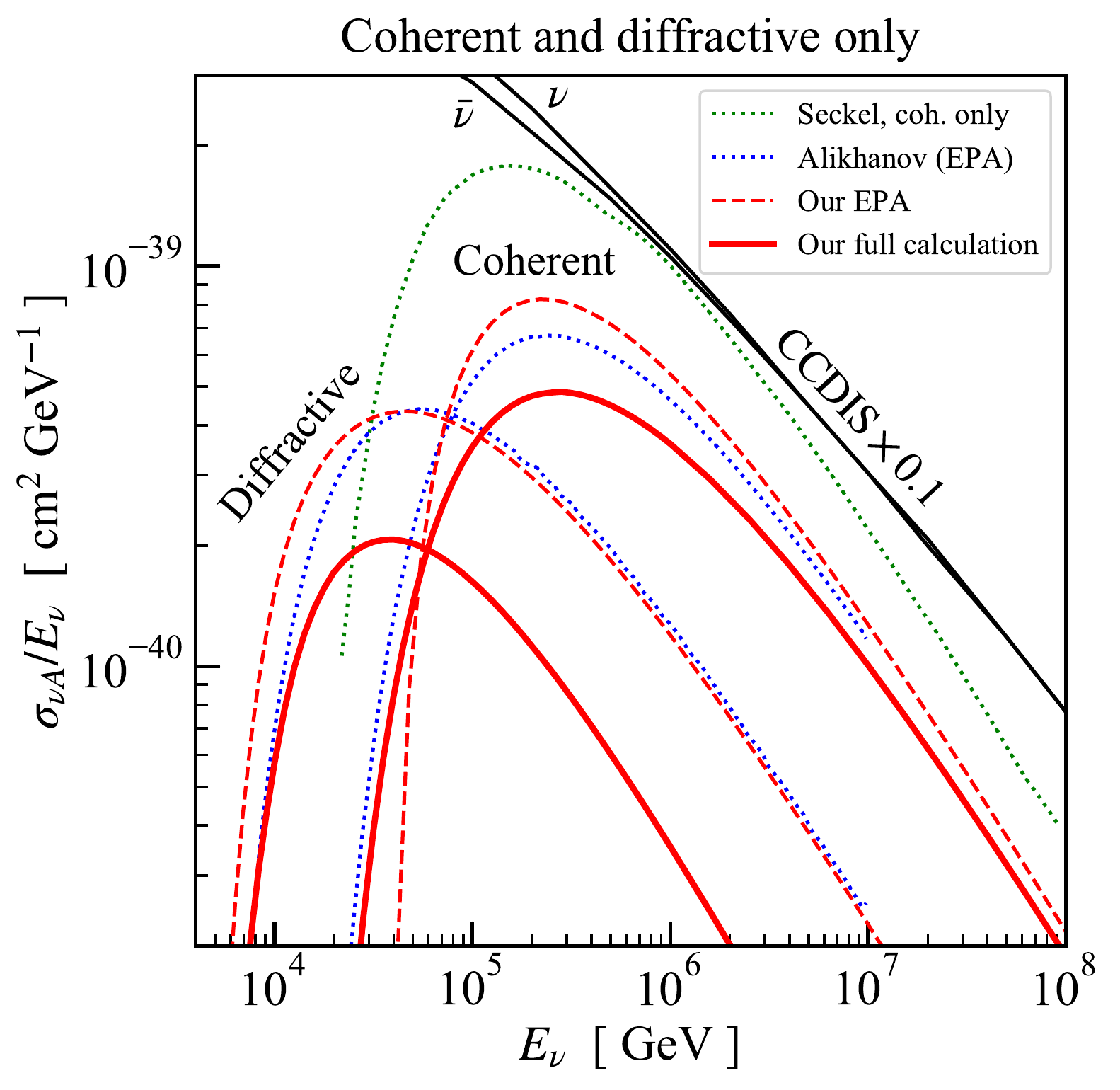}

    \caption{
        Our coherent and diffractive components of $W$-boson production cross sections ({\bf red solid}, from Fig.~\ref{fig_sgmnuA_realW_Coh_Diff} but thicker), for the example of $\nu_e \rightarrow \ell^- W^+$ on $\ce{^{16}O}$ (the flavor with the largest cross section), comparing with our ``EPA + no Pauli blocking'' results ({\bf dashed}) and previous calculations ({\bf dotted}) by Seckel~\cite{Seckel:1997kk} and Alikhanov~\cite{Alikhanov:2014uja, Alikhanov:2015kla}. Left and right bumps are coherent and diffractive components, respectively. Note our results are substantially smaller, which is important.
}
\label{fig_sgmnuA_realW_Coh_Diff_Cmpr}
\end{figure}

{\it The hadronic parts, $h_X^{T/L}(Q^2, \hat{s})$ are dimensionless factors that involve the nuclear/nucleon form factors.} For the coherent regime, we use the Woods-Saxon (nuclear) form factor. For the diffractive regime, we use the nucleon form factors that have a dipole parametrization. More details are given in Ref.~\cite{Ballett:2018uuc}. 

For the diffractive regime, in addition, the Pauli-blocking effects are included by multiplying Eq.~(\ref{eq_sgmnuA_elastic}) by a factor derived from modeling the nucleus as ideal (global) Fermi gas of protons and neutrons with equal density, which is (derived by Ref.~\cite{Bell:1996ms} and used by Refs.~\cite{Lovseth:1971vv, Ballett:2018uuc, Altmannshofer:2019zhy}),
\begin{equation}
f (|\vec{q}|) = \begin{cases} \displaystyle
                    \frac{3}{2} \frac{|\vec{q}|}{2 \, k_F} - \frac{1}{2} \left( \frac{|\vec{q}|}{2 \, k_F} \right)^3 ,\, &\mathrm{if }\;\; |\vec{q}| < 2\, k_F\, ,\\
                    1,\, &\mathrm{if }\;\; |\vec{q}| \geq 2 \, k_F\, ,
                \end{cases}
\end{equation}
where $k_F = 235$~MeV is the Fermi momentum of the gas, which sets the kinetic boundary for the final states, and $|\vec{q}|$ is the magnitude of the transferred 3-momentum in the lab frame, which can be derived to be, for the virtual photon case, $\sqrt{ (Q^2/2M_N)^2 + Q^2}$, where $M_N$ is the mass of the nucleon. This reduces the diffractive cross section by about $50\%$ for protons and $20\%$ for neutrons.

The EPA formalism can be obtained by setting $Q^2=0$ in $\sigma_{\nu \gamma}^{\rm T/L}(Q^2, \hat{s})$ of Eq.~(\ref{eq_sgmnuA_elastic}). This is basically the same as that initially derived by Ref.~\cite{Belusevic:1987cw} and later used by Refs.~\cite{Altmannshofer:2014pba, Magill:2016hgc}. Below, we also show the ``EPA + no Pauli blocking'' results for comparison.

As a validation of our understanding of the formalism, we reproduced the cross section results of Ballett {\it et al.}~\cite{Ballett:2018uuc} using the four-Fermi theory and other same input. Our calculations agree with theirs to within a few percent, with the remaining differences due to numerical precision. (Note that we decompose the phase-space (Appendix~\ref{appdx_trident}) in a different, but equivalent, way from them~\cite{Czyz:1964zz, Ballett:2018uuc}.)

\subsection{$W$-boson production}
\label{sec_elastic_realW}

\subsubsection{Off-shell cross sections, $\sigma^{T/L}_{\nu \gamma}(\hat{s}, Q^2)$ }
\label{sec_elastic_realW_sgmTL}

The process is the same as Eq.~(\ref{eq_realW_nugm}), but replacing the real photon, $\gamma$, by a virtual photon $\gamma^\ast$. We calculate the off-shell cross sections, $\sigma^{T/L}_{\nu \gamma}(\hat{s}, Q^2)$, in the CM frame (consistent with Sec.~\ref{sec_sgmnugamma_realW}), using Eq.~(\ref{eq_sgmnugmTL}). For the leptonic matrix element, the photon virtuality can be included by writing
\begin{subequations}
\begin{align}
    k_1 = \left( \frac{s_{\nu \gamma}+Q^2}{2\sqrt{s_{\nu \gamma}}}, 0, 0, \frac{s_{\nu \gamma}+Q^2}{2\sqrt{s_{\nu \gamma}}} \right)  \, , \\
    q = \left(\frac{s_{\nu \gamma}-Q^2}{2\sqrt{s_{\nu \gamma}}}, 0, 0, -\frac{s_{\nu \gamma}+Q^2}{2\sqrt{s_{\nu \gamma}}} \right)  \, .
\end{align}
\end{subequations}
The 4-momenta of the outgoing particles do not have $Q^2$ dependence, and the phase space is the same as that in Eq.~(\ref{eq_sgmnugm_realW}). When $Q^2=0$, all results return to the real-photon case (Sec.~\ref{sec_sgmnugamma_realW}).

The major features of $\sigma^{T/L}_{\nu \gamma}(\hat{s}, Q^2)$ are the following. First, $\sigma^T_{\nu \gamma}(\hat{s}, Q^2)$ decreases with $Q^2$, especially when $Q^2 \gtrsim m_\ell^2$, because $Q$ enters the denominator of the lepton propagators which suppresses the cross section~\cite{Ballett:2018uuc}. Second, $\sigma^L_{\nu \gamma}(\hat{s}, Q^2)$ increases with $Q^2$, due to the factor $4 Q^2 / \hat{s}^2$, then becomes flat when $Q^2 \gtrsim m_\ell^2$. Third, when $Q^2$ is nearing $\hat{s}-(m_W+m_\ell)^2$, an exponential cutoff happens in both $\sigma^T_{\nu \gamma}$ and $\sigma^L_{\nu \gamma}$, due to running out of phase space ($s_{\nu\gamma} \equiv \hat{s} - Q^2 < (m_W+m_\ell)^2$).

\subsubsection{$\sigma_{\nu A}$ and discussion}
\label{sec_elastic_realW_result}

The coherent ($\sigma_{\nu \rm c}$) and diffractive ($\sigma_{\nu \rm d}$) cross sections are then calculated with Eq.~(\ref{eq_sgmnuA_elastic}), by convolving the leptonic parts with the hadronic parts.

Figure~\ref{fig_sgmnuA_realW_Coh_Diff} shows the cross sections. The features discussed in Sec.~\ref{sec_sgmnugamma_realW} mostly appear here (e.g., $\sigma_{\nu_e A} > \sigma_{\nu_\mu A} > \sigma_{\nu_\tau A}$). The threshold here is effectively set by $E_\nu \sim m_W^2/2Q_{\rm max}^{\rm eff}$, where $Q_{\rm max}^{\rm eff}$ is the effectively maximum $Q$ of the form factors, which is $\sim 0.1$~GeV for $\ce{^16}\rm O$ (coherent) and $\sim 1$~GeV for nucleons (diffractive). Similarly, the typical momentum transfer, $Q$, for each $E_\nu$, is between $\sim m_W^2/2E_\nu$ and $Q_{\rm max}^{\rm eff}$ for both regimes. The sharp peak in $\sigma_{\nu \gamma}$ (Fig.~\ref{fig_sigma_nugm_all}) is here smeared due to convolving with the form factors. The coherent cross sections ($\propto Z^2$) are larger than diffractive ones ($\propto Z$), which is similar to Fig.~\ref{fig_previous_xsecs}, which uses EPA. For both regimes, the transverse part dominates, while the longitudinal part, as shown on the figure, is suppressed by $\sim Q^2/\hat{s}$. For the diffractive regime, the contribution from protons dominates, due to its electric form factor.

\begin{figure}[t!]
\includegraphics[width=\columnwidth]{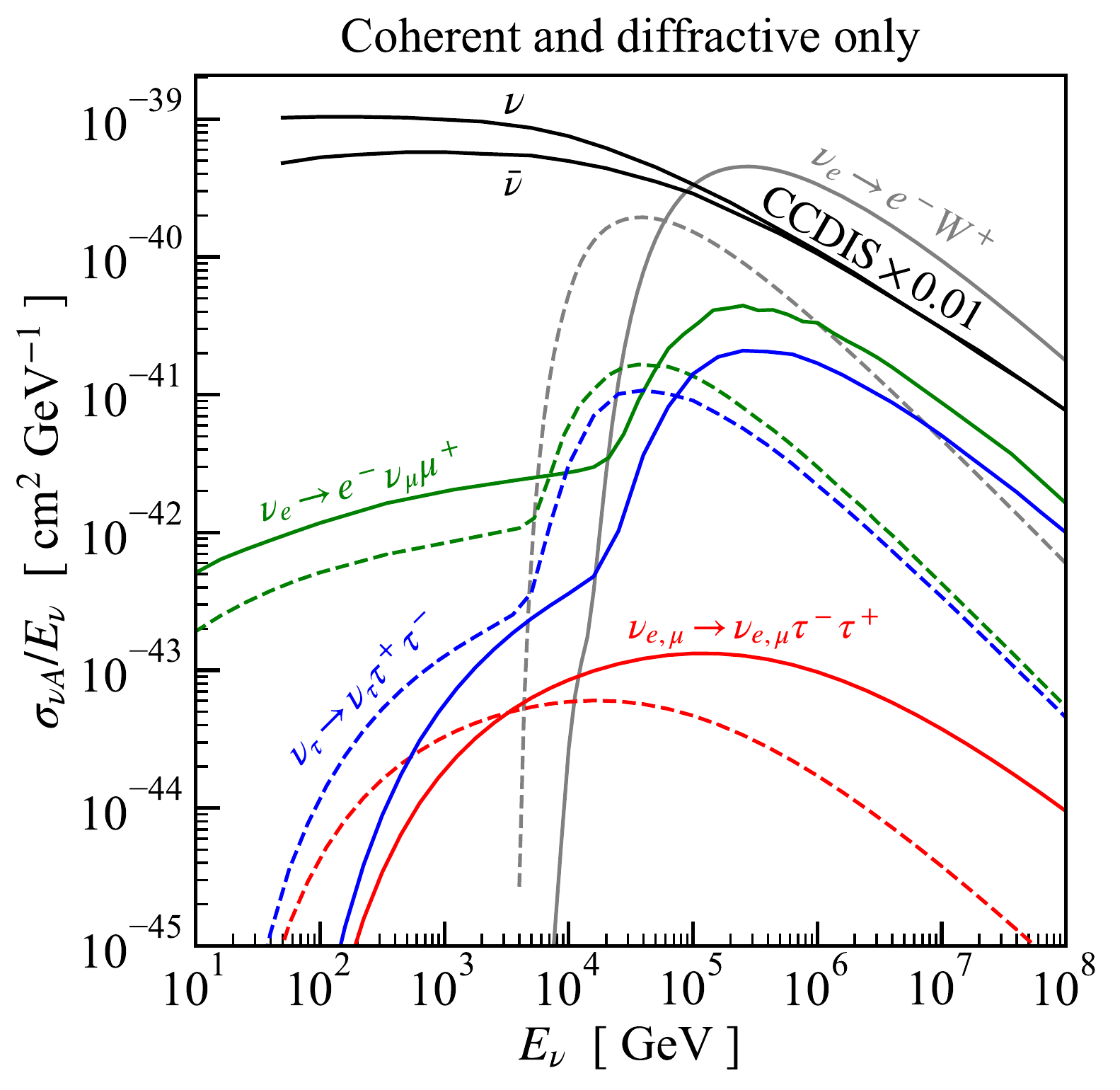}
    \caption{Our coherent ({\bf solid} lines) and diffractive ({\bf dashed} lines) components of trident production cross sections on $\ce{^16}\rm O$. We show one typical channel for each category, i.e., CC, NC and CC+NC, to make the figure simple. For all the channels, see Appendix~\ref{Sec_sgmnuA_trident_Coh_Diff_all}. {\bf Gray} lines are for $\nu_e$-induced $W$-boson production from Fig.~\ref{fig_sgmnuA_realW_Coh_Diff}, shown as a comparison. The corresponding antineutrino cross sections (i.e., obtained by CP-transforming the processes shown) are the same.
}
\label{fig_sgmnuA_trident_Coh_Diff_typical}
\end{figure}

We tested the sensitivity of our cross section results to the choices of form factors.  For the coherent component, we also tried using a Gaussian form factor, $e^{-Q^2/2a^2}$, which is sometimes used for lighter nuclei.  For $\ce{^16}\rm O$, we find that the cross section is changed by $\lesssim 15\%$ for $E_\nu \sim 10^5$~GeV (where the inelastic component dominates anyway) and by $\lesssim 5\%$ for $E_\nu \sim 10^6$~GeV.  For the diffractive component, we explored changing the vector mass in the form factor.  For any reasonable change, the effect on our calculated cross sections is negligible.

Figure~\ref{fig_sgmnuA_realW_Coh_Diff_Cmpr} compares our results with previous ones from Ref.~\cite{Seckel:1997kk} (only the coherent regime was considered) and Refs.~\cite{Alikhanov:2014uja, Alikhanov:2015kla} (EPA was used, and Pauli-blocking effect was not included for the diffractive component), and our ``EPA + no Pauli blocking'' results discussed above. Comparing to Refs.~\cite{Alikhanov:2014uja, Alikhanov:2015kla}, which uses a different EPA formalism, our EPA result is close. Surprisingly, the result from the full calculation is only about half as large, for both coherent and diffractive regimes. {\it This means that the EPA is still not valid at even such high-energy scales.} The reason is that, as discussed in Sec.~\ref{sec_elastic_realW_sgmTL}, the nonzero $Q^2$ suppresses the transverse cross section, $\sigma^T_{\nu \gamma}(\hat{s}, Q^2)$, compared to $\sigma^T_{\nu \gamma}(\hat{s}, Q^2=0)$ used in the EPA. The larger $Q^2$, the larger the suppression. Physically, this is because, although the incoming neutrino is very energetic, the charged particle that directly couples to the photon may not be. The difference between the full calculation and the EPA in the diffractive regime is larger than that in coherent regime is because the nucleon form factor probes larger $Q^2$ than the nuclear form factor, also because the Pauli-blocking effect suppresses the diffractive cross section. Another feature is that, for a specific $Q^2$, the larger the charged-lepton mass, the smaller the suppression, which is small when $Q^2 \lesssim m_\ell^2$. So the difference between full calculation and EPA is smaller for the muon and tau flavors.

\subsection{Trident production}
\label{sec_elastic_trident}

\subsubsection{Off-shell cross sections, $\sigma^{T/L}_{\nu \gamma}(\hat{s}, Q^2)$ }
\label{sec_elastic_trident_sgmTL}

The processes are the same as in Eq.~(\ref{eq_trident_details}), but again replacing the real photon, $\gamma$, by a virtual photon $\gamma^\ast$. Same as above, we work in the CM frame, and both the phase space term $d {\rm PS_3}$ and the leptonic matrix element are modified due to nonzero photon virtuality.

The leptonic matrix element is modified due to the modification of the 4-momenta, the details of which can be found in Appendix~\ref{appdx_trident}.

The phase space integration can be done by decomposing the three-body phase space into two two-body phase spaces~\cite{Murayama_PS, Magill:2016hgc}. The result is the same as Eq.~(\ref{eq_PS3}), but replacing $s_{\nu\gamma}$ by $\hat{s}$. The integration range is now $( (m_1+m_2)^2, \hat{s}-Q^2 )$ for $l$, and 
\begin{equation}
    \left[ l+Q^2, \hat{s}-Q^2 + \left(2 - \frac{l}{\hat{s}-Q^2} \right) Q^2 \right]
\end{equation}
for $t$. See Appendix~\ref{appdx_trident} for details.

The major features due to the nonzero $Q^2$ is the same as those of $W$-boson production (Sec.~\ref{sec_elastic_realW_sgmTL}).

\subsubsection{$\sigma_{\nu A}$ and discussion}
\label{sec_elastic_trident_result}

Figure~\ref{fig_sgmnuA_trident_Coh_Diff_typical} shows the cross sections for the typical channels (for all channels, see Appendix~\ref{Sec_sgmnuA_trident_Coh_Diff_all}). We start from $E_\nu = 10$~GeV, as below this energy, the cross sections have been shown in Refs.~\cite{Ballett:2018uuc, Altmannshofer:2019zhy}. Our results agree with theirs. Same as before, the threshold here is effectively set by $E_\nu \sim (m_1 + m_2)^2/2Q_{\rm max}^{\rm eff}$, which is $\sim 0.1$~GeV for $\ce{^16}\rm O$ and $\sim 1$~GeV for nucleons. The sharp peak in $\sigma_{\nu \gamma}$ (Fig.~\ref{fig_sigma_nugm_all}) is smeared here due to convolving with the form factors. Other features and the physics are the same as those discussed in Sec.~\ref{sec_sgmnugamma_trident} for Fig.~\ref{fig_sigma_nugm_all} and in Sec.~\ref{sec_elastic_realW_result} for Fig.~\ref{fig_sgmnuA_realW_Coh_Diff}.


\section{Neutrino-nucleus cross sections: inelastic regime}
\label{sec_inelastic}

In this section, we calculate the neutrino-nucleus cross sections, $\sigma_{\nu A}$, for $W$-boson and trident production in the inelastic regime, in which the partons of nucleons are probed. 

\subsection{Framework}
\label{sec_inelastic_overview}

The inelastic regime has two contributions, photon-initiated subprocess and quark-initiated subprocess~\cite{Schmidt:2015zda}.

The photon-initiated subprocess is that the hadronic coupling is through a virtual photon, which is similar to Sec.~\ref{sec_elastic}, but with larger photon virtuality $Q^2$. Calculation of this subprocess involves the photon PDF, which describes the photon content of the nucleon (e.g., Refs.~\cite{Gluck:2002fi, Martin:2004dh, Ball:2013hta, Martin:2014nqa, Schmidt:2015zda, Harland-Lang:2016kog, Manohar:2016nzj, Manohar:2017eqh, Bertone:2017bme}). The photon PDF consists of elastic and inelastic components. The elastic component corresponds to the diffractive regime of Sec.~\ref{sec_elastic}, and can be calculated from the nucleon electromagnetic form factors. The inelastic photon PDF consists of nonperturbative and perturbative parts, and the resonance region is included in the former~\cite{Schmidt:2015zda, Manohar:2016nzj, Manohar:2017eqh}. For the $W$-boson production, this component was calculated by Alikhanov~\cite{Alikhanov:2015kla}. {\it For trident, this has never been considered.}

The quark-initiated subprocess is that a quark of a nucleon is explicitly involved as an initial state of the scattering process. The propagator that couples to a quark can be photon, $W$ or $Z$ boson. For $W$-boson production, this was mentioned in Alikhanov~\cite{Alikhanov:2015kla} (diagrams were also shown in its Fig.~8, plus another one from replacing the $Z$ by a photon in the upper middle diagram) but not calculated. For trident production, this was calculated in Ref.~\cite{Magill:2016hgc}.

Those two subprocesses are at the same order though they may not seem to be, as the photon propagator to quark has an additional $\alpha_{\rm EM}$. The reason is that the photon PDF has a factor of $\alpha_{\rm EM}$ implicitly~\cite{Schmidt:2015zda}.

A double-counting problem occurs if summing up the two subprocesses for the total inelastic cross section. This is because the contribution from the photon propagator to the quark of the quark-initiated subprocess is already included in the inelastic photon PDF (perturbative part) of the photon-initiated subprocess. We deal with this problem below.

For the PDF set, we use {\tt CT14qed}~\cite{Schmidt:2015zda, CT14web}, which provides the inelastic photon, quark, and gluon PDFs self-consistently. The inelastic photon PDF of {\tt CT14qed} is modeled as emission from the quarks using quark PDFs and further constrained by comparing with ZEUS data on the DIS process $e p \rightarrow e \gamma + X$~\cite{Chekanov:2009dq}. The quark PDFs are obtained by the usual method and constrained by DIS and other data.

The reasons that we choose {\tt CT14qed} are the following. First, it is the only PDF set that provides the inelastic component of photon PDF only. For the elastic part, we do not use the elastic photon PDF, which is obtained using EPA and does not include the neutron magnetic component form factor, as the treatment in Sec.~\ref{sec_elastic} (diffractive regime) is better. Second, it is also the only PDF set that provides the inelastic photon PDF for both proton and neutron. Finally, though the uncertainty of photon PDF is larger than the later ones by {\tt LUXqeD}~\cite{Manohar:2016nzj, Manohar:2017eqh} and {\tt NNPDF31luxQED}~\cite{Bertone:2017bme}, the central value is very close.

We use {\tt MadGraph (v2.6.4)}~\cite{Alwall:2014hca} to do the calculation, which handles the PDFs and hard processes systematically. We remove kinematic cuts to get the total cross section of both processes. The model we choose in {\tt MadGraph} is ``{\tt sm-lepton\_masses}'', which includes the masses of charged leptons, while the default ``{\tt sm}'' does not. Moreover, this model uses diag(1, 1, 1) for the CKM matrix and ignores the masses of u, d, s, c quarks, which are good approximations for us. Note that for the initial-state neutrinos, which have only a left-handed chirality, we need ``{\tt set polbeam1 = -100}'' (+100 for antineutrinos) to fully polarize the beam, otherwise the cross section  will be mistakenly halved. As a check of above configuration, we calculate the neutrino CCDIS cross sections and the result is consistent with Refs.~\cite{Gandhi:1998ri, Connolly:2011vc, CooperSarkar:2011pa, Chen:2013dza, Block:2014kza, Arguelles:2015wba, Bertone:2018dse} within uncertainties.

\subsection{$W$-boson production}
\label{sec_inelastic_realW}

\begin{figure}[t!]
    \includegraphics[width=\columnwidth]{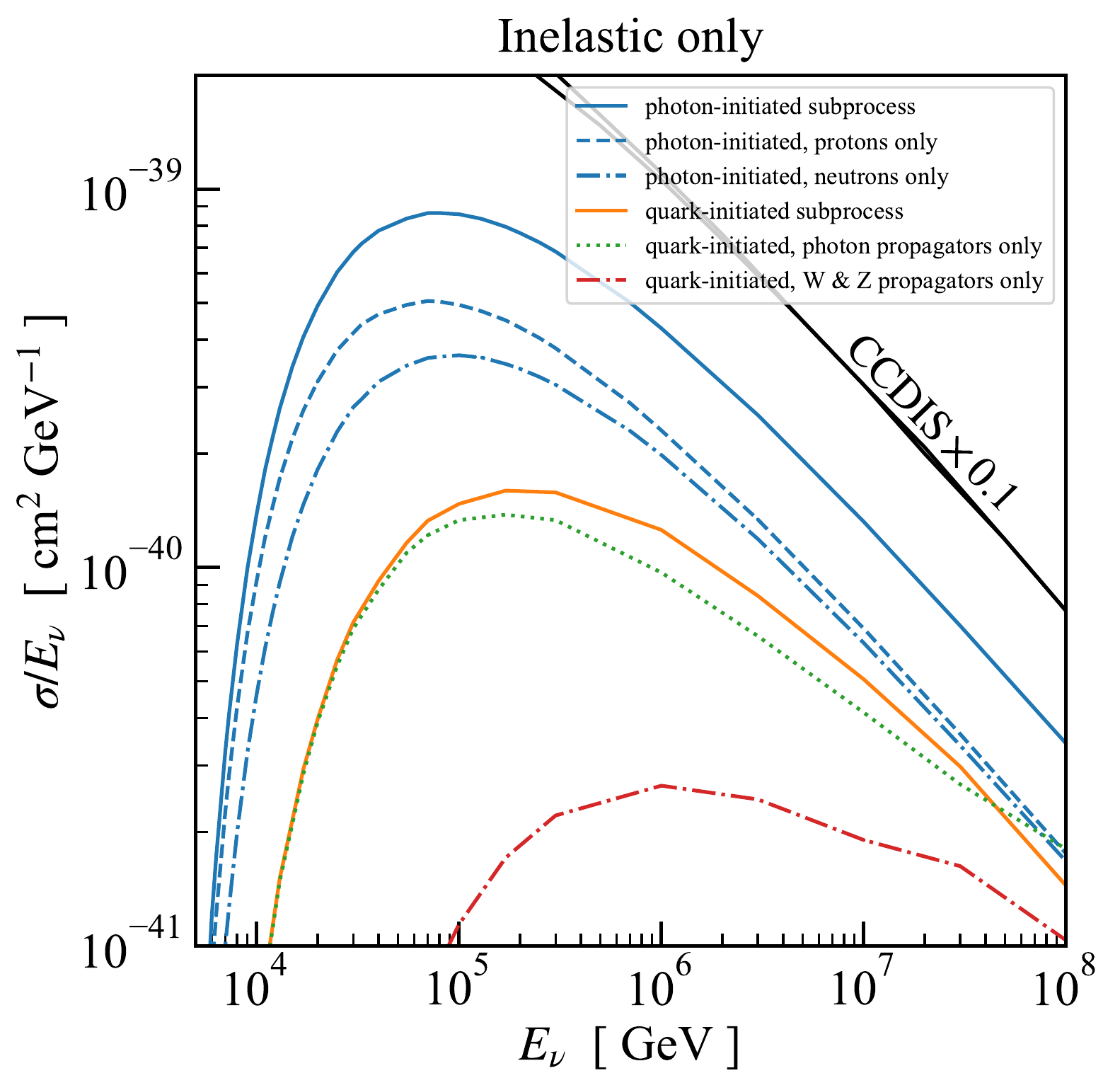}
    \caption{Different components of our inelastic neutrino-nucleus cross sections for $W$-boson production. Only $\nu_e$ is shown to keep the figure simple. For $\nu_\mu$ and $\nu_\tau$, the photon-initiated cross sections are smaller (Fig.~\ref{fig_sgmnuA_realW_inel_all}), while the quark-initiated cross sections are basically the same. See text for details.
    }
\label{fig_sgmnuA_realW_inel_nue}
\end{figure}

\begin{figure}[t!]
    \includegraphics[width=\columnwidth]{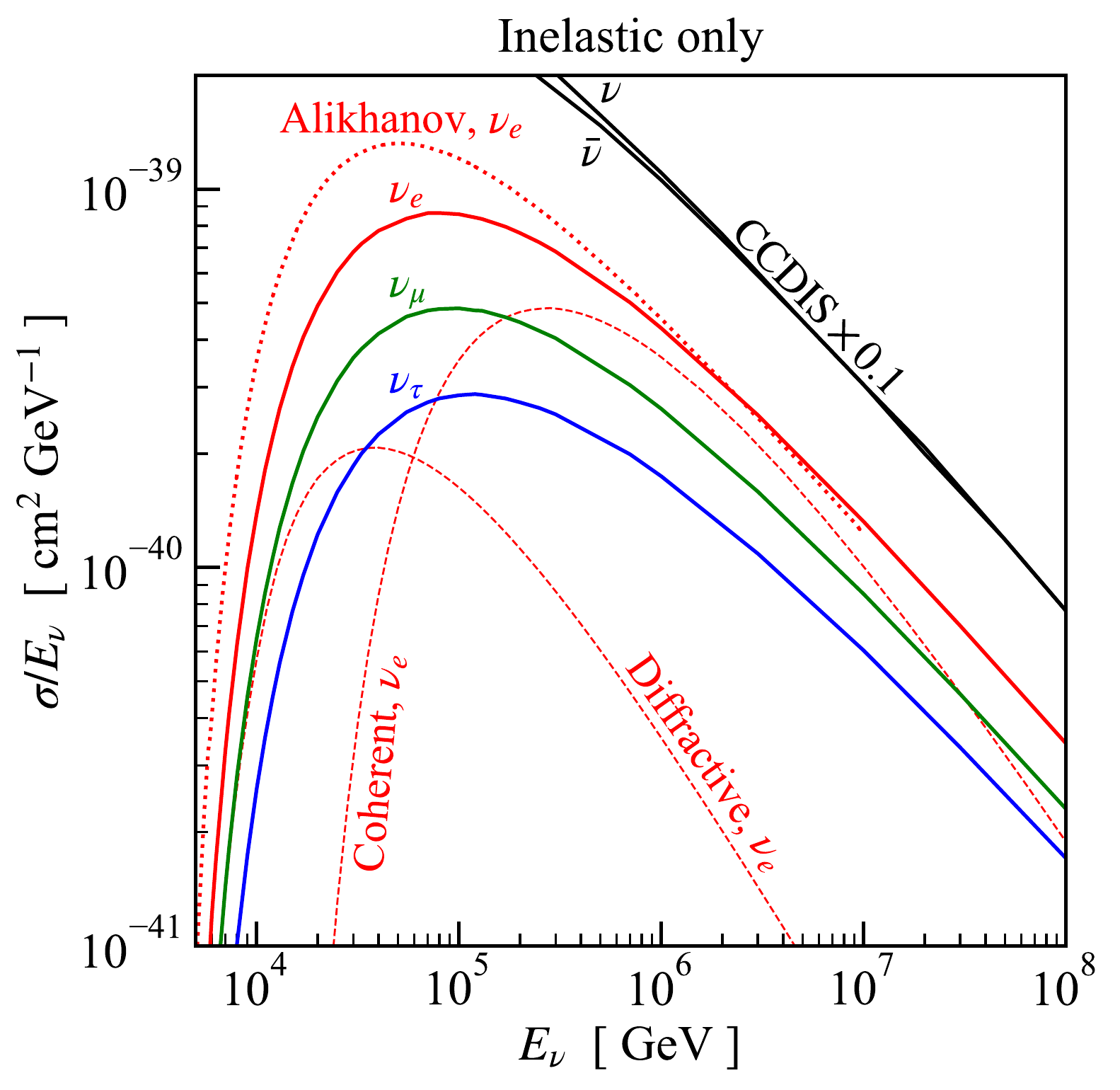}
    \caption{
        Our inelastic neutrino-nucleus cross sections for $W$-boson production on $\ce{^16}\rm O$ (solid lines), for all three flavors. Also shown are previous results from Alikhanov~\cite{Alikhanov:2015kla} and, for comparison, coherent and diffractive cross sections of $\nu_e$ from Fig.~\ref{fig_sgmnuA_realW_Coh_Diff}.
    The corresponding antineutrino cross sections are the same.
}
\label{fig_sgmnuA_realW_inel_all}
\end{figure}

For photon-initiated process, the diagrams are shown in Fig.~\ref{fig_real_W_diagrams}. The factorization and renormalization scales are chosen to be $\sqrt{s_{\nu \gamma}}$. Our choice is consistent and has no ambiguity for both diagrams of Fig.~\ref{fig_real_W_diagrams} compared to $\sqrt{-(k_1 - p_1)^2}$ (motivated by the first diagram) or $\sqrt{-(k_1 - p_2)^2}$ (motivated by the second diagram). The result is only $\simeq 10\%$ larger than that using the default factorization and renormalization scales of {\tt MadGraph}. Changing both scales to $2\sqrt{s_{\nu \gamma}}$ or $\sqrt{s_{\nu \gamma}}/2$ would increase or decrease the cross section by $\sim 15\%$. For quark-initiated subprocess, the diagrams can be found in Fig.~8 of Ref.~\cite{Alikhanov:2015kla}, plus another one from replacing the $Z$ by a photon in its upper middle diagram. 
We use the default factorization and renormalization scales of MadGraph, as using $\sqrt{s_{\nu, \rm quark}}$ causes calculational problems. 

\begin{figure*}[t!]
    \includegraphics[width=\columnwidth]{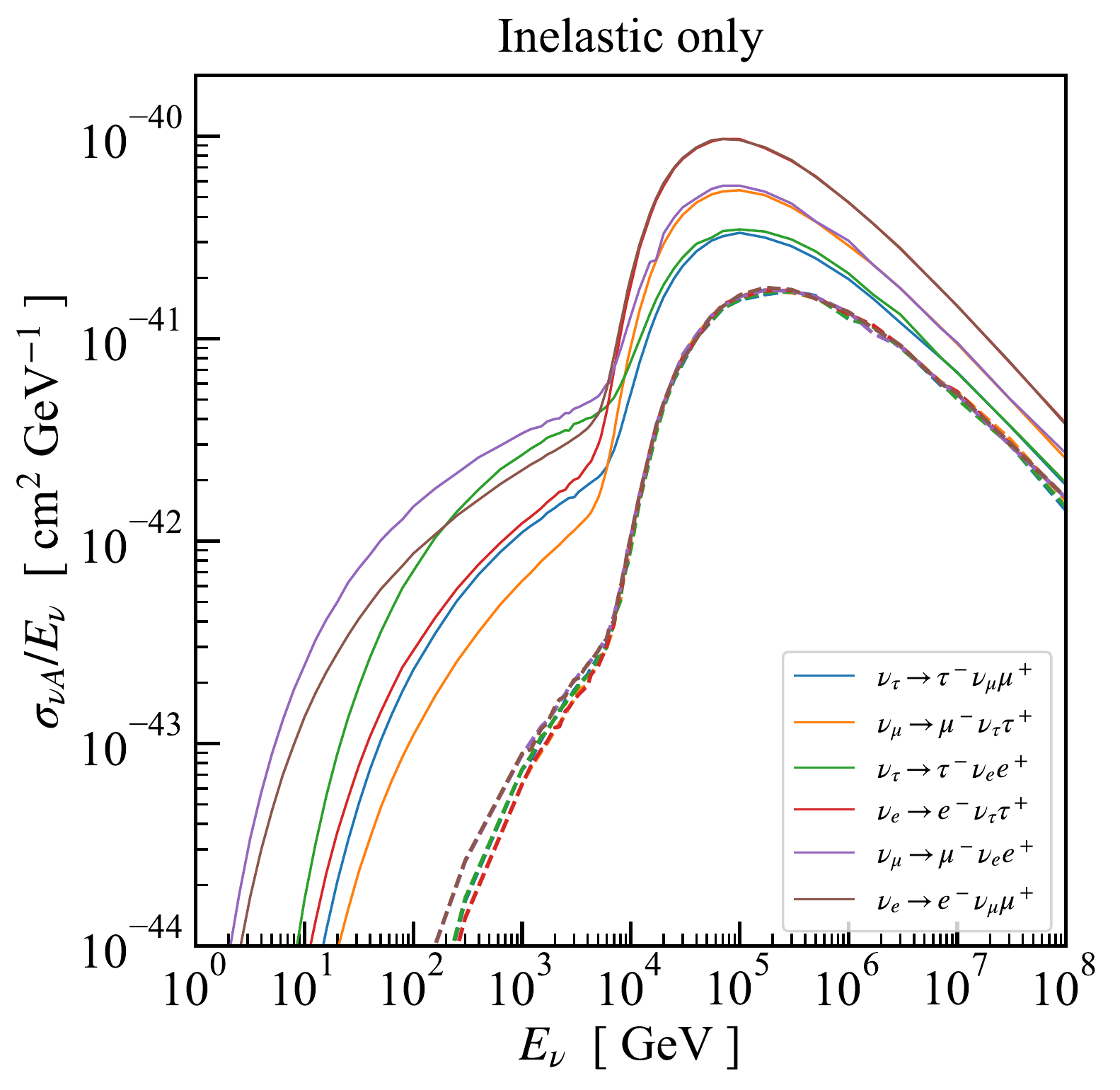}
    \includegraphics[width=\columnwidth]{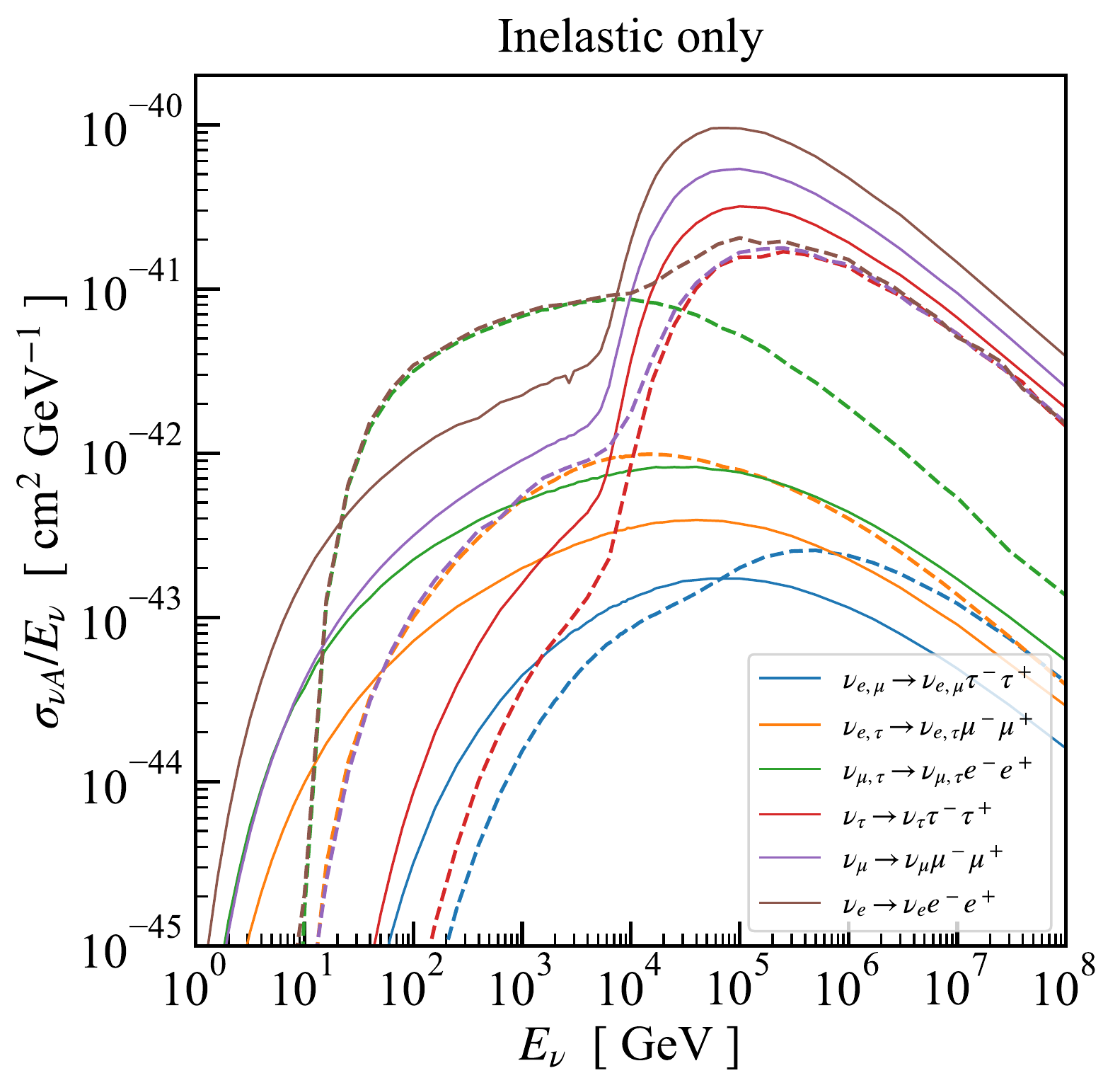}
    \caption{ Our cross sections for trident production in the inelastic regime. {\bf Left}: CC channels. {\bf Right}: CC+NC and NC channels. {\bf Solid}: the photon-initiated subprocess. {\bf Dashed}: quark-initiated subprocess. The corresponding antineutrino cross sections are the same.
}
\label{fig_sgmnuA_trident_inel}
\end{figure*}

Figure~\ref{fig_sgmnuA_realW_inel_nue} shows the results for $\nu_e$. (For $\nu_\mu$ and $\nu_\tau$, the discussion below also applies.) The photon-initiated subprocess is much larger than the quark-initiated process, because softer photons are favored (photons in the low-$Q^2$ region). The contribution from protons is larger than that from neutrons, and their ratio is similar for both subprocesses. For the quark-initiated subprocess, also shown are the contributions from $|\text{photon-propagator diagrams}|^2$ and from the $|W/Z\text{-propagator\ diagrams}|^2$, with the former being much larger than the latter. This indicates the relative importance, though the calculation of each component separately would break gauge invariance, especially above $E_\nu \sim 10^8$~GeV where it is not numerically stable and the mixings between photon and weak bosons are large. However, because the photon-propagator diagrams dominate the quark-initiated subprocess and, as mentioned in the last subsection, are already included in the photon-initiated process (inelastic photon PDF), we can ignore the quark-initiated subprocess in our calculation. (In other words, as long as the solid lines in Fig.~\ref{fig_sgmnuA_realW_inel_all} are much larger than the dot-dashed line in Fig.~\ref{fig_sgmnuA_realW_inel_nue}, the quark-initiated process can be ignored.)
This also avoids the double-counting problem mentioned above. A more complete treatment that includes both subprocesses while avoiding double counting is beyond the scope of this work. One way is to use the $W$ and $Z$ PDFs~\cite{Bauer:2017isx, Bauer:2018arx, Fornal:2018znf}, which may appear in future PDF sets.

Figure~\ref{fig_sgmnuA_realW_inel_all} shows the inelastic cross sections for all three flavors. As before, $\sigma_{\nu_e A} > \sigma_{\nu_\mu A} > \sigma_{\nu_\tau A}$. Also shown are $\nu_e$ coherent and diffractive cross sections, for comparison. The inelastic cross section is the largest for most energies. Our result is smaller than the result of Alikhanov~\cite{Alikhanov:2015kla}. The difference is due to multiple reasons. First, we use much more up-to-date photon PDFs. Second, we use the dynamic scale $\sqrt{s_{\nu \gamma}}$, which is more appropriate, while they used the fixed scale, $m_W$.  Third, we use {\tt MadGraph} which does a full systematic calculation while they used the EPA.

\subsection{Trident production}
\label{sec_inelastic_trident}

Figure~\ref{fig_sgmnuA_trident_inel} shows the cross sections for trident production processes in the inelastic regime. The calculational choice is the same as that for $W$-boson production in the last subsection. Different from before, here we separate CC channels (left) and CC+NC, NC channels (right) into two different panels. {\it Our calculation is the first to include the nonperturbative part of the photon PDF for the trident production in which the resonance region is included}. So here we start from $E_\nu = 1$~GeV. For CC channels, similar to those of $W$-boson production, the photon-initiated subprocess is much larger than the quark-initiated subprocess, and the latter is dominated by photon propagators to the quarks (not shown, for simplicity), which is already included by the former. Therefore, for the total inelastic cross section, we can also ignore the quark-initiated subprocess.

For CC+NC and NC channels, interestingly, the quark-initiated subprocess could be (much) larger than the photon-initiated subprocess.  The reason is that a pair of charged leptons can be split by a virtual photon emitted from the initial- or final-state quark on top of the NC DIS process, which does not happen in the CC channels as the two charged leptons have different flavors. (Radiative corrections through virtual $W$/$Z$ bosons also exist but are suppressed by their large masses.) Therefore, the quark-initiated subprocesses of CC+NC and NC channels are enhanced compared to those of CC channels, and the lighter the final-state charged leptons, the larger the enhancement. For the total inelastic cross section of CC+NC and NC channels, we can sum the two subprocesses up, as the contribution from the double-counting region is negligible.

Our result is different from the DIS cross sections calculated in Ref.~\cite{Magill:2016hgc}. With limited details provided by Ref.~\cite{Magill:2016hgc}, it is hard to trace the exact reason. In addition to the quark PDFs considered by Ref.~\cite{Magill:2016hgc}, we also consider the photon PDFs, which has the nonperturbative part. For the quark-initiated process, we may include more diagrams, including the radiative-correction diagrams mentioned above.


\begin{figure*}[t!]
    \includegraphics[width=\textwidth]{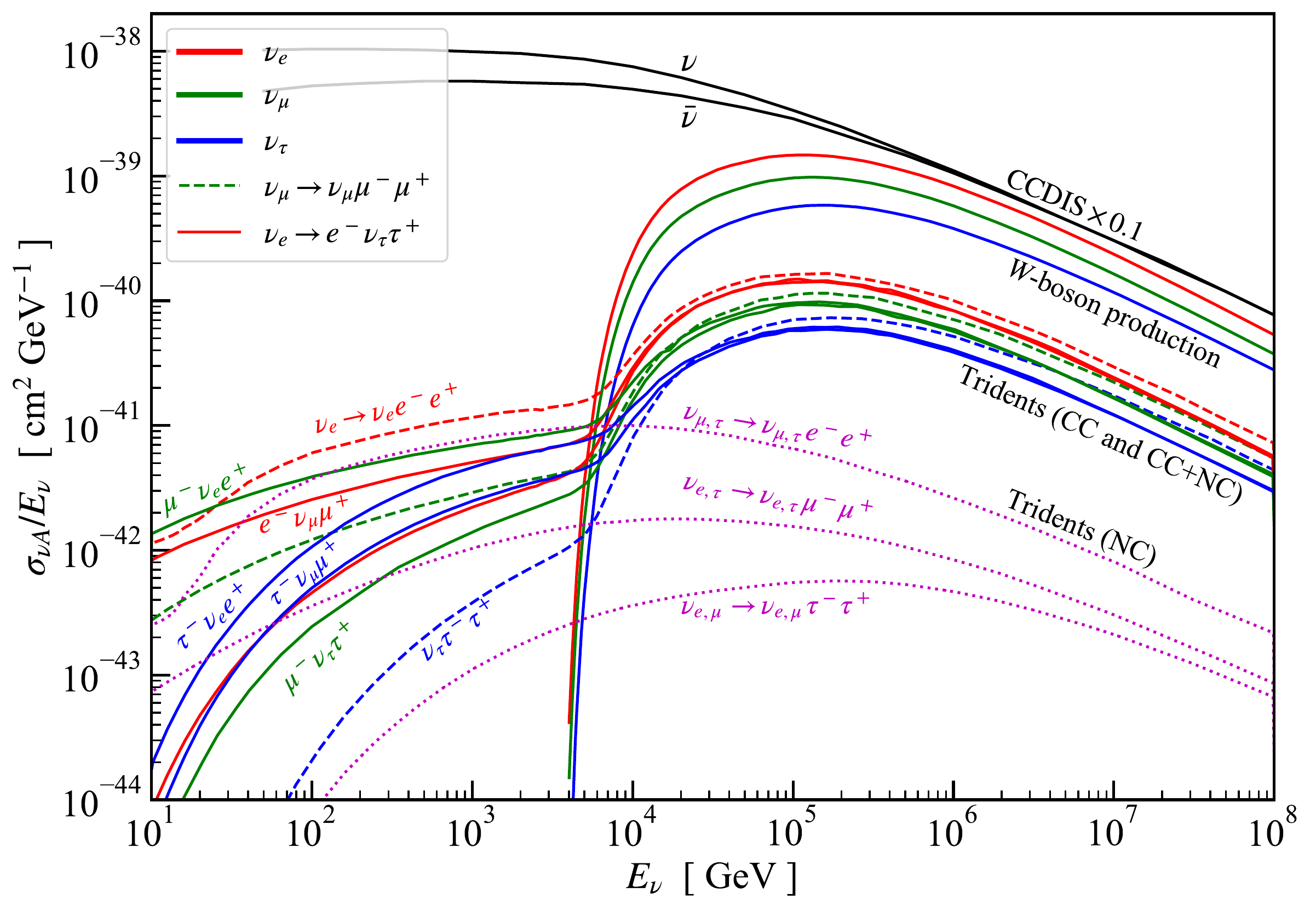}
    \caption{Our {\bf total} cross sections (actually $\sigma_{\nu A}/E_\nu$) for $W$-boson and trident production on $\ce{^16}\rm O$. The colors and line styles are same as in Fig.~\ref{fig_sigma_nugm_all} ({\bf red}, {\bf green}, and {\bf blue} lines are $\nu_e$-, $\nu_\mu$-, and $\nu_\tau$-induced channels, respectively; {\bf solid} lines are CC channels, and {\bf dashed} lines are CC+NC channels; {\bf magenta dotted} lines are NC channels, which depend on only the final-state charged leptons). The trident CC, NC, and CC+NC channels correspond to diagrams (1)--(3), (4)--(5), and (1)--(5) of Fig.~\ref{fig_trident_diagrams}. The corresponding antineutrino cross sections (i.e., obtained by CP-transforming the processes shown) are the same. See text for details.
}
\label{fig_sgmnuA_tot_O16}
\end{figure*}

\begin{figure}[t!]
    \includegraphics[width=\columnwidth]{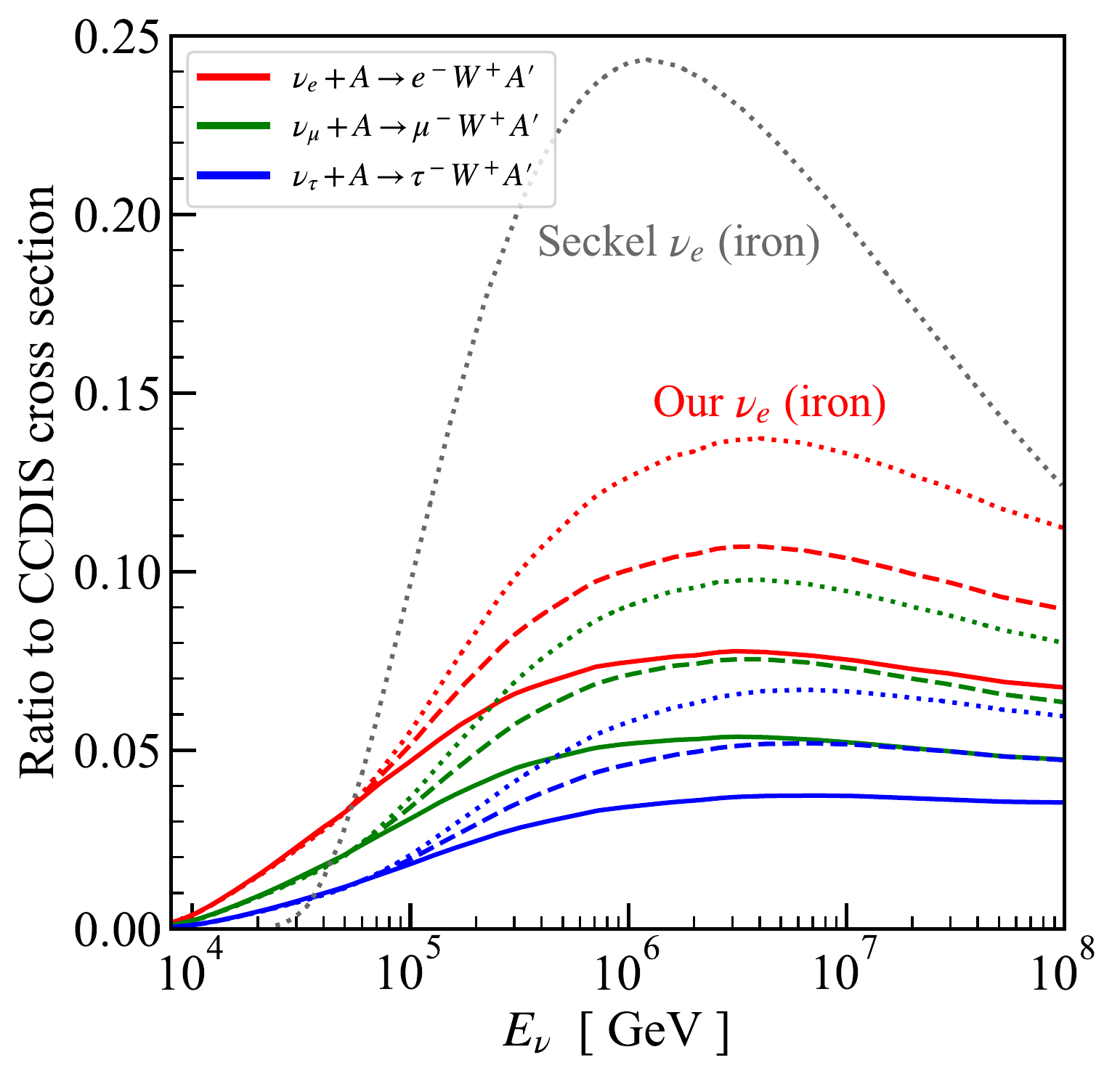}
    \caption{Ratios of the $W$-boson production cross sections to those of CCDIS (($\nu + \bar{\nu}$)/2)~\cite{CooperSarkar:2011pa}. {\bf Solid} lines are for water/ice targets, {\bf dotted} line for iron targets, and {\bf dashed} lines are for the Earth's averaged composition. Color assignment is in the legend. Also shown is the $\nu_e$ (iron) case of Seckel~\cite{Seckel:1997kk}, much larger than ours.
}
\label{fig_sgmnuA_ratio_O16}
\end{figure}

\section{Total cross sections of $W$-boson and trident production, and ratio to CCDIS}
\label{sec_total}

Figure~\ref{fig_sgmnuA_tot_O16} shows the total cross sections ($\sigma_{\nu A}/E_\nu$) with $\ce{^16}\rm O$ for $W$-boson and trident production, which summarize our calculations in previous sections.  We divide out $E_{\nu}$, the dominant trend, to highlight the deviations over the wide range of the x axis. Specifically, the total cross sections are obtained by summing up the coherent, diffractive and inelastic components, for each interaction channel.  For the $W$-boson production and trident CC channels, the inelastic cross section is the photon-initiated subprocess, while for trident CC+NC and NC channels, it is from summing up both photon- and quark-initiated subprocesses. Reasons are discussed in Sec.~\ref{sec_inelastic_trident}. 

We estimate the uncertainties of the $W$-boson and trident cross sections. For the coherent regime, it is about $6\%$, coming from higher order electroweak corrections (dominant) and nuclear form factors (see Ref.~\cite{Altmannshofer:2019zhy} for more details). For the diffractive regime, the nuclear effects lead to larger uncertainties.  Our calculations include the Pauli-blocking factor derived from ideal Fermi gas model for nucleus and ignore other subdominant effects.  Therefore, to be conservative, we assume 30\% uncertainty, following Ref.~\cite{Altmannshofer:2019zhy}. (Further details on the sensitivity to the choices of nuclear and nucleon form factors are given above.) For the inelastic regime, the uncertainty mainly comes from choosing the factorization and renormalization scale, which is $\simeq 15\%$, and mixing between photon and weak boson which matters more at higher energy, for which we give $\simeq 25\%$ to be conservative (Sec.~\ref{sec_inelastic}). There is no study of the nuclear uncertainties of photon PDF, but they should be subdominant especially for light nucleus, like $\ce{^16}\rm O$, considering that of the quark PDFs are subdominant~\cite{Kovarik:2015cma, Khanpour:2016pph, Wang:2016mzo, Eskola:2016oht}. Their combination gives $\simeq 30\%$ for the inelastic regime. Combining the three regimes in quadrature, the uncertainty is estimated to be $\simeq 15\%$.

Figure~\ref{fig_sgmnuA_ratio_O16} shows the ratios of the important channels to the neutrino CCDIS cross sections ($( \sigma^{\rm CCDIS}_\nu + \sigma^{\rm CCDIS}_{\bar{\nu}} )/2$)~\cite{CooperSarkar:2011pa}, for different targets (right panel), including water/ice (for neutrino detectors), iron and the Earth's averaged composition (for neutrino propagation). The coherent cross sections ($\propto Z^2$) for different isotopes are calculated in the same way as for $\ce{^16}\rm O$. For diffractive and inelastic cross sections, one can just rescale by atomic and mass numbers from $\ce{^16}\rm O$, as the cross section on single nucleon is independent of nucleus in above formalism (the Pauli-blocking factor in the diffractive cross section derived from ideal Fermi-gas model~\cite{Lovseth:1971vv, Ballett:2018uuc, Altmannshofer:2019zhy} is nucleus independent). For hydrogen ($\ce{^1}\rm H$) only, there is no coherent component, and the Pauli-blocking factor should not be included. The CCDIS cross sections for different isotopes are calculated by multiplying their mass number by the CCDIS cross section on isoscalar nucleon target~\cite{CooperSarkar:2011pa}. The bias caused by the fact that some isotopes have a slightly different number of protons and neutrons and by nuclear effects on PDFs is negligible.

The maximum ratios of $W$-boson production to CCDIS are $\simeq 7.5\%$ ($\nu_e$), $\simeq 5\%$ ($\nu_\mu$), and $\simeq 3.5\%$ ($\nu_\tau$) on water/ice target, respectively. For trident production (not shown on the figure), the CC+NC and NC channels are large enough to matter, i.e., $\simeq 0.75\%$ ($\nu_e$-induced), $0.5\%$ ($\nu_\mu$-induced), and $0.35\%$ ($\nu_\tau$-induced). This is a factor $\simeq 0.1$ of corresponding $W$-boson production channels, as the dominant contribution at these energies are from the $W$-boson production followed by $W \rightarrow \nu_\ell + \ell$ (Sec.~\ref{sec_sgmnugamma_trident}). These trident channels and also hadronic decay of those $W$ bosons could produce distinct signatures in the high-energy neutrino detectors like IceCube~\cite{Aartsen:2015knd}, KM3NeT~\cite{Adrian-Martinez:2016fdl}, and especially the forthcoming IceCube-Gen2~\cite{Blaufuss:2015muc}.  In IceCube, there could already be a few $W$-boson production events. IceCube-Gen2 will have much larger yields. The detectability and implications are detailed in our companion paper~\cite{Beacom:2019pzs}. (For clarification, the $W$-resonance enhancement part of the trident, which dominates above $\sim 10^4$~GeV, is the $W$-boson production followed by leptonic decay of the $W$ boson, so they would not be studied separately for detection.)

The maximum ratios of $W$-boson production to CCDIS are $\simeq 14\%$ ($\nu_e$), $\simeq 10\%$ ($\nu_\mu$), and $\simeq 7\%$ ($\nu_\tau$) on iron, and $\simeq 11\%$ ($\nu_e$), $\simeq 7.5\%$ ($\nu_\mu$), and $\simeq 5\%$ ($\nu_\tau$) on the Earth's averaged composition. (The larger the charge number of a nucleus, the larger the ratio is, due to the coherent component $\propto Z^2$.) Trident CC and CC+NC channels (not shown) are $\simeq 0.1$ of these numbers, same as above.  ({\it As a comparison, the ratio for $\nu_e$-iron case by Seckel~\cite{Seckel:1997kk} is 25\%, much larger than ours (14\%).}) This affects the absorption rate of high-energy neutrinos when propagating through the Earth, which affects the measurement of neutrino cross sections by IceCube~\cite{Hooper:2002yq, Borriello:2007cs, Klein:2013xoa, Aartsen:2017kpd, Bustamante:2017xuy}.

\section{Conclusions}
\label{sec_concl}

The interactions of neutrinos with elementary particles, nucleons, and nuclei are a cornerstone of the Standard Model, and a crucial input for studying neutrino mixing, neutrino astrophysics, and new physics. Above $E_\nu \sim 5$~GeV, the neutrino-nucleus cross section is dominated by deep inelastic scattering, in which neutrinos couple via weak bosons to the quarks. However, additionally, there are two processes where the hadronic coupling is through a virtual photon: $W$-boson and trident production, the cross sections of which increase rapidly and become relevant at TeV--PeV energies.

In this paper, we do a complete calculation of the $W$-boson and trident production processes. We significantly improve the completeness and precision of prior calculations. We start by giving a systematic review of both processes, pointing out the improvements that can/should be made compared to previous calculations (Sec.~\ref{sec_review} and Table~\ref{tab_summary_calc}). 

Our results can be put into three major categories. 
\bi
\item {\it The neutrino-real photon cross sections for these two processes over a wide energy range (Sec.~\ref{sec_sgmnugamma}).}
This sets the foundation for our neutrino-nucleus cross section calculation and for discussing the underlying physics (Sec.~\ref{sec_sgmnugamma}). For trident production, there are three different categories of interaction channels (Eq.~(\ref{eq_trident_details})): CC, CC+NC, and NC, arising from different groups of diagrams (Fig.~\ref{fig_trident_diagrams}). Interestingly, for the CC and CC+NC channels, above $s_{\nu\gamma} = m_W^2$, the cross sections are enhanced by two orders of magnitude (Fig.~\ref{fig_sigma_nugm_all}). The reason is that the $s$-channel like $W$-boson propagators in the trident diagrams (Fig.~\ref{fig_trident_diagrams}) can be produced on shell (i.e., $W$-boson production in Fig.~\ref{fig_real_W_diagrams}) and then decay leptonically. {\it This indicates the unification of these two processes.}

\item {\it A first complete calculation of neutrino-nucleus $W$-boson production cross sections.} 
    For the neutrino-nucleus cross sections, we handle them in different regimes: elastic (Sec.~\ref{sec_elastic}; including coherent and diffractive) and inelastic (Sec.~\ref{sec_inelastic}; including both photon- and quark-initiated subprocesses). 
{\it For $W$-boson production, we show that the previously used equivalent photon approximation~\cite{Alikhanov:2014uja, Alikhanov:2015kla} is not good for its cross section calculation.} For the $\nu_e$-induced channel, our results are about half of those in Refs.~\cite{Alikhanov:2014uja, Alikhanov:2015kla} for both coherent and diffractive regimes, and $\sim 1/4$ of Ref.~\cite{Seckel:1997kk} for coherent regime. 
The reasons for this are largely understood. A significant factor is the nonzero photon virtuality, $Q^2$, which suppresses the cross section (EPA assumes $Q^2=0$ as used by previous works). Also, the Pauli-blocking effect suppresses the diffractive cross section.
For the inelastic regime, we use an up-to-date photon PDF and more reasonably dynamic factorization scale, $\sqrt{s_{\nu \gamma}}$. Moreover, we do a first calculation of the quark-initiated subprocess and find that they can be neglected below $\simeq 10^8$~GeV.

\item {\it A first calculation of neutrino-nucleus trident production cross sections at TeV--PeV energies.} 
The full Standard Model is used in order to also study the TeV--PeV behavior, compared to the four-Fermi theory used by previous work. The equivalent photon approximation is also avoided. Moreover, we do a more careful treatment of the inelastic regime, and, importantly, we for the first time use the inelastic photon PDF~\cite{Schmidt:2015zda, CT14web} for trident calculation, which includes the resonance region as a component of the nonperturbative part of the inelastic photon PDF.
More improvements are detailed in the previous sections.
\ei 

These cross section are large enough to matter (Sec.~\ref{sec_total}).  For a water/ice target, the $W$-boson production cross sections are $\simeq 7.5\%$ ($\nu_e$), $\simeq 5\%$ ($\nu_\mu$), and $\simeq 3.5\%$ ($\nu_\tau$) of CCDIS~\cite{CooperSarkar:2011pa}. (For the corresponding CC and CC+NC trident channels, they are $\simeq 0.1$ times the numbers above.)  This means these processes are detectable, or will be detectable, by high-energy neutrino detectors like IceCube~\cite{Aartsen:2015knd}, KM3NeT~\cite{Adrian-Martinez:2016fdl}, and especially the forthcoming IceCube-Gen2~\cite{Blaufuss:2015muc}, with distinct signatures~\cite{Beacom:2019pzs}. For the iron target or the Earth's averaged composition, the $W$-boson production cross sections are $14\%$/$11\%$ ($\nu_e$), $10\%$/$7.5\%$ ($\nu_\mu$), and $7\%$/$5\%$ ($\nu_\tau$) of CCDIS~\cite{CooperSarkar:2011pa}.  This affects the absorption rate of high-energy neutrinos during propagation. Moreover, the DIS cross sections extracted from in-Earth absorption as seen by IceCube~\cite{Hooper:2002yq, Borriello:2007cs, Klein:2013xoa, Aartsen:2017kpd, Bustamante:2017xuy} contain a contribution from W-boson production.  Note the fact that the cross section affects the absorption rate exponentially may make these processes even more important than that shown by the numbers above.

This paper sets the theoretical framework and calculates the cross sections of these processes. In our companion paper~\cite{Beacom:2019pzs}, we discuss the phenomenological consequences, including the effects mentioned above and other aspects of high-energy neutrino astrophysics that these processes make a difference, such as neutrino cross-section and spectrum measurement, flavor ratio determination, neutrino mixing, and new physics.


\section*{Acknowledgments}

We are grateful for helpful discussions with Brian Batell, Eric Braaten, Mauricio Bustamante, Richard Furnstahl, Bin Guo, Liping He, Matheus Hostert, Junichiro Kawamura, Matthew Kistler, Yuri Kovchegov, Gordan Krnjaic, Shirley Li, Pedro Machado, Aneesh Manohar, Kenny Ng, Alexander Pukhov, Stuart Raby, Subir Sarkar, Juri Smirnov, Carl Schmidt, Xilin Zhang, and especially Spencer Klein, Olivier Mattelaer, Sergio Palomares-Ruiz, Yuber Perez-Gonzalez, Ryan Plestid, Mary Hall Reno, David Seckel, and Keping Xie.  

We used {\tt FeynCalc}~\cite{Mertig:1990an, Shtabovenko:2016sxi}, {\tt MadGraph}~\cite{Alwall:2014hca} and {\tt CalcHEP}~\cite{Belyaev:2012qa} for some calculations. 

This work was supported by NSF grant PHY-1714479 to JFB. BZ was also supported in part by a University Fellowship from The Ohio State University. 

\onecolumngrid
\newpage
\appendix

\section{$W$-boson production: amplitudes in the Standard Model}
\label{appdx_realW_amplitudes}

The leptonic amplitudes of the diagrams of Fig.~\ref{fig_real_W_diagrams}, in the unitarity gauge, are
\begin{subequations}
\begin{align}
    {\mathrm L}_1^\mu = &
    \frac{e g_W}{2\sqrt{2}} 
    \bar{u}\left(p_1\right) \gamma^\mu \frac{\left(\slashed{k}-\slashed{p_2}\right)+m_\ell}{\left(k-p_2\right)^2-m_\ell^2} \gamma^\nu (1-\gamma^5) u\left(k\right) \times \epsilon_\nu(W) \,,
    \\ 
    {\mathrm L}_2^\mu = &
    - \frac{e g_W}{2\sqrt{2}} 
    \bar{u}\left(p_1\right) \gamma^\nu \left(1-\gamma^{5}\right) u\left(k\right)  \times
    \frac{g_{\mu \nu}-\frac{\left(k-p_1\right)_{\mu}\left(k-p_1\right)_{\nu}}{m_{W}^2}}{\left(k-p_1\right)^2-m_W^2} \nonumber \\
    & \hspace{5cm} \times
    \left[ g^{\rho\lambda}(q+p_2)^\mu + g^{\lambda\mu} (p_1-p_2-k)^\rho + g^{\mu\rho}(k-p_1-q)^\lambda \right] 
    \epsilon_\lambda(W) \,,
\end{align}
\end{subequations}
where the $m_\ell$ and $m_W$ are masses of the charged lepton ($p_1$) and the $W$ boson ($p_2$), respectively, and $\epsilon(W)$ the polarization vector of the $W$ boson, for which we have
\begin{equation}
    \sum_{i=1}^4 \epsilon^\nu_i(W) \epsilon^\nu_i(W) = - g^{\mu \nu} + \frac{p_2^\mu p_2^\nu}{m_W^2} \,.
\end{equation}
The amplitudes above are used for both Sec.~\ref{sec_sgmnugamma_realW} ($\mathcal{M}^{\rm WBP} = \epsilon_\mu( {\mathrm L}_1^\mu - {\mathrm L}_2^\mu )$, where $\epsilon^\mu$ is the photon polarization vector) and Sec.~\ref{sec_elastic_realW} (${\mathrm L}^\mu = {\mathrm L}_1^\mu - {\mathrm L}_2^\mu$).

\newpage
\section{Trident production: amplitudes in the Standard Model}
\label{appdx_trident_amplitudes}
The leptonic amplitudes of the diagrams of Fig.~\ref{fig_trident_diagrams}, in the unitarity gauge, are

\begin{subequations}
\begin{align}
    {\mathrm L}_1^\mu = &
    -\frac{e g_W^2}{8} \bar{u}\left(k_{2}\right) \gamma^{\beta} \left(1-\gamma^{5}\right) v\left(p_{1}\right)  \times
    \frac{g_{\alpha \beta}-\frac{\left(p_1+k_2\right)_{\alpha}\left(p_1+k_2\right)_{\beta}}{m_{W}^2}}{\left(p_1+k_2\right)^2-m_W^2+i m_W \Gamma_W} \times
    \bar{u}\left(p_2\right) \gamma^\mu \frac{\left(\slashed{p_2}-\slashed{q}\right)+m_2}{\left(p_2-q\right)^2-m_2^2} \gamma^\alpha (1-\gamma^5) u\left(k_1\right) \, , 
    \\[10pt] 
    {\mathrm L}_2^\mu = &
    \frac{e g_W^2}{8} \bar{u}\left(p_{2}\right) \gamma^\alpha \left(1-\gamma^{5}\right) u\left(k_1\right) \times
    \frac{g_{\alpha \beta}-\frac{\left(k_1-p_2\right)_{\alpha}\left(k_1-p_2\right)_{\beta}}{m_{W}^2} }{\left(k_1-p_2\right)^2-m_W^2} 
    \nonumber \\
    & \times 
    \left[ g^{\gamma\beta}(-p_1-k_2-k_1+p_2)^\mu + g^{\beta\mu} (k_1-p_2-q)^\gamma + g^{\mu\gamma}(q+p_1+k_2)^\beta \right] 
    \times
    \frac{g_{\gamma \delta}-\frac{\left(p_1+k_2\right)_{\gamma}\left(p_1+k_2\right)_{\delta}}{m_{W}^2}}{\left(p_1+k_2\right)^2-m_W^2 + i m_W \Gamma_W} \nonumber \\
    & \times
    \bar{u}\left(k_2\right) \gamma^\delta (1-\gamma^5) v\left(p_1\right) \, , 
    \\[10pt]
    {\mathrm L}_3^\mu = &
    -\frac{e g_W^2}{4} \bar{u}\left(p_{2}\right) \gamma^{\alpha} \left(1-\gamma^{5}\right) u\left(k_{1}\right)  \times
    \frac{g_{\alpha \beta}-\frac{\left(k_1-p_2\right)_{\alpha}\left(k_1-p_2\right)_{\beta}}{m_{W}^2}}{\left(k_1-p_2\right)^2-m_W^2} 
    \times
    \bar{u}\left(k_2\right) \gamma^{\beta}(1-\gamma^5) \frac{\left(\slashed{q}-\slashed{p_1}\right)+m_1}{\left(q-p_1\right)^2-m_1^2} \gamma^{\mu} v\left(p_1\right) 
    \, ,
    \\[10pt]
    {\mathrm L}_4^\mu = &
    -\frac{e g_Z^2}{4} \bar{u}\left(k_{2}\right) \gamma^{\alpha} \left(\frac{1}{2}-\frac{\gamma^{5}}{2}\right) u\left(k_{1}\right)  \times
    \frac{g_{\alpha \beta}-\frac{\left(k_1-k_2\right)_{\alpha}\left(k_1-k_2\right)_{\beta}}{m_{Z}^2}}{\left(k_1-k_2\right)^2-m_Z^2} 
    \nonumber \\     & \hspace{5cm} 
    \times
    \bar{u}\left(p_2\right) \gamma^{\mu} \frac{\left(\slashed{p}_2-\slashed{q}\right)+m_2}{\left(p_2-q\right)^2-m_2^2} \gamma^{\beta}\left(-\frac{1}{2}+2 \sin ^2 \theta_{W}+\frac{1}{2} \gamma^{5}\right) v\left(p_1\right) 
    \, ,\\[10pt]
    {\mathrm L}_5^\mu = &
    -\frac{e g_Z^2}{4} \bar{u}\left(k_{2}\right) \gamma^{\alpha} \left(\frac{1}{2}-\frac{\gamma^{5}}{2}\right) u\left(k_{1}\right)  \times
    \frac{g_{\alpha \beta}-\frac{\left(k_1-k_2\right)_{\alpha}\left(k_1-k_2\right)_{\beta}}{m_{Z}^2}}{\left(k_1-k_2\right)^2-m_Z^2} 
    \nonumber \\
    & \hspace{5cm} \times
    \bar{u}\left(p_2\right) \gamma^{\beta} \left(-\frac{1}{2}+2 \sin ^2 \theta_{W}+\frac{1}{2} \gamma^{5}\right) \frac{\left(\slashed{q}-\slashed{p_1}\right)+m_1}{\left(q-p_1\right)^2-m_1^2} \gamma^{\mu} v\left(p_1\right) 
    \, .
\end{align}
\end{subequations}
Note that for $\mathrm{L}^\mu_1$ and $\mathrm{L}^\mu_2$, the width term of the $W$ propagator, $i m_W \Gamma_W$, should be included. 
These are used for both Sec.~\ref{sec_sgmnugamma_trident} ($
\mathcal{M}^{\rm Tri}_i = \epsilon_\mu ( 
({\mathrm L}_1^\mu - {\mathrm L}_2^\mu + {\mathrm L}_3^\mu) - ( {\mathrm L}_4^\mu + {\mathrm L}_5^\mu) 
) 
$, same as Eq.~(\ref{eq_trident_relsign})) and Sec.~\ref{sec_elastic_trident} (${\mathrm L}_i^\mu = ({\mathrm L}_1^\mu - {\mathrm L}_2^\mu + {\mathrm L}_3^\mu) - ({\mathrm L}_4^\mu + {\mathrm L}_5^\mu)$).

\newpage
\section{Trident production: kinematics and phase space for $\sigma^{T/L}_{\nu \gamma}(\hat{s}, Q^2)$}
\label{appdx_trident}

Following Refs.~\cite{Vysotsky:2002ix, Magill:2016hgc}, we give more details of the kinematics and the three-body phase space of $\sigma^{T/L}_{\nu \gamma}(\hat{s}, Q^2)$ for trident production, and derive the case for the virtual photon.  The momenta are labeled in Fig.~\ref{fig_trident_diagrams}.

In the CM frame of the $\nu$-A interaction and treating the two charged leptons together ($p \equiv p_1 + p_2$, i.e., the total momentum of $\ell^+$ and $\ell^-$), we can write the 4-momenta by 
\begin{subequations}
\label{eq_4momentum_A1}
\begin{align}
    k_{1} =& \frac{s+Q^2}{2 \sqrt{s}} \left( 1, \sin\theta, 0,-\cos\theta \right) \,, \\
    q     =& \frac{s+Q^2}{2 \sqrt{s}} \left(\frac{s-Q^2}{s+Q^2}, -\sin\theta, 0, \cos\theta \right) \,, \\
    k_{2} =& \frac{s-l}{2 \sqrt{s}} \left(1,0,0,-1\right) \,, \\
    p     =& \left( \frac{s+l}{2\sqrt{s}}, 0, 0, \frac{s-l}{2\sqrt{s}} \right) \,,
\end{align}
\end{subequations}
where $Q^2 \equiv -q^2$ is the photon virtuality, 
$l \equiv p^2$, 
$s \equiv s_{\nu \gamma} \equiv (k_1 + q)^2$ (for simplicity, we use $s \equiv s_{\nu \gamma}$ in this section only), 
and $\theta$ is the angle of the incoming particles with respect to the direction of $p$, which is chosen to be the $z$ axis.

Define another variable, 
\begin{equation}
    t \equiv 2q \cdot (k_1-k_2) =  \frac{l \left[Q^2 (cos\theta-1)+s+s cos\theta \right] + s \left[ Q^2 (3-cos\theta) + s-s cos\theta \right] }{2 s} \,.
\label{eq_app_t}
\end{equation}
This relation allows us to rewrite $\sin\theta$ and $\cos\theta$ in terms of $t$, which is Lorentz invariant, then putting back into Eq.(~\ref{eq_4momentum_A1}), we obtain
\begin{subequations}
\begin{align}
    k_{1} =& \left( \frac{Q^2+s}{2 \sqrt{s}}, \frac{ \sqrt{(l+Q^2-t)(l Q^2+s(-2 Q^2-s+t))}}{s-l}, 0,-\frac{l(s-Q^2)+s(3 Q^2+s-2 t)}{2\sqrt{s}(s-l)} \right)  \,, \\
    q     =&   \left(\frac{s-Q^2}{2 \sqrt{s}}, \frac{ \sqrt{(l+Q^2-t)(l Q^2+s(-2 Q^2-s+t))}}{l-s}, 0, \frac{l(s-Q^2)+s(3(Q^2+s-2 t))}{2\sqrt{s}(s-l)} \right) \,, \\
    k_{2} =& \frac{s-l}{2 \sqrt{s}} \left(1,0,0,-1\right) \,, \\
    p     =& \left( \frac{l+s}{2 \sqrt{s}}, 0,0, \frac{s-l}{2 \sqrt{s}} \right) \,.
\end{align}
\end{subequations}

To find the expression of $p_1$ and $p_2$, it is easier to go to the rest frame of $p$. We do a Lorentz transformation to boost to this frame, using $\beta = \frac{s+l}{s-l} $, $\gamma = \frac{s+l}{2\sqrt{sl}}$ for the transformation matrix,
\begin{subequations}
\begin{align}
    k'_{1} =&  \left( \frac{l+2 Q^2+s-t}{2 \sqrt{l}}, \frac{\sqrt{(l+Q^2-t)(l Q^2+s(-2 Q^2-s+t))}}{s-l}, 0, \frac{\left(l^{2}-l t+s(2 Q^2+s-t)\right)}{2 \sqrt{l}(l-s)}\right) \,, \\
    q'     =& \left( \frac{t-2 Q^2}{2 \sqrt{l}}, \frac{\sqrt{(l+Q^2-t)(l Q^2+s(-2 Q^2-s+t))}}{l-s}, 0,-\frac{(2 l s-l t+2 Q^2 s-s t)}{2 \sqrt{l}(l-s)} \right)   \,, \\
    k'_{2} =& \frac{s-l}{2 \sqrt{s}} \left(1,0,0,-1\right) \,, \\
    p'     =& ( \sqrt{l}, 0,0,0 ) \,.
\end{align}
\end{subequations}

The situation can be further simplified if we work in the frame where $q'$ is along the $z$ axis. So we need to do a rotation of above. The rotation angle, $\eta_q$, can be determined by 
\begin{align}
    \sin \eta_q  =  \frac{q'[2]}{\sqrt{q'[2]^2 + q'[4]^2}} \,, \text{       and         } 
    \cos \eta_q  =  \frac{q'[4]}{\sqrt{q'[2]^2 + q'[4]^2}} \,.
\end{align}

So, finally, we have
\begin{subequations}
\begin{align}
    k''_{1} &= \left(
    \frac{l+2 Q^{2}+s-t}{2 \sqrt{l}}, -\sqrt{\frac{(l+Q^2-t)(l Q^2+s(-2 Q^2-s+t))}{4 l Q^2+(t-2 Q^2)^{2}}}, 0,-\frac{(l(4 Q^2+2 s-t)+(2 Q^2-t)(2 Q^2+s-t))}{2 l \sqrt{\frac{\left(t-2 Q^2\right)^{2}}{l}+4 Q^2}} 
    \right) \,, \\
    q''     &= \left(\frac{t-2 Q^2}{2 \sqrt{l}}, 0,0, \frac{1}{2} \sqrt{\frac{(t-2 Q^2)^{2}}{l}+4 Q^2}\right)   \,,
    \\
    k''_{2} &=  \left(
    \frac{s-l}{2 \sqrt{l}},
    -\frac{\sqrt{(l+Q^2-t) (l Q^2+s (-2 Q^2-s+t))}}{\sqrt{(t-2 Q^2)^2+4 l Q^2}},
    0,
    -\frac{(2 l s-l t+2 Q^2 s-s t)}{2 \sqrt{(t-2 Q^2)^2+4 l Q^2}}
    \right) \,,
    \\
    p''     &= ( \sqrt{l}, 0,0,0 ) \,.
\end{align}
\end{subequations}

In this frame, the $p_{1,2}$ can be written as
\begin{subequations}
\label{eq_4momentum_p1p2}
\begin{align}
    p''_{1} &= \left(E_{1}, + \rho \sin \theta^{\prime \prime} \cos \phi^{\prime \prime}, + \rho \sin \theta^{\prime \prime} \sin \phi^{\prime \prime}, + \rho \cos \theta^{\prime \prime}\right) \,, \\
    p''_{2} &= \left(E_{2}, - \rho \sin \theta^{\prime \prime} \cos \phi^{\prime \prime}, - \rho \sin \theta^{\prime \prime} \sin \phi^{\prime \prime}, - \rho \cos \theta^{\prime \prime}\right) \,,
\end{align}
\end{subequations}
where $\theta''$ and $\phi''$ are the angles with respect to the photon, $q''$, in the current frame, $E_{1,2} = \sqrt{\rho^2 + m_{1,2}^2}$, and
\begin{equation}
    \rho^{2} = \frac{l^{2}-2 l\left(m_{1}^{2}+m_{2}^{2}\right)+\left(m_{1}^{2}-m_{2}^{2}\right)^{2}}{4 l} \,.
\end{equation}

\vspace{2cm}

The three-body phase space can be done by decomposing it into two two-body phase spaces~\cite{Murayama_PS, Magill:2016hgc} (each is independently Lorentz invariant),
\begin{equation}
    d{\rm PS_3}(k_2, p_1, p_2) = \frac{d l}{2 \pi}\, d{\rm PS_2}(k_2, p)\,  d{\rm PS_2}(p_1, p_2) \,
\end{equation}
with 
\begin{equation}
    d{\rm PS_2}(x_1, x_2) = \overline{\beta}(x_1, x_2) \frac{d \Omega}{32 \pi^2} \,
\end{equation}
which is frame independent, and
\begin{equation}
    \overline{\beta}(x_1, x_2) = \sqrt{ 1 - \frac{2(x_1^2+x_2^2)}{(x_1+x_2)^2} + \frac{(x_1^2-x_2^2)^2}{(x_1+x_2)^4}} \, .
\label{eq_beta_x1x2}
\end{equation}

The $d{\rm PS_2}(k_2, p)$, in the CM frame, can be written as
\begin{equation}
    d{\rm PS_2}(k_2, p) = \overline{\beta}(k_2, p) \frac{d \Omega}{32 \pi^2} \, = \frac{s-l}{2s} \frac{d\Omega}{16 \pi^2} \,,
\end{equation}
and
\begin{equation}
    \frac{d\Omega}{16 \pi^2} = \frac{d\cos\theta d\phi}{16 \pi^2} = \frac{1}{8\pi} \frac{s}{s+Q^2} \frac{2}{s-l} dt \,,
\end{equation}
which is Lorentz invariant, where the first step uses the azimuthal symmetry of the system in the CM frame, and the second step uses Eq.~(\ref{eq_app_t}).

The $d{\rm PS_2}(p_1, p_2)$, in the rest frame of $p$, can be written as
\begin{equation}
    d{\rm PS_2}(p_1, p_2) = \overline{\beta}(p_1, p_2) \frac{d \Omega''}{32 \pi^2} \,
\end{equation}
where $\overline{\beta}(p_1, p_2)$ can be derived using Eq.~(\ref{eq_beta_x1x2}), i.e.,
\begin{equation}
     \overline{\beta}(p_1, p_2) \equiv \overline{\beta}(l) = \sqrt{ 1 - \frac{2(m_1^2+m_2^2)}{l} + \frac{(m_1^2-m_2^2)^2}{l^2} } \, ,
\end{equation}
i.e., Eq.~(\ref{eq_PS3_beta}), and $d \Omega'' = d\cos\theta'' d\phi''$ is the solid angle with respect to the photon in the rest frame of $p$ (c.f. Eq.~(\ref{eq_4momentum_p1p2})).

Putting above together, we get the phase space of the off-shell cross section, $\sigma^{T/L}_{\nu \gamma}(\hat{s}, Q^2)$, for the trident production, 
\begin{equation}
    d {\rm PS_3} = \frac{1}{2} \frac{1}{(4 \pi)^2} 
    \frac{d l}{2 \pi} 
    \overline{\beta}(l) 
    \frac{d t}{2 (s_{\nu \gamma} + Q^2)} 
    \frac{d\Omega''}{4 \pi} \,,
    \text{ or }  
    d {\rm PS_3} = \frac{1}{2} \frac{1}{(4 \pi)^2} 
    \frac{d l}{2 \pi} 
    \overline{\beta}(l) 
    \frac{d t}{2 \hat{s}} 
    \frac{d\Omega''}{4 \pi} \,.
\label{eq_PS3_vph}
\end{equation}
This is the same as the phase space for the real photon case, Eq.~(\ref{eq_PS3}), but replacing the $s_{\nu \gamma}$ by $\hat{s} \equiv s_{\nu \gamma} + Q^2$.
The integration range of $l$ is now $\left[ (m_1+m_2)^2, s_{\nu \gamma} \right]$ or $\left[ (m_1+m_2)^2, \hat{s}-Q^2 \right]$.
And the integration range of $t$ can be obtain from Eq.~(\ref{eq_app_t}), 
which gives 
\begin{equation}
    \left[ l+Q^2, s_{\nu\gamma} + \left( 2-\frac{l}{s_{\nu\gamma}} \right) Q^2 \right], \text{ or } 
    \left[ l+Q^2, \hat{s} - Q^2 + \left( 2 - \frac{l}{\hat{s}-Q^2} \right)Q^2 \right]\,.
\end{equation}

When $Q^2 = 0$, all the above return to the on-shell photon case.

\newpage
\section{Trident production: coherent and diffractive cross sections for all channels}
\label{Sec_sgmnuA_trident_Coh_Diff_all}
\begin{figure*}[h!]
\includegraphics[width=\textwidth]{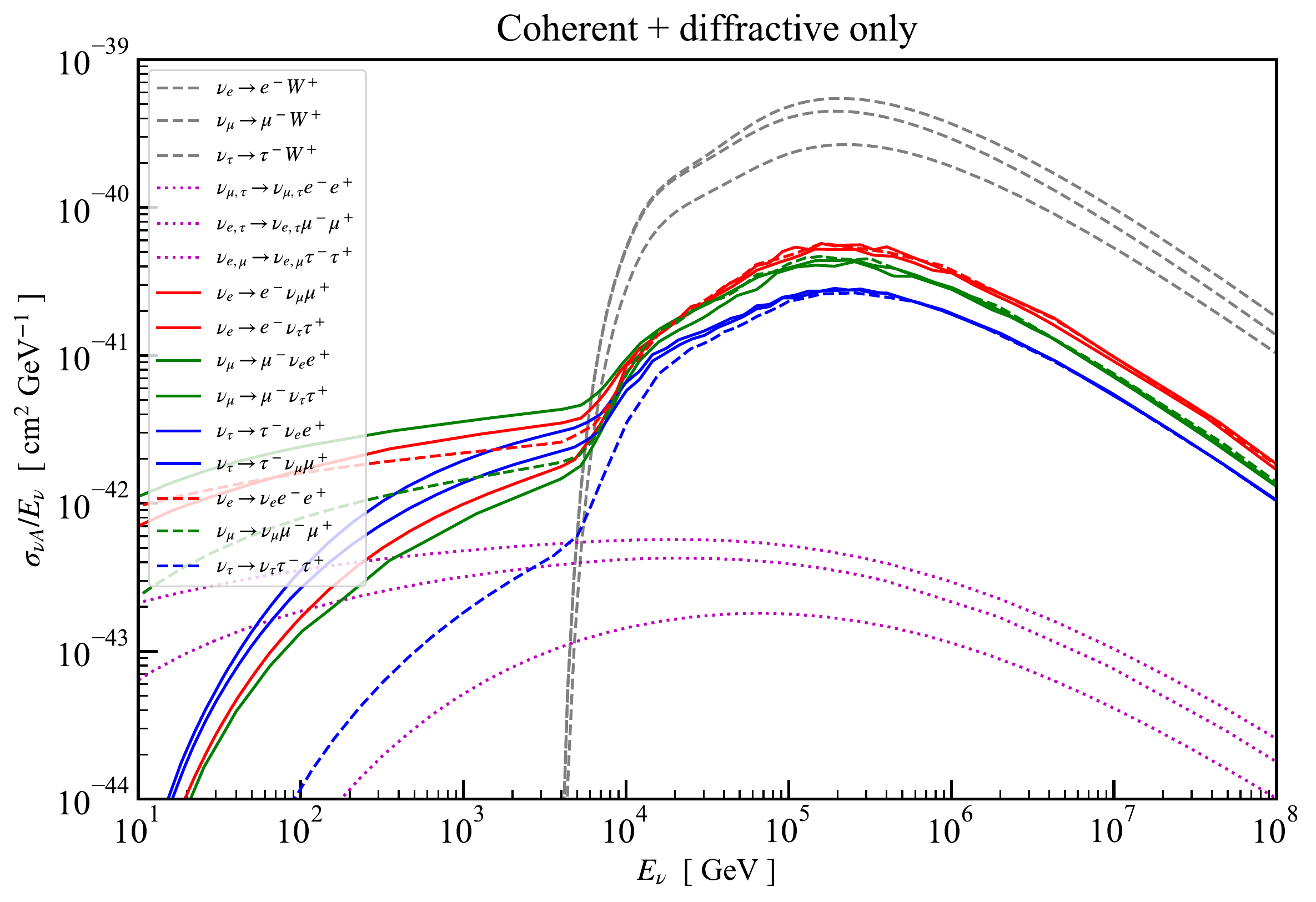}
    \caption{Our elastic cross sections for all trident channels. We add the coherent and diffractive components together to simplify the figure.
    The colors and line styles are same as in Fig.~\ref{fig_sigma_nugm_all}
    ({\bf red}, {\bf green}, and {\bf blue} lines are $\nu_e$-, $\nu_\mu$-, and $\nu_\tau$-induced channels, respectively; {\bf solid} lines are CC channels, and {\bf dashed} lines are CC+NC channels;
    {\bf magenta dotted} lines are NC channels, which depend on only the final-state charged leptons).
    The trident CC, NC, and CC+NC channels correspond to diagrams (1)--(3), (4)--(5), and (1)--(5) of Fig.~\ref{fig_trident_diagrams}.
    {\bf Gray dashed} lines are the coherent and diffractive cross sections for $W$-boson production from Fig.~\ref{fig_sgmnuA_realW_Coh_Diff}, shown as a comparison.
    For antineutrinos, which have the same corresponding cross sections, take the CP transformation of the channel labels. See text for details.
}
\label{fig_sgmnuA_trident_Coh_Diff_all}
\end{figure*}

Figure~\ref{fig_sgmnuA_trident_Coh_Diff_all} shows our elastic (coherent + diffractive components) cross sections for all trident channels.

\newpage
\twocolumngrid

\bibliography{references}

\begin{thebibliography}{116}%
\makeatletter
\providecommand \@ifxundefined [1]{%
 \@ifx{#1\undefined}
}%
\providecommand \@ifnum [1]{%
 \ifnum #1\expandafter \@firstoftwo
 \else \expandafter \@secondoftwo
 \fi
}%
\providecommand \@ifx [1]{%
 \ifx #1\expandafter \@firstoftwo
 \else \expandafter \@secondoftwo
 \fi
}%
\providecommand \natexlab [1]{#1}%
\providecommand \enquote  [1]{``#1''}%
\providecommand \bibnamefont  [1]{#1}%
\providecommand \bibfnamefont [1]{#1}%
\providecommand \citenamefont [1]{#1}%
\providecommand \href@noop [0]{\@secondoftwo}%
\providecommand \href [0]{\begingroup \@sanitize@url \@href}%
\providecommand \@href[1]{\@@startlink{#1}\@@href}%
\providecommand \@@href[1]{\endgroup#1\@@endlink}%
\providecommand \@sanitize@url [0]{\catcode `\\12\catcode `\$12\catcode
  `\&12\catcode `\#12\catcode `\^12\catcode `\_12\catcode `\%12\relax}%
\providecommand \@@startlink[1]{}%
\providecommand \@@endlink[0]{}%
\providecommand \url  [0]{\begingroup\@sanitize@url \@url }%
\providecommand \@url [1]{\endgroup\@href {#1}{\urlprefix }}%
\providecommand \urlprefix  [0]{URL }%
\providecommand \Eprint [0]{\href }%
\providecommand \doibase [0]{http://dx.doi.org/}%
\providecommand \selectlanguage [0]{\@gobble}%
\providecommand \bibinfo  [0]{\@secondoftwo}%
\providecommand \bibfield  [0]{\@secondoftwo}%
\providecommand \translation [1]{[#1]}%
\providecommand \BibitemOpen [0]{}%
\providecommand \bibitemStop [0]{}%
\providecommand \bibitemNoStop [0]{.\EOS\space}%
\providecommand \EOS [0]{\spacefactor3000\relax}%
\providecommand \BibitemShut  [1]{\csname bibitem#1\endcsname}%
\let\auto@bib@innerbib\@empty
\bibitem [{\citenamefont {Zhou}\ and\ \citenamefont
  {Beacom}(2019)}]{Beacom:2019pzs}%
  \BibitemOpen
  \bibfield  {author} {\bibinfo {author} {\bibfnamefont {Bei}\ \bibnamefont
  {Zhou}}\ and\ \bibinfo {author} {\bibfnamefont {John~F.}\ \bibnamefont
  {Beacom}},\ }\bibfield  {title} {\enquote {\bibinfo {title} {{W-boson and
  trident production in TeV--PeV neutrino observatories}},}\ }\href@noop {} {\
  (\bibinfo {year} {2019})},\ \Eprint {http://arxiv.org/abs/1910.10720}
  {arXiv:1910.10720 [hep-ph]} \BibitemShut {NoStop}%
\bibitem [{\citenamefont {Llewellyn~Smith}(1972)}]{LlewellynSmith:1971uhs}%
  \BibitemOpen
  \bibfield  {author} {\bibinfo {author} {\bibfnamefont {C.~H.}\ \bibnamefont
  {Llewellyn~Smith}},\ }\bibfield  {title} {\enquote {\bibinfo {title}
  {{Neutrino Reactions at Accelerator Energies}},}\ }\bibfield  {booktitle}
  {\emph {\bibinfo {booktitle} {{Gauge Theories and Neutrino Physics, Jacob,
  1978:0175}}},\ }\href {\doibase 10.1016/0370-1573(72)90010-5} {\bibfield
  {journal} {\bibinfo  {journal} {Phys. Rept.}\ }\textbf {\bibinfo {volume}
  {3}},\ \bibinfo {pages} {261--379} (\bibinfo {year} {1972})}\BibitemShut
  {NoStop}%
\bibitem [{\citenamefont {Hayato}(2002)}]{Hayato:2002sd}%
  \BibitemOpen
  \bibfield  {author} {\bibinfo {author} {\bibfnamefont {Y.}~\bibnamefont
  {Hayato}},\ }\bibfield  {title} {\enquote {\bibinfo {title} {{NEUT}},}\
  }\bibfield  {booktitle} {\emph {\bibinfo {booktitle} {{Proceedings, 1st
  International Workshop on Neutrino-nucleus interactions in the few GeV region
  (NuInt 01): Tsukuba, Japan, December 13-16, 2001}}},\ }\href {\doibase
  10.1016/S0920-5632(02)01759-0} {\bibfield  {journal} {\bibinfo  {journal}
  {Nucl. Phys. Proc. Suppl.}\ }\textbf {\bibinfo {volume} {112}},\ \bibinfo
  {pages} {171--176} (\bibinfo {year} {2002})}\BibitemShut {NoStop}%
\bibitem [{\citenamefont {Andreopoulos}\ \emph {et~al.}(2010)\citenamefont
  {Andreopoulos} \emph {et~al.}}]{Andreopoulos:2009rq}%
  \BibitemOpen
  \bibfield  {author} {\bibinfo {author} {\bibfnamefont {C.}~\bibnamefont
  {Andreopoulos}} \emph {et~al.},\ }\bibfield  {title} {\enquote {\bibinfo
  {title} {{The GENIE Neutrino Monte Carlo Generator}},}\ }\href {\doibase
  10.1016/j.nima.2009.12.009} {\bibfield  {journal} {\bibinfo  {journal} {Nucl.
  Instrum. Meth.}\ }\textbf {\bibinfo {volume} {A614}},\ \bibinfo {pages}
  {87--104} (\bibinfo {year} {2010})},\ \Eprint
  {http://arxiv.org/abs/0905.2517} {arXiv:0905.2517 [hep-ph]} \BibitemShut
  {NoStop}%
\bibitem [{\citenamefont {Buss}\ \emph {et~al.}(2012)\citenamefont {Buss},
  \citenamefont {Gaitanos}, \citenamefont {Gallmeister}, \citenamefont {van
  Hees}, \citenamefont {Kaskulov}, \citenamefont {Lalakulich}, \citenamefont
  {Larionov}, \citenamefont {Leitner}, \citenamefont {Weil},\ and\
  \citenamefont {Mosel}}]{Buss:2011mx}%
  \BibitemOpen
  \bibfield  {author} {\bibinfo {author} {\bibfnamefont {O.}~\bibnamefont
  {Buss}}, \bibinfo {author} {\bibfnamefont {T.}~\bibnamefont {Gaitanos}},
  \bibinfo {author} {\bibfnamefont {K.}~\bibnamefont {Gallmeister}}, \bibinfo
  {author} {\bibfnamefont {H.}~\bibnamefont {van Hees}}, \bibinfo {author}
  {\bibfnamefont {M.}~\bibnamefont {Kaskulov}}, \bibinfo {author}
  {\bibfnamefont {O.}~\bibnamefont {Lalakulich}}, \bibinfo {author}
  {\bibfnamefont {A.~B.}\ \bibnamefont {Larionov}}, \bibinfo {author}
  {\bibfnamefont {T.}~\bibnamefont {Leitner}}, \bibinfo {author} {\bibfnamefont
  {J.}~\bibnamefont {Weil}}, \ and\ \bibinfo {author} {\bibfnamefont
  {U.}~\bibnamefont {Mosel}},\ }\bibfield  {title} {\enquote {\bibinfo {title}
  {{Transport-theoretical Description of Nuclear Reactions}},}\ }\href
  {\doibase 10.1016/j.physrep.2011.12.001} {\bibfield  {journal} {\bibinfo
  {journal} {Phys. Rept.}\ }\textbf {\bibinfo {volume} {512}},\ \bibinfo
  {pages} {1--124} (\bibinfo {year} {2012})},\ \Eprint
  {http://arxiv.org/abs/1106.1344} {arXiv:1106.1344 [hep-ph]} \BibitemShut
  {NoStop}%
\bibitem [{\citenamefont {Lalakulich}\ \emph {et~al.}(2011)\citenamefont
  {Lalakulich}, \citenamefont {Gallmeister},\ and\ \citenamefont
  {Mosel}}]{Lalakulich:2011eh}%
  \BibitemOpen
  \bibfield  {author} {\bibinfo {author} {\bibfnamefont {O.}~\bibnamefont
  {Lalakulich}}, \bibinfo {author} {\bibfnamefont {K.}~\bibnamefont
  {Gallmeister}}, \ and\ \bibinfo {author} {\bibfnamefont {U.}~\bibnamefont
  {Mosel}},\ }\bibfield  {title} {\enquote {\bibinfo {title} {{Neutrino Nucleus
  Reactions within the GiBUU Model}},}\ }\bibfield  {booktitle} {\emph
  {\bibinfo {booktitle} {{Proceedings, 13th International Workshop on Neutrino
  Factories, Superbeams and Beta beams (NuFact11): Geneva, Switzerland, August
  1-6, 2011}}},\ }\href {\doibase 10.1088/1742-6596/408/1/012053} {\  (\bibinfo
  {year} {2011}),\ 10.1088/1742-6596/408/1/012053},\ \bibinfo {note} {[J. Phys.
  Conf. Ser.408,012053(2013)]},\ \Eprint {http://arxiv.org/abs/1110.0674}
  {arXiv:1110.0674 [hep-ph]} \BibitemShut {NoStop}%
\bibitem [{\citenamefont {Golan}\ \emph {et~al.}(2012)\citenamefont {Golan},
  \citenamefont {Sobczyk},\ and\ \citenamefont {Zmuda}}]{Golan:2012rfa}%
  \BibitemOpen
  \bibfield  {author} {\bibinfo {author} {\bibfnamefont {T.}~\bibnamefont
  {Golan}}, \bibinfo {author} {\bibfnamefont {J.~T.}\ \bibnamefont {Sobczyk}},
  \ and\ \bibinfo {author} {\bibfnamefont {J.}~\bibnamefont {Zmuda}},\
  }\bibfield  {title} {\enquote {\bibinfo {title} {{NuWro: the Wroclaw Monte
  Carlo Generator of Neutrino Interactions}},}\ }\bibfield  {booktitle} {\emph
  {\bibinfo {booktitle} {{Proceedings, 24th International Conference on
  Neutrino physics and astrophysics (Neutrino 2010): Athens, Greece, June
  14-19, 2010}}},\ }\href {\doibase 10.1016/j.nuclphysbps.2012.09.136}
  {\bibfield  {journal} {\bibinfo  {journal} {Nucl. Phys. Proc. Suppl.}\
  }\textbf {\bibinfo {volume} {229-232}},\ \bibinfo {pages} {499--499}
  (\bibinfo {year} {2012})}\BibitemShut {NoStop}%
\bibitem [{\citenamefont {Formaggio}\ and\ \citenamefont
  {Zeller}(2012)}]{Formaggio:2013kya}%
  \BibitemOpen
  \bibfield  {author} {\bibinfo {author} {\bibfnamefont {J.~A.}\ \bibnamefont
  {Formaggio}}\ and\ \bibinfo {author} {\bibfnamefont {G.~P.}\ \bibnamefont
  {Zeller}},\ }\bibfield  {title} {\enquote {\bibinfo {title} {{From eV to EeV:
  Neutrino Cross Sections Across Energy Scales}},}\ }\href {\doibase
  10.1103/RevModPhys.84.1307} {\bibfield  {journal} {\bibinfo  {journal} {Rev.
  Mod. Phys.}\ }\textbf {\bibinfo {volume} {84}},\ \bibinfo {pages}
  {1307--1341} (\bibinfo {year} {2012})},\ \Eprint
  {http://arxiv.org/abs/1305.7513} {arXiv:1305.7513 [hep-ex]} \BibitemShut
  {NoStop}%
\bibitem [{\citenamefont {Cornet}\ \emph {et~al.}(2001)\citenamefont {Cornet},
  \citenamefont {Illana},\ and\ \citenamefont {Masip}}]{Cornet:2001gy}%
  \BibitemOpen
  \bibfield  {author} {\bibinfo {author} {\bibfnamefont {F.}~\bibnamefont
  {Cornet}}, \bibinfo {author} {\bibfnamefont {Jose~I.}\ \bibnamefont
  {Illana}}, \ and\ \bibinfo {author} {\bibfnamefont {M.}~\bibnamefont
  {Masip}},\ }\bibfield  {title} {\enquote {\bibinfo {title} {{TeV strings and
  the neutrino nucleon cross-section at ultrahigh-energies}},}\ }\href
  {\doibase 10.1103/PhysRevLett.86.4235} {\bibfield  {journal} {\bibinfo
  {journal} {Phys. Rev. Lett.}\ }\textbf {\bibinfo {volume} {86}},\ \bibinfo
  {pages} {4235--4238} (\bibinfo {year} {2001})},\ \Eprint
  {http://arxiv.org/abs/hep-ph/0102065} {arXiv:hep-ph/0102065 [hep-ph]}
  \BibitemShut {NoStop}%
\bibitem [{\citenamefont {Alvarez-Muniz}\ \emph {et~al.}(2002)\citenamefont
  {Alvarez-Muniz}, \citenamefont {Feng}, \citenamefont {Halzen}, \citenamefont
  {Han},\ and\ \citenamefont {Hooper}}]{AlvarezMuniz:2002ga}%
  \BibitemOpen
  \bibfield  {author} {\bibinfo {author} {\bibfnamefont {Jaime}\ \bibnamefont
  {Alvarez-Muniz}}, \bibinfo {author} {\bibfnamefont {Jonathan~L.}\
  \bibnamefont {Feng}}, \bibinfo {author} {\bibfnamefont {Francis}\
  \bibnamefont {Halzen}}, \bibinfo {author} {\bibfnamefont {Tao}\ \bibnamefont
  {Han}}, \ and\ \bibinfo {author} {\bibfnamefont {Dan}\ \bibnamefont
  {Hooper}},\ }\bibfield  {title} {\enquote {\bibinfo {title} {{Detecting
  microscopic black holes with neutrino telescopes}},}\ }\href {\doibase
  10.1103/PhysRevD.65.124015} {\bibfield  {journal} {\bibinfo  {journal} {Phys.
  Rev.}\ }\textbf {\bibinfo {volume} {D65}},\ \bibinfo {pages} {124015}
  (\bibinfo {year} {2002})},\ \Eprint {http://arxiv.org/abs/hep-ph/0202081}
  {arXiv:hep-ph/0202081 [hep-ph]} \BibitemShut {NoStop}%
\bibitem [{\citenamefont {Altmannshofer}\ \emph {et~al.}(2014)\citenamefont
  {Altmannshofer}, \citenamefont {Gori}, \citenamefont {Pospelov},\ and\
  \citenamefont {Yavin}}]{Altmannshofer:2014pba}%
  \BibitemOpen
  \bibfield  {author} {\bibinfo {author} {\bibfnamefont {Wolfgang}\
  \bibnamefont {Altmannshofer}}, \bibinfo {author} {\bibfnamefont {Stefania}\
  \bibnamefont {Gori}}, \bibinfo {author} {\bibfnamefont {Maxim}\ \bibnamefont
  {Pospelov}}, \ and\ \bibinfo {author} {\bibfnamefont {Itay}\ \bibnamefont
  {Yavin}},\ }\bibfield  {title} {\enquote {\bibinfo {title} {{Neutrino Trident
  Production: A Powerful Probe of New Physics with Neutrino Beams}},}\ }\href
  {\doibase 10.1103/PhysRevLett.113.091801} {\bibfield  {journal} {\bibinfo
  {journal} {Phys. Rev. Lett.}\ }\textbf {\bibinfo {volume} {113}},\ \bibinfo
  {pages} {091801} (\bibinfo {year} {2014})},\ \Eprint
  {http://arxiv.org/abs/1406.2332} {arXiv:1406.2332 [hep-ph]} \BibitemShut
  {NoStop}%
\bibitem [{\citenamefont {Coloma}\ \emph {et~al.}(2017)\citenamefont {Coloma},
  \citenamefont {Machado}, \citenamefont {Martinez-Soler},\ and\ \citenamefont
  {Shoemaker}}]{Coloma:2017ppo}%
  \BibitemOpen
  \bibfield  {author} {\bibinfo {author} {\bibfnamefont {Pilar}\ \bibnamefont
  {Coloma}}, \bibinfo {author} {\bibfnamefont {Pedro A.~N.}\ \bibnamefont
  {Machado}}, \bibinfo {author} {\bibfnamefont {Ivan}\ \bibnamefont
  {Martinez-Soler}}, \ and\ \bibinfo {author} {\bibfnamefont {Ian~M.}\
  \bibnamefont {Shoemaker}},\ }\bibfield  {title} {\enquote {\bibinfo {title}
  {{Double-Cascade Events from New Physics in Icecube}},}\ }\href {\doibase
  10.1103/PhysRevLett.119.201804} {\bibfield  {journal} {\bibinfo  {journal}
  {Phys. Rev. Lett.}\ }\textbf {\bibinfo {volume} {119}},\ \bibinfo {pages}
  {201804} (\bibinfo {year} {2017})},\ \Eprint
  {http://arxiv.org/abs/1707.08573} {arXiv:1707.08573 [hep-ph]} \BibitemShut
  {NoStop}%
\bibitem [{\citenamefont {Bertuzzo}\ \emph {et~al.}(2018)\citenamefont
  {Bertuzzo}, \citenamefont {Jana}, \citenamefont {Machado},\ and\
  \citenamefont {Zukanovich~Funchal}}]{Bertuzzo:2018itn}%
  \BibitemOpen
  \bibfield  {author} {\bibinfo {author} {\bibfnamefont {Enrico}\ \bibnamefont
  {Bertuzzo}}, \bibinfo {author} {\bibfnamefont {Sudip}\ \bibnamefont {Jana}},
  \bibinfo {author} {\bibfnamefont {Pedro A.~N.}\ \bibnamefont {Machado}}, \
  and\ \bibinfo {author} {\bibfnamefont {Renata}\ \bibnamefont
  {Zukanovich~Funchal}},\ }\bibfield  {title} {\enquote {\bibinfo {title}
  {{Dark Neutrino Portal to Explain MiniBooNE excess}},}\ }\href {\doibase
  10.1103/PhysRevLett.121.241801} {\bibfield  {journal} {\bibinfo  {journal}
  {Phys. Rev. Lett.}\ }\textbf {\bibinfo {volume} {121}},\ \bibinfo {pages}
  {241801} (\bibinfo {year} {2018})},\ \Eprint
  {http://arxiv.org/abs/1807.09877} {arXiv:1807.09877 [hep-ph]} \BibitemShut
  {NoStop}%
\bibitem [{\citenamefont {Capozzi}\ \emph {et~al.}(2018)\citenamefont
  {Capozzi}, \citenamefont {Li}, \citenamefont {Zhu},\ and\ \citenamefont
  {Beacom}}]{Capozzi:2018dat}%
  \BibitemOpen
  \bibfield  {author} {\bibinfo {author} {\bibfnamefont {Francesco}\
  \bibnamefont {Capozzi}}, \bibinfo {author} {\bibfnamefont {Shirley~Weishi}\
  \bibnamefont {Li}}, \bibinfo {author} {\bibfnamefont {Guanying}\ \bibnamefont
  {Zhu}}, \ and\ \bibinfo {author} {\bibfnamefont {John~F.}\ \bibnamefont
  {Beacom}},\ }\bibfield  {title} {\enquote {\bibinfo {title} {{DUNE as the
  Next-Generation Solar Neutrino Experiment}},}\ }\href@noop {} {\  (\bibinfo
  {year} {2018})},\ \Eprint {http://arxiv.org/abs/1808.08232} {arXiv:1808.08232
  [hep-ph]} \BibitemShut {NoStop}%
\bibitem [{\citenamefont {Machado}\ \emph {et~al.}(2019)\citenamefont
  {Machado}, \citenamefont {Palamara},\ and\ \citenamefont
  {Schmitz}}]{Machado:2019oxb}%
  \BibitemOpen
  \bibfield  {author} {\bibinfo {author} {\bibfnamefont {Pedro~AN}\
  \bibnamefont {Machado}}, \bibinfo {author} {\bibfnamefont {Ornella}\
  \bibnamefont {Palamara}}, \ and\ \bibinfo {author} {\bibfnamefont {David~W}\
  \bibnamefont {Schmitz}},\ }\bibfield  {title} {\enquote {\bibinfo {title}
  {{The Short-Baseline Neutrino Program at Fermilab}},}\ }\href@noop {}
  {\bibfield  {journal} {\bibinfo  {journal} {Ann. Rev. Nucl. Part. Sci.}\
  }\textbf {\bibinfo {volume} {69}} (\bibinfo {year} {2019})},\ \Eprint
  {http://arxiv.org/abs/1903.04608} {arXiv:1903.04608 [hep-ex]} \BibitemShut
  {NoStop}%
\bibitem [{\citenamefont {Gaisser}\ \emph {et~al.}(1995)\citenamefont
  {Gaisser}, \citenamefont {Halzen},\ and\ \citenamefont
  {Stanev}}]{Gaisser:1994yf}%
  \BibitemOpen
  \bibfield  {author} {\bibinfo {author} {\bibfnamefont {Thomas~K.}\
  \bibnamefont {Gaisser}}, \bibinfo {author} {\bibfnamefont {Francis}\
  \bibnamefont {Halzen}}, \ and\ \bibinfo {author} {\bibfnamefont {Todor}\
  \bibnamefont {Stanev}},\ }\bibfield  {title} {\enquote {\bibinfo {title}
  {{Particle astrophysics with high-energy neutrinos}},}\ }\href {\doibase
  10.1016/0370-1573(95)00003-Y} {\bibfield  {journal} {\bibinfo  {journal}
  {Phys. Rept.}\ }\textbf {\bibinfo {volume} {258}},\ \bibinfo {pages}
  {173--236} (\bibinfo {year} {1995})},\ \bibinfo {note} {[Erratum: Phys.
  Rept.271,355(1996)]},\ \Eprint {http://arxiv.org/abs/hep-ph/9410384}
  {arXiv:hep-ph/9410384 [hep-ph]} \BibitemShut {NoStop}%
\bibitem [{\citenamefont {Gandhi}\ \emph {et~al.}(1996)\citenamefont {Gandhi},
  \citenamefont {Quigg}, \citenamefont {Reno},\ and\ \citenamefont
  {Sarcevic}}]{Gandhi:1995tf}%
  \BibitemOpen
  \bibfield  {author} {\bibinfo {author} {\bibfnamefont {Raj}\ \bibnamefont
  {Gandhi}}, \bibinfo {author} {\bibfnamefont {Chris}\ \bibnamefont {Quigg}},
  \bibinfo {author} {\bibfnamefont {Mary~Hall}\ \bibnamefont {Reno}}, \ and\
  \bibinfo {author} {\bibfnamefont {Ina}\ \bibnamefont {Sarcevic}},\ }\bibfield
   {title} {\enquote {\bibinfo {title} {{Ultrahigh-energy neutrino
  interactions}},}\ }\href {\doibase 10.1016/0927-6505(96)00008-4} {\bibfield
  {journal} {\bibinfo  {journal} {Astropart. Phys.}\ }\textbf {\bibinfo
  {volume} {5}},\ \bibinfo {pages} {81--110} (\bibinfo {year} {1996})},\
  \Eprint {http://arxiv.org/abs/hep-ph/9512364} {arXiv:hep-ph/9512364 [hep-ph]}
  \BibitemShut {NoStop}%
\bibitem [{\citenamefont {Bahcall}(1997)}]{Bahcall:1997eg}%
  \BibitemOpen
  \bibfield  {author} {\bibinfo {author} {\bibfnamefont {John~N.}\ \bibnamefont
  {Bahcall}},\ }\bibfield  {title} {\enquote {\bibinfo {title} {{Gallium solar
  neutrino experiments: Absorption cross-sections, neutrino spectra, and
  predicted event rates}},}\ }\href {\doibase 10.1103/PhysRevC.56.3391}
  {\bibfield  {journal} {\bibinfo  {journal} {Phys. Rev.}\ }\textbf {\bibinfo
  {volume} {C56}},\ \bibinfo {pages} {3391--3409} (\bibinfo {year} {1997})},\
  \Eprint {http://arxiv.org/abs/hep-ph/9710491} {arXiv:hep-ph/9710491 [hep-ph]}
  \BibitemShut {NoStop}%
\bibitem [{\citenamefont {Gandhi}\ \emph {et~al.}(1998)\citenamefont {Gandhi},
  \citenamefont {Quigg}, \citenamefont {Reno},\ and\ \citenamefont
  {Sarcevic}}]{Gandhi:1998ri}%
  \BibitemOpen
  \bibfield  {author} {\bibinfo {author} {\bibfnamefont {Raj}\ \bibnamefont
  {Gandhi}}, \bibinfo {author} {\bibfnamefont {Chris}\ \bibnamefont {Quigg}},
  \bibinfo {author} {\bibfnamefont {Mary~Hall}\ \bibnamefont {Reno}}, \ and\
  \bibinfo {author} {\bibfnamefont {Ina}\ \bibnamefont {Sarcevic}},\ }\bibfield
   {title} {\enquote {\bibinfo {title} {{Neutrino interactions at
  ultrahigh-energies}},}\ }\href {\doibase 10.1103/PhysRevD.58.093009}
  {\bibfield  {journal} {\bibinfo  {journal} {Phys. Rev.}\ }\textbf {\bibinfo
  {volume} {D58}},\ \bibinfo {pages} {093009} (\bibinfo {year} {1998})},\
  \Eprint {http://arxiv.org/abs/hep-ph/9807264} {arXiv:hep-ph/9807264 [hep-ph]}
  \BibitemShut {NoStop}%
\bibitem [{\citenamefont {Langanke}\ \emph {et~al.}(2004)\citenamefont
  {Langanke}, \citenamefont {Martinez-Pinedo}, \citenamefont {von
  Neumann-Cosel},\ and\ \citenamefont {Richter}}]{Langanke:2004vx}%
  \BibitemOpen
  \bibfield  {author} {\bibinfo {author} {\bibfnamefont {K.}~\bibnamefont
  {Langanke}}, \bibinfo {author} {\bibfnamefont {G.}~\bibnamefont
  {Martinez-Pinedo}}, \bibinfo {author} {\bibfnamefont {P.}~\bibnamefont {von
  Neumann-Cosel}}, \ and\ \bibinfo {author} {\bibfnamefont {A.}~\bibnamefont
  {Richter}},\ }\bibfield  {title} {\enquote {\bibinfo {title} {{Supernova
  inelastic neutrino nucleus cross-sections from high resolution electron
  scattering experiments and shell model calculations}},}\ }\href {\doibase
  10.1103/PhysRevLett.93.202501} {\bibfield  {journal} {\bibinfo  {journal}
  {Phys. Rev. Lett.}\ }\textbf {\bibinfo {volume} {93}},\ \bibinfo {pages}
  {202501} (\bibinfo {year} {2004})},\ \Eprint
  {http://arxiv.org/abs/nucl-th/0402001} {arXiv:nucl-th/0402001 [nucl-th]}
  \BibitemShut {NoStop}%
\bibitem [{\citenamefont {Cocco}\ \emph {et~al.}(2007)\citenamefont {Cocco},
  \citenamefont {Mangano},\ and\ \citenamefont {Messina}}]{Cocco:2007za}%
  \BibitemOpen
  \bibfield  {author} {\bibinfo {author} {\bibfnamefont {Alfredo~G.}\
  \bibnamefont {Cocco}}, \bibinfo {author} {\bibfnamefont {Gianpiero}\
  \bibnamefont {Mangano}}, \ and\ \bibinfo {author} {\bibfnamefont {Marcello}\
  \bibnamefont {Messina}},\ }\bibfield  {title} {\enquote {\bibinfo {title}
  {{Probing low energy neutrino backgrounds with neutrino capture on beta
  decaying nuclei}},}\ }\href {\doibase 10.1088/1475-7516/2007/06/015}
  {\bibfield  {journal} {\bibinfo  {journal} {JCAP}\ }\textbf {\bibinfo
  {volume} {0706}},\ \bibinfo {pages} {015} (\bibinfo {year} {2007})},\ \Eprint
  {http://arxiv.org/abs/hep-ph/0703075} {arXiv:hep-ph/0703075 [hep-ph]}
  \BibitemShut {NoStop}%
\bibitem [{\citenamefont {Yoshida}\ \emph {et~al.}(2008)\citenamefont
  {Yoshida}, \citenamefont {Suzuki}, \citenamefont {Chiba}, \citenamefont
  {Kajino}, \citenamefont {Yokomakura}, \citenamefont {Kimura}, \citenamefont
  {Takamura},\ and\ \citenamefont {Hartmann}}]{Yoshida:2008zb}%
  \BibitemOpen
  \bibfield  {author} {\bibinfo {author} {\bibfnamefont {Takashi}\ \bibnamefont
  {Yoshida}}, \bibinfo {author} {\bibfnamefont {Toshio}\ \bibnamefont
  {Suzuki}}, \bibinfo {author} {\bibfnamefont {Satoshi}\ \bibnamefont {Chiba}},
  \bibinfo {author} {\bibfnamefont {Toshitaka}\ \bibnamefont {Kajino}},
  \bibinfo {author} {\bibfnamefont {Hidekazu}\ \bibnamefont {Yokomakura}},
  \bibinfo {author} {\bibfnamefont {Keiichi}\ \bibnamefont {Kimura}}, \bibinfo
  {author} {\bibfnamefont {Akira}\ \bibnamefont {Takamura}}, \ and\ \bibinfo
  {author} {\bibfnamefont {Dieter~H.}\ \bibnamefont {Hartmann}},\ }\bibfield
  {title} {\enquote {\bibinfo {title} {{Neutrino-Nucleus Reaction Cross
  Sections for Light Element Synthesis in Supernova Explosions}},}\ }\href
  {\doibase 10.1086/591266} {\bibfield  {journal} {\bibinfo  {journal}
  {Astrophys. J.}\ }\textbf {\bibinfo {volume} {686}},\ \bibinfo {pages}
  {448--466} (\bibinfo {year} {2008})},\ \Eprint
  {http://arxiv.org/abs/0807.2723} {arXiv:0807.2723 [astro-ph]} \BibitemShut
  {NoStop}%
\bibitem [{\citenamefont {Cooper-Sarkar}\ \emph {et~al.}(2011)\citenamefont
  {Cooper-Sarkar}, \citenamefont {Mertsch},\ and\ \citenamefont
  {Sarkar}}]{CooperSarkar:2011pa}%
  \BibitemOpen
  \bibfield  {author} {\bibinfo {author} {\bibfnamefont {Amanda}\ \bibnamefont
  {Cooper-Sarkar}}, \bibinfo {author} {\bibfnamefont {Philipp}\ \bibnamefont
  {Mertsch}}, \ and\ \bibinfo {author} {\bibfnamefont {Subir}\ \bibnamefont
  {Sarkar}},\ }\bibfield  {title} {\enquote {\bibinfo {title} {{The high energy
  neutrino cross-section in the Standard Model and its uncertainty}},}\ }\href
  {\doibase 10.1007/JHEP08(2011)042} {\bibfield  {journal} {\bibinfo  {journal}
  {JHEP}\ }\textbf {\bibinfo {volume} {08}},\ \bibinfo {pages} {042} (\bibinfo
  {year} {2011})},\ \Eprint {http://arxiv.org/abs/1106.3723} {arXiv:1106.3723
  [hep-ph]} \BibitemShut {NoStop}%
\bibitem [{\citenamefont {Connolly}\ \emph {et~al.}(2011)\citenamefont
  {Connolly}, \citenamefont {Thorne},\ and\ \citenamefont
  {Waters}}]{Connolly:2011vc}%
  \BibitemOpen
  \bibfield  {author} {\bibinfo {author} {\bibfnamefont {Amy}\ \bibnamefont
  {Connolly}}, \bibinfo {author} {\bibfnamefont {Robert~S.}\ \bibnamefont
  {Thorne}}, \ and\ \bibinfo {author} {\bibfnamefont {David}\ \bibnamefont
  {Waters}},\ }\bibfield  {title} {\enquote {\bibinfo {title} {{Calculation of
  High Energy Neutrino-Nucleon Cross Sections and Uncertainties Using the MSTW
  Parton Distribution Functions and Implications for Future Experiments}},}\
  }\href {\doibase 10.1103/PhysRevD.83.113009} {\bibfield  {journal} {\bibinfo
  {journal} {Phys. Rev.}\ }\textbf {\bibinfo {volume} {D83}},\ \bibinfo {pages}
  {113009} (\bibinfo {year} {2011})},\ \Eprint {http://arxiv.org/abs/1102.0691}
  {arXiv:1102.0691 [hep-ph]} \BibitemShut {NoStop}%
\bibitem [{\citenamefont {Laha}\ \emph {et~al.}(2013)\citenamefont {Laha},
  \citenamefont {Beacom}, \citenamefont {Dasgupta}, \citenamefont {Horiuchi},\
  and\ \citenamefont {Murase}}]{Laha:2013eev}%
  \BibitemOpen
  \bibfield  {author} {\bibinfo {author} {\bibfnamefont {Ranjan}\ \bibnamefont
  {Laha}}, \bibinfo {author} {\bibfnamefont {John~F.}\ \bibnamefont {Beacom}},
  \bibinfo {author} {\bibfnamefont {Basudeb}\ \bibnamefont {Dasgupta}},
  \bibinfo {author} {\bibfnamefont {Shunsaku}\ \bibnamefont {Horiuchi}}, \ and\
  \bibinfo {author} {\bibfnamefont {Kohta}\ \bibnamefont {Murase}},\ }\bibfield
   {title} {\enquote {\bibinfo {title} {{Demystifying the PeV Cascades in
  IceCube: Less (Energy) is More (Events)}},}\ }\href {\doibase
  10.1103/PhysRevD.88.043009} {\bibfield  {journal} {\bibinfo  {journal} {Phys.
  Rev.}\ }\textbf {\bibinfo {volume} {D88}},\ \bibinfo {pages} {043009}
  (\bibinfo {year} {2013})},\ \Eprint {http://arxiv.org/abs/1306.2309}
  {arXiv:1306.2309 [astro-ph.HE]} \BibitemShut {NoStop}%
\bibitem [{\citenamefont {Chen}\ \emph {et~al.}(2014)\citenamefont {Chen},
  \citenamefont {Bhupal~Dev},\ and\ \citenamefont {Soni}}]{Chen:2013dza}%
  \BibitemOpen
  \bibfield  {author} {\bibinfo {author} {\bibfnamefont {Chien-Yi}\
  \bibnamefont {Chen}}, \bibinfo {author} {\bibfnamefont {P.~S.}\ \bibnamefont
  {Bhupal~Dev}}, \ and\ \bibinfo {author} {\bibfnamefont {Amarjit}\
  \bibnamefont {Soni}},\ }\bibfield  {title} {\enquote {\bibinfo {title}
  {{Standard model explanation of the ultrahigh energy neutrino events at
  IceCube}},}\ }\href {\doibase 10.1103/PhysRevD.89.033012} {\bibfield
  {journal} {\bibinfo  {journal} {Phys. Rev.}\ }\textbf {\bibinfo {volume}
  {D89}},\ \bibinfo {pages} {033012} (\bibinfo {year} {2014})},\ \Eprint
  {http://arxiv.org/abs/1309.1764} {arXiv:1309.1764 [hep-ph]} \BibitemShut
  {NoStop}%
\bibitem [{\citenamefont {Seligman}(1997)}]{Seligman:1997fe}%
  \BibitemOpen
  \bibfield  {author} {\bibinfo {author} {\bibfnamefont {William~Glenn}\
  \bibnamefont {Seligman}},\ }\emph {\bibinfo {title} {{A Next-to-Leading Order
  QCD Analysis of Neutrino - Iron Structure Functions at the Tevatron}}},\
  \href {\doibase 10.2172/1421736} {Ph.D. thesis},\ \bibinfo  {school} {Nevis
  Labs, Columbia U.} (\bibinfo {year} {1997})\BibitemShut {NoStop}%
\bibitem [{\citenamefont {Tzanov}\ \emph {et~al.}(2006)\citenamefont {Tzanov}
  \emph {et~al.}}]{Tzanov:2005kr}%
  \BibitemOpen
  \bibfield  {author} {\bibinfo {author} {\bibfnamefont {M.}~\bibnamefont
  {Tzanov}} \emph {et~al.} (\bibinfo {collaboration} {NuTeV}),\ }\bibfield
  {title} {\enquote {\bibinfo {title} {{Precise measurement of neutrino and
  anti-neutrino differential cross sections}},}\ }\href {\doibase
  10.1103/PhysRevD.74.012008} {\bibfield  {journal} {\bibinfo  {journal} {Phys.
  Rev.}\ }\textbf {\bibinfo {volume} {D74}},\ \bibinfo {pages} {012008}
  (\bibinfo {year} {2006})},\ \Eprint {http://arxiv.org/abs/hep-ex/0509010}
  {arXiv:hep-ex/0509010 [hep-ex]} \BibitemShut {NoStop}%
\bibitem [{\citenamefont {Tanabashi}\ \emph {et~al.}(2018)\citenamefont
  {Tanabashi} \emph {et~al.}}]{Tanabashi:2018oca}%
  \BibitemOpen
  \bibfield  {author} {\bibinfo {author} {\bibfnamefont {M.}~\bibnamefont
  {Tanabashi}} \emph {et~al.} (\bibinfo {collaboration} {Particle Data
  Group}),\ }\bibfield  {title} {\enquote {\bibinfo {title} {{Review of
  Particle Physics}},}\ }\href {\doibase 10.1103/PhysRevD.98.030001} {\bibfield
   {journal} {\bibinfo  {journal} {Phys. Rev.}\ }\textbf {\bibinfo {volume}
  {D98}},\ \bibinfo {pages} {030001} (\bibinfo {year} {2018})}\BibitemShut
  {NoStop}%
\bibitem [{\citenamefont {Aartsen}\ \emph {et~al.}(2014)\citenamefont {Aartsen}
  \emph {et~al.}}]{Aartsen:2014gkd}%
  \BibitemOpen
  \bibfield  {author} {\bibinfo {author} {\bibfnamefont {M.~G.}\ \bibnamefont
  {Aartsen}} \emph {et~al.} (\bibinfo {collaboration} {IceCube}),\ }\bibfield
  {title} {\enquote {\bibinfo {title} {{Observation of High-Energy
  Astrophysical Neutrinos in Three Years of IceCube Data}},}\ }\href {\doibase
  10.1103/PhysRevLett.113.101101} {\bibfield  {journal} {\bibinfo  {journal}
  {Phys. Rev. Lett.}\ }\textbf {\bibinfo {volume} {113}},\ \bibinfo {pages}
  {101101} (\bibinfo {year} {2014})},\ \Eprint {http://arxiv.org/abs/1405.5303}
  {arXiv:1405.5303 [astro-ph.HE]} \BibitemShut {NoStop}%
\bibitem [{\citenamefont {Aartsen}\ \emph {et~al.}(2015)\citenamefont {Aartsen}
  \emph {et~al.}}]{Aartsen:2015knd}%
  \BibitemOpen
  \bibfield  {author} {\bibinfo {author} {\bibfnamefont {M.~G.}\ \bibnamefont
  {Aartsen}} \emph {et~al.} (\bibinfo {collaboration} {IceCube}),\ }\bibfield
  {title} {\enquote {\bibinfo {title} {{A combined maximum-likelihood analysis
  of the high-energy astrophysical neutrino flux measured with IceCube}},}\
  }\href {\doibase 10.1088/0004-637X/809/1/98} {\bibfield  {journal} {\bibinfo
  {journal} {Astrophys. J.}\ }\textbf {\bibinfo {volume} {809}},\ \bibinfo
  {pages} {98} (\bibinfo {year} {2015})},\ \Eprint
  {http://arxiv.org/abs/1507.03991} {arXiv:1507.03991 [astro-ph.HE]}
  \BibitemShut {NoStop}%
\bibitem [{\citenamefont {Hooper}(2002)}]{Hooper:2002yq}%
  \BibitemOpen
  \bibfield  {author} {\bibinfo {author} {\bibfnamefont {Dan}\ \bibnamefont
  {Hooper}},\ }\bibfield  {title} {\enquote {\bibinfo {title} {{Measuring
  high-energy neutrino nucleon cross-sections with future neutrino
  telescopes}},}\ }\href {\doibase 10.1103/PhysRevD.65.097303} {\bibfield
  {journal} {\bibinfo  {journal} {Phys. Rev.}\ }\textbf {\bibinfo {volume}
  {D65}},\ \bibinfo {pages} {097303} (\bibinfo {year} {2002})},\ \Eprint
  {http://arxiv.org/abs/hep-ph/0203239} {arXiv:hep-ph/0203239 [hep-ph]}
  \BibitemShut {NoStop}%
\bibitem [{\citenamefont {Borriello}\ \emph {et~al.}(2008)\citenamefont
  {Borriello}, \citenamefont {Cuoco}, \citenamefont {Mangano}, \citenamefont
  {Miele}, \citenamefont {Pastor}, \citenamefont {Pisanti},\ and\ \citenamefont
  {Serpico}}]{Borriello:2007cs}%
  \BibitemOpen
  \bibfield  {author} {\bibinfo {author} {\bibfnamefont {E.}~\bibnamefont
  {Borriello}}, \bibinfo {author} {\bibfnamefont {A.}~\bibnamefont {Cuoco}},
  \bibinfo {author} {\bibfnamefont {G.}~\bibnamefont {Mangano}}, \bibinfo
  {author} {\bibfnamefont {G.}~\bibnamefont {Miele}}, \bibinfo {author}
  {\bibfnamefont {S.}~\bibnamefont {Pastor}}, \bibinfo {author} {\bibfnamefont
  {O.}~\bibnamefont {Pisanti}}, \ and\ \bibinfo {author} {\bibfnamefont
  {P.~D.}\ \bibnamefont {Serpico}},\ }\bibfield  {title} {\enquote {\bibinfo
  {title} {{Disentangling neutrino-nucleon cross section and high energy
  neutrino flux with a km$^3$ neutrino telescope}},}\ }\href {\doibase
  10.1103/PhysRevD.77.045019} {\bibfield  {journal} {\bibinfo  {journal} {Phys.
  Rev.}\ }\textbf {\bibinfo {volume} {D77}},\ \bibinfo {pages} {045019}
  (\bibinfo {year} {2008})},\ \Eprint {http://arxiv.org/abs/0711.0152}
  {arXiv:0711.0152 [astro-ph]} \BibitemShut {NoStop}%
\bibitem [{\citenamefont {Klein}\ and\ \citenamefont
  {Connolly}(2013)}]{Klein:2013xoa}%
  \BibitemOpen
  \bibfield  {author} {\bibinfo {author} {\bibfnamefont {Spencer~R.}\
  \bibnamefont {Klein}}\ and\ \bibinfo {author} {\bibfnamefont {Amy}\
  \bibnamefont {Connolly}},\ }\bibfield  {title} {\enquote {\bibinfo {title}
  {{Neutrino Absorption in the Earth, Neutrino Cross-Sections, and New
  Physics}},}\ }in\ \href
  {http://www.slac.stanford.edu/econf/C1307292/docs/submittedArxivFiles/1304.4891.pdf}
  {\emph {\bibinfo {booktitle} {{Proceedings, 2013 Community Summer Study on
  the Future of U.S. Particle Physics: Snowmass on the Mississippi (CSS2013):
  Minneapolis, MN, USA, July 29-August 6, 2013}}}}\ (\bibinfo {year} {2013})\
  \Eprint {http://arxiv.org/abs/1304.4891} {arXiv:1304.4891 [astro-ph.HE]}
  \BibitemShut {NoStop}%
\bibitem [{\citenamefont {Aartsen}\ \emph {et~al.}(2017)\citenamefont {Aartsen}
  \emph {et~al.}}]{Aartsen:2017kpd}%
  \BibitemOpen
  \bibfield  {author} {\bibinfo {author} {\bibfnamefont {M.~G.}\ \bibnamefont
  {Aartsen}} \emph {et~al.} (\bibinfo {collaboration} {IceCube}),\ }\bibfield
  {title} {\enquote {\bibinfo {title} {{Measurement of the multi-TeV neutrino
  cross section with IceCube using Earth absorption}},}\ }\href {\doibase
  10.1038/nature24459} {\bibfield  {journal} {\bibinfo  {journal} {Nature}\
  }\textbf {\bibinfo {volume} {551}},\ \bibinfo {pages} {596--600} (\bibinfo
  {year} {2017})},\ \Eprint {http://arxiv.org/abs/1711.08119} {arXiv:1711.08119
  [hep-ex]} \BibitemShut {NoStop}%
\bibitem [{\citenamefont {Bustamante}\ and\ \citenamefont
  {Connolly}(2019)}]{Bustamante:2017xuy}%
  \BibitemOpen
  \bibfield  {author} {\bibinfo {author} {\bibfnamefont {Mauricio}\
  \bibnamefont {Bustamante}}\ and\ \bibinfo {author} {\bibfnamefont {Amy}\
  \bibnamefont {Connolly}},\ }\bibfield  {title} {\enquote {\bibinfo {title}
  {{Extracting the Energy-Dependent Neutrino-Nucleon Cross Section above 10 TeV
  Using IceCube Showers}},}\ }\href {\doibase 10.1103/PhysRevLett.122.041101}
  {\bibfield  {journal} {\bibinfo  {journal} {Phys. Rev. Lett.}\ }\textbf
  {\bibinfo {volume} {122}},\ \bibinfo {pages} {041101} (\bibinfo {year}
  {2019})},\ \Eprint {http://arxiv.org/abs/1711.11043} {arXiv:1711.11043
  [astro-ph.HE]} \BibitemShut {NoStop}%
\bibitem [{\citenamefont {Padovani}\ and\ \citenamefont
  {Resconi}(2014)}]{Padovani:2014bha}%
  \BibitemOpen
  \bibfield  {author} {\bibinfo {author} {\bibfnamefont {P.}~\bibnamefont
  {Padovani}}\ and\ \bibinfo {author} {\bibfnamefont {E.}~\bibnamefont
  {Resconi}},\ }\bibfield  {title} {\enquote {\bibinfo {title} {{Are both BL
  Lacs and pulsar wind nebulae the astrophysical counterparts of IceCube
  neutrino events?}}}\ }\href {\doibase 10.1093/mnras/stu1166} {\bibfield
  {journal} {\bibinfo  {journal} {Mon. Not. Roy. Astron. Soc.}\ }\textbf
  {\bibinfo {volume} {443}},\ \bibinfo {pages} {474--484} (\bibinfo {year}
  {2014})},\ \Eprint {http://arxiv.org/abs/1406.0376} {arXiv:1406.0376
  [astro-ph.HE]} \BibitemShut {NoStop}%
\bibitem [{\citenamefont {Tamborra}\ \emph {et~al.}(2014)\citenamefont
  {Tamborra}, \citenamefont {Ando},\ and\ \citenamefont
  {Murase}}]{Tamborra:2014xia}%
  \BibitemOpen
  \bibfield  {author} {\bibinfo {author} {\bibfnamefont {Irene}\ \bibnamefont
  {Tamborra}}, \bibinfo {author} {\bibfnamefont {Shin'ichiro}\ \bibnamefont
  {Ando}}, \ and\ \bibinfo {author} {\bibfnamefont {Kohta}\ \bibnamefont
  {Murase}},\ }\bibfield  {title} {\enquote {\bibinfo {title} {{Star-forming
  galaxies as the origin of diffuse high-energy backgrounds: Gamma-ray and
  neutrino connections, and implications for starburst history}},}\ }\href
  {\doibase 10.1088/1475-7516/2014/09/043} {\bibfield  {journal} {\bibinfo
  {journal} {JCAP}\ }\textbf {\bibinfo {volume} {1409}},\ \bibinfo {pages}
  {043} (\bibinfo {year} {2014})},\ \Eprint {http://arxiv.org/abs/1404.1189}
  {arXiv:1404.1189 [astro-ph.HE]} \BibitemShut {NoStop}%
\bibitem [{\citenamefont {Murase}\ \emph {et~al.}(2014)\citenamefont {Murase},
  \citenamefont {Inoue},\ and\ \citenamefont {Dermer}}]{Murase:2014foa}%
  \BibitemOpen
  \bibfield  {author} {\bibinfo {author} {\bibfnamefont {Kohta}\ \bibnamefont
  {Murase}}, \bibinfo {author} {\bibfnamefont {Yoshiyuki}\ \bibnamefont
  {Inoue}}, \ and\ \bibinfo {author} {\bibfnamefont {Charles~D.}\ \bibnamefont
  {Dermer}},\ }\bibfield  {title} {\enquote {\bibinfo {title} {{Diffuse
  Neutrino Intensity from the Inner Jets of Active Galactic Nuclei: Impacts of
  External Photon Fields and the Blazar Sequence}},}\ }\href {\doibase
  10.1103/PhysRevD.90.023007} {\bibfield  {journal} {\bibinfo  {journal} {Phys.
  Rev.}\ }\textbf {\bibinfo {volume} {D90}},\ \bibinfo {pages} {023007}
  (\bibinfo {year} {2014})},\ \Eprint {http://arxiv.org/abs/1403.4089}
  {arXiv:1403.4089 [astro-ph.HE]} \BibitemShut {NoStop}%
\bibitem [{\citenamefont {Ng}\ and\ \citenamefont {Beacom}(2014)}]{Ng:2014pca}%
  \BibitemOpen
  \bibfield  {author} {\bibinfo {author} {\bibfnamefont {Kenny C.~Y.}\
  \bibnamefont {Ng}}\ and\ \bibinfo {author} {\bibfnamefont {John~F.}\
  \bibnamefont {Beacom}},\ }\bibfield  {title} {\enquote {\bibinfo {title}
  {{Cosmic neutrino cascades from secret neutrino interactions}},}\ }\href
  {\doibase 10.1103/PhysRevD.90.065035, 10.1103/PhysRevD.90.089904} {\bibfield
  {journal} {\bibinfo  {journal} {Phys. Rev.}\ }\textbf {\bibinfo {volume}
  {D90}},\ \bibinfo {pages} {065035} (\bibinfo {year} {2014})},\ \bibinfo
  {note} {[Erratum: Phys. Rev.D90,no.8,089904(2014)]},\ \Eprint
  {http://arxiv.org/abs/1404.2288} {arXiv:1404.2288 [astro-ph.HE]} \BibitemShut
  {NoStop}%
\bibitem [{\citenamefont {Ioka}\ and\ \citenamefont
  {Murase}(2014)}]{Ioka:2014kca}%
  \BibitemOpen
  \bibfield  {author} {\bibinfo {author} {\bibfnamefont {Kunihto}\ \bibnamefont
  {Ioka}}\ and\ \bibinfo {author} {\bibfnamefont {Kohta}\ \bibnamefont
  {Murase}},\ }\bibfield  {title} {\enquote {\bibinfo {title} {{IceCube PeVøEeV
  neutrinos and secret interactions of neutrinos}},}\ }\href {\doibase
  10.1093/ptep/ptu090} {\bibfield  {journal} {\bibinfo  {journal} {PTEP}\
  }\textbf {\bibinfo {volume} {2014}},\ \bibinfo {pages} {061E01} (\bibinfo
  {year} {2014})},\ \Eprint {http://arxiv.org/abs/1404.2279} {arXiv:1404.2279
  [astro-ph.HE]} \BibitemShut {NoStop}%
\bibitem [{\citenamefont {Rott}\ \emph {et~al.}(2015)\citenamefont {Rott},
  \citenamefont {Kohri},\ and\ \citenamefont {Park}}]{Rott:2014kfa}%
  \BibitemOpen
  \bibfield  {author} {\bibinfo {author} {\bibfnamefont {Carsten}\ \bibnamefont
  {Rott}}, \bibinfo {author} {\bibfnamefont {Kazunori}\ \bibnamefont {Kohri}},
  \ and\ \bibinfo {author} {\bibfnamefont {Seong~Chan}\ \bibnamefont {Park}},\
  }\bibfield  {title} {\enquote {\bibinfo {title} {{Superheavy dark matter and
  IceCube neutrino signals: Bounds on decaying dark matter}},}\ }\href
  {\doibase 10.1103/PhysRevD.92.023529} {\bibfield  {journal} {\bibinfo
  {journal} {Phys. Rev.}\ }\textbf {\bibinfo {volume} {D92}},\ \bibinfo {pages}
  {023529} (\bibinfo {year} {2015})},\ \Eprint {http://arxiv.org/abs/1408.4575}
  {arXiv:1408.4575 [hep-ph]} \BibitemShut {NoStop}%
\bibitem [{\citenamefont {Gauld}\ \emph {et~al.}(2016)\citenamefont {Gauld},
  \citenamefont {Rojo}, \citenamefont {Rottoli}, \citenamefont {Sarkar},\ and\
  \citenamefont {Talbert}}]{Gauld:2015kvh}%
  \BibitemOpen
  \bibfield  {author} {\bibinfo {author} {\bibfnamefont {Rhorry}\ \bibnamefont
  {Gauld}}, \bibinfo {author} {\bibfnamefont {Juan}\ \bibnamefont {Rojo}},
  \bibinfo {author} {\bibfnamefont {Luca}\ \bibnamefont {Rottoli}}, \bibinfo
  {author} {\bibfnamefont {Subir}\ \bibnamefont {Sarkar}}, \ and\ \bibinfo
  {author} {\bibfnamefont {Jim}\ \bibnamefont {Talbert}},\ }\bibfield  {title}
  {\enquote {\bibinfo {title} {{The prompt atmospheric neutrino flux in the
  light of LHCb}},}\ }\href {\doibase 10.1007/JHEP02(2016)130} {\bibfield
  {journal} {\bibinfo  {journal} {JHEP}\ }\textbf {\bibinfo {volume} {02}},\
  \bibinfo {pages} {130} (\bibinfo {year} {2016})},\ \Eprint
  {http://arxiv.org/abs/1511.06346} {arXiv:1511.06346 [hep-ph]} \BibitemShut
  {NoStop}%
\bibitem [{\citenamefont {Murase}\ \emph {et~al.}(2016)\citenamefont {Murase},
  \citenamefont {Guetta},\ and\ \citenamefont {Ahlers}}]{Murase:2015xka}%
  \BibitemOpen
  \bibfield  {author} {\bibinfo {author} {\bibfnamefont {Kohta}\ \bibnamefont
  {Murase}}, \bibinfo {author} {\bibfnamefont {Dafne}\ \bibnamefont {Guetta}},
  \ and\ \bibinfo {author} {\bibfnamefont {Markus}\ \bibnamefont {Ahlers}},\
  }\bibfield  {title} {\enquote {\bibinfo {title} {{Hidden Cosmic-Ray
  Accelerators as an Origin of TeV-PeV Cosmic Neutrinos}},}\ }\href {\doibase
  10.1103/PhysRevLett.116.071101} {\bibfield  {journal} {\bibinfo  {journal}
  {Phys. Rev. Lett.}\ }\textbf {\bibinfo {volume} {116}},\ \bibinfo {pages}
  {071101} (\bibinfo {year} {2016})},\ \Eprint
  {http://arxiv.org/abs/1509.00805} {arXiv:1509.00805 [astro-ph.HE]}
  \BibitemShut {NoStop}%
\bibitem [{\citenamefont {Murase}\ \emph {et~al.}(2015)\citenamefont {Murase},
  \citenamefont {Laha}, \citenamefont {Ando},\ and\ \citenamefont
  {Ahlers}}]{Murase:2015gea}%
  \BibitemOpen
  \bibfield  {author} {\bibinfo {author} {\bibfnamefont {Kohta}\ \bibnamefont
  {Murase}}, \bibinfo {author} {\bibfnamefont {Ranjan}\ \bibnamefont {Laha}},
  \bibinfo {author} {\bibfnamefont {Shin'ichiro}\ \bibnamefont {Ando}}, \ and\
  \bibinfo {author} {\bibfnamefont {Markus}\ \bibnamefont {Ahlers}},\
  }\bibfield  {title} {\enquote {\bibinfo {title} {{Testing the Dark Matter
  Scenario for PeV Neutrinos Observed in IceCube}},}\ }\href {\doibase
  10.1103/PhysRevLett.115.071301} {\bibfield  {journal} {\bibinfo  {journal}
  {Phys. Rev. Lett.}\ }\textbf {\bibinfo {volume} {115}},\ \bibinfo {pages}
  {071301} (\bibinfo {year} {2015})},\ \Eprint
  {http://arxiv.org/abs/1503.04663} {arXiv:1503.04663 [hep-ph]} \BibitemShut
  {NoStop}%
\bibitem [{\citenamefont {Bustamante}\ \emph {et~al.}(2015)\citenamefont
  {Bustamante}, \citenamefont {Beacom},\ and\ \citenamefont
  {Winter}}]{Bustamante:2015waa}%
  \BibitemOpen
  \bibfield  {author} {\bibinfo {author} {\bibfnamefont {Mauricio}\
  \bibnamefont {Bustamante}}, \bibinfo {author} {\bibfnamefont {John~F.}\
  \bibnamefont {Beacom}}, \ and\ \bibinfo {author} {\bibfnamefont {Walter}\
  \bibnamefont {Winter}},\ }\bibfield  {title} {\enquote {\bibinfo {title}
  {{Theoretically palatable flavor combinations of astrophysical neutrinos}},}\
  }\href {\doibase 10.1103/PhysRevLett.115.161302} {\bibfield  {journal}
  {\bibinfo  {journal} {Phys. Rev. Lett.}\ }\textbf {\bibinfo {volume} {115}},\
  \bibinfo {pages} {161302} (\bibinfo {year} {2015})},\ \Eprint
  {http://arxiv.org/abs/1506.02645} {arXiv:1506.02645 [astro-ph.HE]}
  \BibitemShut {NoStop}%
\bibitem [{\citenamefont {Bhattacharya}\ \emph {et~al.}(2015)\citenamefont
  {Bhattacharya}, \citenamefont {Enberg}, \citenamefont {Reno}, \citenamefont
  {Sarcevic},\ and\ \citenamefont {Stasto}}]{Bhattacharya:2015jpa}%
  \BibitemOpen
  \bibfield  {author} {\bibinfo {author} {\bibfnamefont {Atri}\ \bibnamefont
  {Bhattacharya}}, \bibinfo {author} {\bibfnamefont {Rikard}\ \bibnamefont
  {Enberg}}, \bibinfo {author} {\bibfnamefont {Mary~Hall}\ \bibnamefont
  {Reno}}, \bibinfo {author} {\bibfnamefont {Ina}\ \bibnamefont {Sarcevic}}, \
  and\ \bibinfo {author} {\bibfnamefont {Anna}\ \bibnamefont {Stasto}},\
  }\bibfield  {title} {\enquote {\bibinfo {title} {{Perturbative charm
  production and the prompt atmospheric neutrino flux in light of RHIC and
  LHC}},}\ }\href {\doibase 10.1007/JHEP06(2015)110} {\bibfield  {journal}
  {\bibinfo  {journal} {JHEP}\ }\textbf {\bibinfo {volume} {06}},\ \bibinfo
  {pages} {110} (\bibinfo {year} {2015})},\ \Eprint
  {http://arxiv.org/abs/1502.01076} {arXiv:1502.01076 [hep-ph]} \BibitemShut
  {NoStop}%
\bibitem [{\citenamefont {Bechtol}\ \emph {et~al.}(2017)\citenamefont
  {Bechtol}, \citenamefont {Ahlers}, \citenamefont {Di~Mauro}, \citenamefont
  {Ajello},\ and\ \citenamefont {Vandenbroucke}}]{Bechtol:2015uqb}%
  \BibitemOpen
  \bibfield  {author} {\bibinfo {author} {\bibfnamefont {Keith}\ \bibnamefont
  {Bechtol}}, \bibinfo {author} {\bibfnamefont {Markus}\ \bibnamefont
  {Ahlers}}, \bibinfo {author} {\bibfnamefont {Mattia}\ \bibnamefont
  {Di~Mauro}}, \bibinfo {author} {\bibfnamefont {Marco}\ \bibnamefont
  {Ajello}}, \ and\ \bibinfo {author} {\bibfnamefont {Justin}\ \bibnamefont
  {Vandenbroucke}},\ }\bibfield  {title} {\enquote {\bibinfo {title} {{Evidence
  against star-forming galaxies as the dominant source of IceCube
  neutrinos}},}\ }\href {\doibase 10.3847/1538-4357/836/1/47} {\bibfield
  {journal} {\bibinfo  {journal} {Astrophys. J.}\ }\textbf {\bibinfo {volume}
  {836}},\ \bibinfo {pages} {47} (\bibinfo {year} {2017})},\ \Eprint
  {http://arxiv.org/abs/1511.00688} {arXiv:1511.00688 [astro-ph.HE]}
  \BibitemShut {NoStop}%
\bibitem [{\citenamefont {Bustamante}\ \emph {et~al.}(2017)\citenamefont
  {Bustamante}, \citenamefont {Beacom},\ and\ \citenamefont
  {Murase}}]{Bustamante:2016ciw}%
  \BibitemOpen
  \bibfield  {author} {\bibinfo {author} {\bibfnamefont {Mauricio}\
  \bibnamefont {Bustamante}}, \bibinfo {author} {\bibfnamefont {John~F.}\
  \bibnamefont {Beacom}}, \ and\ \bibinfo {author} {\bibfnamefont {Kohta}\
  \bibnamefont {Murase}},\ }\bibfield  {title} {\enquote {\bibinfo {title}
  {{Testing decay of astrophysical neutrinos with incomplete information}},}\
  }\href {\doibase 10.1103/PhysRevD.95.063013} {\bibfield  {journal} {\bibinfo
  {journal} {Phys. Rev.}\ }\textbf {\bibinfo {volume} {D95}},\ \bibinfo {pages}
  {063013} (\bibinfo {year} {2017})},\ \Eprint
  {http://arxiv.org/abs/1610.02096} {arXiv:1610.02096 [astro-ph.HE]}
  \BibitemShut {NoStop}%
\bibitem [{\citenamefont {Kadler}\ \emph {et~al.}(2016)\citenamefont {Kadler}
  \emph {et~al.}}]{Kadler:2016ygj}%
  \BibitemOpen
  \bibfield  {author} {\bibinfo {author} {\bibfnamefont {M.}~\bibnamefont
  {Kadler}} \emph {et~al.},\ }\bibfield  {title} {\enquote {\bibinfo {title}
  {{Coincidence of a high-fluence blazar outburst with a PeV-energy neutrino
  event}},}\ }\href {\doibase 10.1038/nphys3715, 10.1038/NPHYS3715} {\bibfield
  {journal} {\bibinfo  {journal} {Nature Phys.}\ }\textbf {\bibinfo {volume}
  {12}},\ \bibinfo {pages} {807--814} (\bibinfo {year} {2016})},\ \Eprint
  {http://arxiv.org/abs/1602.02012} {arXiv:1602.02012 [astro-ph.HE]}
  \BibitemShut {NoStop}%
\bibitem [{\citenamefont {Blaufuss}\ \emph {et~al.}(2016)\citenamefont
  {Blaufuss}, \citenamefont {Kopper},\ and\ \citenamefont
  {Haack}}]{Blaufuss:2015muc}%
  \BibitemOpen
  \bibfield  {author} {\bibinfo {author} {\bibfnamefont {Erik}\ \bibnamefont
  {Blaufuss}}, \bibinfo {author} {\bibfnamefont {C.}~\bibnamefont {Kopper}}, \
  and\ \bibinfo {author} {\bibfnamefont {C.}~\bibnamefont {Haack}} (\bibinfo
  {collaboration} {IceCube-Gen2}),\ }\bibfield  {title} {\enquote {\bibinfo
  {title} {{The IceCube-Gen2 High Energy Array}},}\ }\bibfield  {booktitle}
  {\emph {\bibinfo {booktitle} {{Proceedings, 34th International Cosmic Ray
  Conference (ICRC 2015): The Hague, The Netherlands, July 30-August 6,
  2015}}},\ }\href {\doibase 10.22323/1.236.1146} {\bibfield  {journal}
  {\bibinfo  {journal} {PoS}\ }\textbf {\bibinfo {volume} {ICRC2015}},\
  \bibinfo {pages} {1146} (\bibinfo {year} {2016})}\BibitemShut {NoStop}%
\bibitem [{\citenamefont {Adrian-Martinez}\ \emph {et~al.}(2016)\citenamefont
  {Adrian-Martinez} \emph {et~al.}}]{Adrian-Martinez:2016fdl}%
  \BibitemOpen
  \bibfield  {author} {\bibinfo {author} {\bibfnamefont {S.}~\bibnamefont
  {Adrian-Martinez}} \emph {et~al.} (\bibinfo {collaboration} {KM3Net}),\
  }\bibfield  {title} {\enquote {\bibinfo {title} {{Letter of intent for KM3NeT
  2.0}},}\ }\href {\doibase 10.1088/0954-3899/43/8/084001} {\bibfield
  {journal} {\bibinfo  {journal} {J. Phys.}\ }\textbf {\bibinfo {volume}
  {G43}},\ \bibinfo {pages} {084001} (\bibinfo {year} {2016})},\ \Eprint
  {http://arxiv.org/abs/1601.07459} {arXiv:1601.07459 [astro-ph.IM]}
  \BibitemShut {NoStop}%
\bibitem [{\citenamefont {Seckel}(1998)}]{Seckel:1997kk}%
  \BibitemOpen
  \bibfield  {author} {\bibinfo {author} {\bibfnamefont {D.}~\bibnamefont
  {Seckel}},\ }\bibfield  {title} {\enquote {\bibinfo {title} {{Neutrino photon
  reactions in astrophysics and cosmology}},}\ }\href {\doibase
  10.1103/PhysRevLett.80.900} {\bibfield  {journal} {\bibinfo  {journal} {Phys.
  Rev. Lett.}\ }\textbf {\bibinfo {volume} {80}},\ \bibinfo {pages} {900--903}
  (\bibinfo {year} {1998})},\ \Eprint {http://arxiv.org/abs/hep-ph/9709290}
  {arXiv:hep-ph/9709290 [hep-ph]} \BibitemShut {NoStop}%
\bibitem [{\citenamefont {Alikhanov}(2015)}]{Alikhanov:2014uja}%
  \BibitemOpen
  \bibfield  {author} {\bibinfo {author} {\bibfnamefont {I.}~\bibnamefont
  {Alikhanov}},\ }\bibfield  {title} {\enquote {\bibinfo {title} {{The Glashow
  resonance in neutrino¿photon scattering}},}\ }\bibfield  {booktitle} {\emph
  {\bibinfo {booktitle} {{Proceedings, 18th International Seminar on High
  Energy Physics (Quarks 2014): Suzdal, Russia, June 2-8, 2014}}},\ }\href
  {\doibase 10.1016/j.physletb.2014.12.056} {\bibfield  {journal} {\bibinfo
  {journal} {Phys. Lett.}\ }\textbf {\bibinfo {volume} {B741}},\ \bibinfo
  {pages} {295--300} (\bibinfo {year} {2015})},\ \Eprint
  {http://arxiv.org/abs/1402.6678} {arXiv:1402.6678 [hep-ph]} \BibitemShut
  {NoStop}%
\bibitem [{\citenamefont {Alikhanov}(2016)}]{Alikhanov:2015kla}%
  \BibitemOpen
  \bibfield  {author} {\bibinfo {author} {\bibfnamefont {I.}~\bibnamefont
  {Alikhanov}},\ }\bibfield  {title} {\enquote {\bibinfo {title} {{Hidden
  Glashow resonance in neutrino-nucleus collisions}},}\ }\href {\doibase
  10.1016/j.physletb.2016.03.009} {\bibfield  {journal} {\bibinfo  {journal}
  {Phys. Lett.}\ }\textbf {\bibinfo {volume} {B756}},\ \bibinfo {pages}
  {247--253} (\bibinfo {year} {2016})},\ \Eprint
  {http://arxiv.org/abs/1503.08817} {arXiv:1503.08817 [hep-ph]} \BibitemShut
  {NoStop}%
\bibitem [{\citenamefont {Stachurska}(2019)}]{Stachurska:2019srh}%
  \BibitemOpen
  \bibfield  {author} {\bibinfo {author} {\bibfnamefont {Juliana}\ \bibnamefont
  {Stachurska}} (\bibinfo {collaboration} {IceCube}),\ }\bibfield  {title}
  {\enquote {\bibinfo {title} {{IceCube High Energy Starting Events at 7.5
  Years -- New Measurements of Flux and Flavor}},}\ }\bibfield  {booktitle}
  {\emph {\bibinfo {booktitle} {{Proceedings, 8th Very Large Volume Neutrino
  Telescope Workshop (VLVnT-2018): Dubna, Russia, October 2-4, 2018}}},\ }\href
  {\doibase 10.1051/epjconf/201920702005} {\bibfield  {journal} {\bibinfo
  {journal} {EPJ Web Conf.}\ }\textbf {\bibinfo {volume} {207}},\ \bibinfo
  {pages} {02005} (\bibinfo {year} {2019})},\ \Eprint
  {http://arxiv.org/abs/1905.04237} {arXiv:1905.04237 [hep-ex]} \BibitemShut
  {NoStop}%
\bibitem [{\citenamefont {Schneider}(2019)}]{Schneider:2019ayi}%
  \BibitemOpen
  \bibfield  {author} {\bibinfo {author} {\bibfnamefont {Austin}\ \bibnamefont
  {Schneider}},\ }\bibfield  {title} {\enquote {\bibinfo {title}
  {{Characterization of the Astrophysical Diffuse Neutrino Flux with IceCube
  High-Energy Starting Events}},}\ }in\ \href@noop {} {\emph {\bibinfo
  {booktitle} {{36th International Cosmic Ray Conference (ICRC 2019) Madison,
  Wisconsin, USA, July 24-August 1, 2019}}}}\ (\bibinfo {year} {2019})\ \Eprint
  {http://arxiv.org/abs/1907.11266} {arXiv:1907.11266 [astro-ph.HE]}
  \BibitemShut {NoStop}%
\bibitem [{\citenamefont {Magill}\ and\ \citenamefont
  {Plestid}(2017)}]{Magill:2016hgc}%
  \BibitemOpen
  \bibfield  {author} {\bibinfo {author} {\bibfnamefont {Gabriel}\ \bibnamefont
  {Magill}}\ and\ \bibinfo {author} {\bibfnamefont {Ryan}\ \bibnamefont
  {Plestid}},\ }\bibfield  {title} {\enquote {\bibinfo {title} {{Neutrino
  Trident Production at the Intensity Frontier}},}\ }\href {\doibase
  10.1103/PhysRevD.95.073004} {\bibfield  {journal} {\bibinfo  {journal} {Phys.
  Rev.}\ }\textbf {\bibinfo {volume} {D95}},\ \bibinfo {pages} {073004}
  (\bibinfo {year} {2017})},\ \Eprint {http://arxiv.org/abs/1612.05642}
  {arXiv:1612.05642 [hep-ph]} \BibitemShut {NoStop}%
\bibitem [{\citenamefont {Ge}\ \emph {et~al.}(2017)\citenamefont {Ge},
  \citenamefont {Lindner},\ and\ \citenamefont {Rodejohann}}]{Ge:2017poy}%
  \BibitemOpen
  \bibfield  {author} {\bibinfo {author} {\bibfnamefont {Shao-Feng}\
  \bibnamefont {Ge}}, \bibinfo {author} {\bibfnamefont {Manfred}\ \bibnamefont
  {Lindner}}, \ and\ \bibinfo {author} {\bibfnamefont {Werner}\ \bibnamefont
  {Rodejohann}},\ }\bibfield  {title} {\enquote {\bibinfo {title} {{Atmospheric
  Trident Production for Probing New Physics}},}\ }\href {\doibase
  10.1016/j.physletb.2017.06.020} {\bibfield  {journal} {\bibinfo  {journal}
  {Phys. Lett.}\ }\textbf {\bibinfo {volume} {B772}},\ \bibinfo {pages}
  {164--168} (\bibinfo {year} {2017})},\ \Eprint
  {http://arxiv.org/abs/1702.02617} {arXiv:1702.02617 [hep-ph]} \BibitemShut
  {NoStop}%
\bibitem [{\citenamefont {Ballett}\ \emph {et~al.}(2019)\citenamefont
  {Ballett}, \citenamefont {Hostert}, \citenamefont {Pascoli}, \citenamefont
  {Perez-Gonzalez}, \citenamefont {Tabrizi},\ and\ \citenamefont
  {Zukanovich~Funchal}}]{Ballett:2018uuc}%
  \BibitemOpen
  \bibfield  {author} {\bibinfo {author} {\bibfnamefont {Peter}\ \bibnamefont
  {Ballett}}, \bibinfo {author} {\bibfnamefont {Matheus}\ \bibnamefont
  {Hostert}}, \bibinfo {author} {\bibfnamefont {Silvia}\ \bibnamefont
  {Pascoli}}, \bibinfo {author} {\bibfnamefont {Yuber~F.}\ \bibnamefont
  {Perez-Gonzalez}}, \bibinfo {author} {\bibfnamefont {Zahra}\ \bibnamefont
  {Tabrizi}}, \ and\ \bibinfo {author} {\bibfnamefont {Renata}\ \bibnamefont
  {Zukanovich~Funchal}},\ }\bibfield  {title} {\enquote {\bibinfo {title}
  {{Neutrino Trident Scattering at Near Detectors}},}\ }\href {\doibase
  10.1007/JHEP01(2019)119} {\bibfield  {journal} {\bibinfo  {journal} {JHEP}\
  }\textbf {\bibinfo {volume} {01}},\ \bibinfo {pages} {119} (\bibinfo {year}
  {2019})},\ \Eprint {http://arxiv.org/abs/1807.10973} {arXiv:1807.10973
  [hep-ph]} \BibitemShut {NoStop}%
\bibitem [{\citenamefont {Altmannshofer}\ \emph {et~al.}(2019)\citenamefont
  {Altmannshofer}, \citenamefont {Gori}, \citenamefont {Martín-Albo},
  \citenamefont {Sousa},\ and\ \citenamefont
  {Wallbank}}]{Altmannshofer:2019zhy}%
  \BibitemOpen
  \bibfield  {author} {\bibinfo {author} {\bibfnamefont {Wolfgang}\
  \bibnamefont {Altmannshofer}}, \bibinfo {author} {\bibfnamefont {Stefania}\
  \bibnamefont {Gori}}, \bibinfo {author} {\bibfnamefont {Justo}\ \bibnamefont
  {Martín-Albo}}, \bibinfo {author} {\bibfnamefont {Alexandre}\ \bibnamefont
  {Sousa}}, \ and\ \bibinfo {author} {\bibfnamefont {Michael}\ \bibnamefont
  {Wallbank}},\ }\bibfield  {title} {\enquote {\bibinfo {title} {{Neutrino
  Tridents at DUNE}},}\ }\href@noop {} {\  (\bibinfo {year} {2019})},\ \Eprint
  {http://arxiv.org/abs/1902.06765} {arXiv:1902.06765 [hep-ph]} \BibitemShut
  {NoStop}%
\bibitem [{\citenamefont {Gauld}(2019)}]{Gauld:2019pgt}%
  \BibitemOpen
  \bibfield  {author} {\bibinfo {author} {\bibfnamefont {Rhorry}\ \bibnamefont
  {Gauld}},\ }\bibfield  {title} {\enquote {\bibinfo {title} {{Precise
  predictions for multi-${\rm TeV}$ and ${\rm PeV}$ energy neutrino scattering
  rates}},}\ }\href@noop {} {\  (\bibinfo {year} {2019})},\ \Eprint
  {http://arxiv.org/abs/1905.03792} {arXiv:1905.03792 [hep-ph]} \BibitemShut
  {NoStop}%
\bibitem [{\citenamefont {Fermi}(1924)}]{Fermi:1924tc}%
  \BibitemOpen
  \bibfield  {author} {\bibinfo {author} {\bibfnamefont {E.}~\bibnamefont
  {Fermi}},\ }\bibfield  {title} {\enquote {\bibinfo {title} {{On the Theory of
  the impact between atoms and electrically charged particles}},}\ }\href
  {\doibase 10.1007/BF03184853} {\bibfield  {journal} {\bibinfo  {journal} {Z.
  Phys.}\ }\textbf {\bibinfo {volume} {29}},\ \bibinfo {pages} {315--327}
  (\bibinfo {year} {1924})}\BibitemShut {NoStop}%
\bibitem [{\citenamefont {von Weizsacker}(1934)}]{vonWeizsacker:1934nji}%
  \BibitemOpen
  \bibfield  {author} {\bibinfo {author} {\bibfnamefont {C.~F.}\ \bibnamefont
  {von Weizsacker}},\ }\bibfield  {title} {\enquote {\bibinfo {title}
  {{Radiation emitted in collisions of very fast electrons}},}\ }\href
  {\doibase 10.1007/BF01333110} {\bibfield  {journal} {\bibinfo  {journal} {Z.
  Phys.}\ }\textbf {\bibinfo {volume} {88}},\ \bibinfo {pages} {612--625}
  (\bibinfo {year} {1934})}\BibitemShut {NoStop}%
\bibitem [{\citenamefont {Williams}(1934)}]{Williams:1934ad}%
  \BibitemOpen
  \bibfield  {author} {\bibinfo {author} {\bibfnamefont {E.~J.}\ \bibnamefont
  {Williams}},\ }\bibfield  {title} {\enquote {\bibinfo {title} {{Nature of the
  high-energy particles of penetrating radiation and status of ionization and
  radiation formulae}},}\ }\href {\doibase 10.1103/PhysRev.45.729} {\bibfield
  {journal} {\bibinfo  {journal} {Phys. Rev.}\ }\textbf {\bibinfo {volume}
  {45}},\ \bibinfo {pages} {729--730} (\bibinfo {year} {1934})}\BibitemShut
  {NoStop}%
\bibitem [{\citenamefont {Kozhushner}\ and\ \citenamefont
  {Shabalin}(1961)}]{Kozhushner1961}%
  \BibitemOpen
  \bibfield  {author} {\bibinfo {author} {\bibfnamefont {M.A.}\ \bibnamefont
  {Kozhushner}}\ and\ \bibinfo {author} {\bibfnamefont {E.P.}\ \bibnamefont
  {Shabalin}},\ }\bibfield  {title} {\enquote {\bibinfo {title} {Production of
  lepton particle pairs on a coulomb center},}\ }\href@noop {} {\bibfield
  {journal} {\bibinfo  {journal} {Soviet Journal of Experimental and
  Theoretical Physics}\ }\textbf {\bibinfo {volume} {41}},\ \bibinfo {pages}
  {949} (\bibinfo {year} {1961})}\BibitemShut {NoStop}%
\bibitem [{\citenamefont {{Shabalin}}(1963)}]{Shabalin1963}%
  \BibitemOpen
  \bibfield  {author} {\bibinfo {author} {\bibfnamefont {E.~P.}\ \bibnamefont
  {{Shabalin}}},\ }\bibfield  {title} {\enquote {\bibinfo {title} {{The
  {$\mu$}$^{+}${$\mu$}$^{-}$ and e$^{+}$e$^{-}$ Pair Production Cross Sections
  for Neutrinos Scattered by Nuclei}},}\ }\href@noop {} {\bibfield  {journal}
  {\bibinfo  {journal} {Soviet Journal of Experimental and Theoretical
  Physics}\ }\textbf {\bibinfo {volume} {16}},\ \bibinfo {pages} {125}
  (\bibinfo {year} {1963})}\BibitemShut {NoStop}%
\bibitem [{\citenamefont {Czyz}\ \emph {et~al.}(1964)\citenamefont {Czyz},
  \citenamefont {Sheppey},\ and\ \citenamefont {Walecka}}]{Czyz:1964zz}%
  \BibitemOpen
  \bibfield  {author} {\bibinfo {author} {\bibfnamefont {W.}~\bibnamefont
  {Czyz}}, \bibinfo {author} {\bibfnamefont {G.~C.}\ \bibnamefont {Sheppey}}, \
  and\ \bibinfo {author} {\bibfnamefont {J.~D.}\ \bibnamefont {Walecka}},\
  }\bibfield  {title} {\enquote {\bibinfo {title} {{Neutrino production of
  lepton pairs through the point four-fermion interaction}},}\ }\href {\doibase
  10.1007/BF02734586} {\bibfield  {journal} {\bibinfo  {journal} {Nuovo Cim.}\
  }\textbf {\bibinfo {volume} {34}},\ \bibinfo {pages} {404--435} (\bibinfo
  {year} {1964})}\BibitemShut {NoStop}%
\bibitem [{\citenamefont {Schmidt}\ \emph {et~al.}(2016)\citenamefont
  {Schmidt}, \citenamefont {Pumplin}, \citenamefont {Stump},\ and\
  \citenamefont {Yuan}}]{Schmidt:2015zda}%
  \BibitemOpen
  \bibfield  {author} {\bibinfo {author} {\bibfnamefont {Carl}\ \bibnamefont
  {Schmidt}}, \bibinfo {author} {\bibfnamefont {Jon}\ \bibnamefont {Pumplin}},
  \bibinfo {author} {\bibfnamefont {Daniel}\ \bibnamefont {Stump}}, \ and\
  \bibinfo {author} {\bibfnamefont {C.~P.}\ \bibnamefont {Yuan}},\ }\bibfield
  {title} {\enquote {\bibinfo {title} {{CT14QED parton distribution functions
  from isolated photon production in deep inelastic scattering}},}\ }\href
  {\doibase 10.1103/PhysRevD.93.114015} {\bibfield  {journal} {\bibinfo
  {journal} {Phys. Rev.}\ }\textbf {\bibinfo {volume} {D93}},\ \bibinfo {pages}
  {114015} (\bibinfo {year} {2016})},\ \Eprint
  {http://arxiv.org/abs/1509.02905} {arXiv:1509.02905 [hep-ph]} \BibitemShut
  {NoStop}%
\bibitem [{\citenamefont {Lee}\ and\ \citenamefont {Yang}(1960)}]{Lee:1960qv}%
  \BibitemOpen
  \bibfield  {author} {\bibinfo {author} {\bibfnamefont {T.~D.}\ \bibnamefont
  {Lee}}\ and\ \bibinfo {author} {\bibfnamefont {Chen-Ning}\ \bibnamefont
  {Yang}},\ }\bibfield  {title} {\enquote {\bibinfo {title} {{THEORETICAL
  DISCUSSIONS ON POSSIBLE HIGH-ENERGY NEUTRINO EXPERIMENTS}},}\ }\href
  {\doibase 10.1103/PhysRevLett.4.307} {\bibfield  {journal} {\bibinfo
  {journal} {Phys. Rev. Lett.}\ }\textbf {\bibinfo {volume} {4}},\ \bibinfo
  {pages} {307--311} (\bibinfo {year} {1960})}\BibitemShut {NoStop}%
\bibitem [{\citenamefont {Lee}\ \emph {et~al.}(1961)\citenamefont {Lee},
  \citenamefont {Yang},\ and\ \citenamefont {Markstein}}]{Lee:1961jj}%
  \BibitemOpen
  \bibfield  {author} {\bibinfo {author} {\bibfnamefont {T.~D.}\ \bibnamefont
  {Lee}}, \bibinfo {author} {\bibfnamefont {Chen-Ning}\ \bibnamefont {Yang}}, \
  and\ \bibinfo {author} {\bibfnamefont {P.}~\bibnamefont {Markstein}},\
  }\bibfield  {title} {\enquote {\bibinfo {title} {{Production Cross-section of
  Intermediate Bosons by Neutrinos in the Coulomb Field of Protons and
  Iron}},}\ }\href {\doibase 10.1103/PhysRevLett.7.429} {\bibfield  {journal}
  {\bibinfo  {journal} {Phys. Rev. Lett.}\ }\textbf {\bibinfo {volume} {7}},\
  \bibinfo {pages} {429--433} (\bibinfo {year} {1961})}\BibitemShut {NoStop}%
\bibitem [{\citenamefont {Bell}\ and\ \citenamefont
  {Veltman}(1963{\natexlab{a}})}]{Bell:1996ms}%
  \BibitemOpen
  \bibfield  {author} {\bibinfo {author} {\bibfnamefont {J.~S.}\ \bibnamefont
  {Bell}}\ and\ \bibinfo {author} {\bibfnamefont {M.~J.~G.}\ \bibnamefont
  {Veltman}},\ }\bibfield  {title} {\enquote {\bibinfo {title} {{Intermediate
  boson production by neutrinos}},}\ }\href {\doibase
  10.1016/S0375-9601(63)80045-6} {\bibfield  {journal} {\bibinfo  {journal}
  {Phys. Lett.}\ }\textbf {\bibinfo {volume} {5}},\ \bibinfo {pages} {94--96}
  (\bibinfo {year} {1963}{\natexlab{a}})}\BibitemShut {NoStop}%
\bibitem [{\citenamefont {Bell}\ and\ \citenamefont
  {Veltman}(1963{\natexlab{b}})}]{Bell:1996mr}%
  \BibitemOpen
  \bibfield  {author} {\bibinfo {author} {\bibfnamefont {J.~S.}\ \bibnamefont
  {Bell}}\ and\ \bibinfo {author} {\bibfnamefont {M.~J.~G.}\ \bibnamefont
  {Veltman}},\ }\bibfield  {title} {\enquote {\bibinfo {title} {{Polarisation
  of vector bosons produced by neutrinos}},}\ }\href {\doibase
  10.1016/S0375-9601(63)92358-2} {\bibfield  {journal} {\bibinfo  {journal}
  {Phys. Lett.}\ }\textbf {\bibinfo {volume} {5}},\ \bibinfo {pages} {151--152}
  (\bibinfo {year} {1963}{\natexlab{b}})}\BibitemShut {NoStop}%
\bibitem [{\citenamefont {Brown}(1971)}]{Brown:1971qr}%
  \BibitemOpen
  \bibfield  {author} {\bibinfo {author} {\bibfnamefont {R.~W.}\ \bibnamefont
  {Brown}},\ }\bibfield  {title} {\enquote {\bibinfo {title} {{Intermediate
  boson. i. theoretical production cross-sections in high-energy neutrino and
  muon experiments}},}\ }\href {\doibase 10.1103/PhysRevD.3.207} {\bibfield
  {journal} {\bibinfo  {journal} {Phys. Rev.}\ }\textbf {\bibinfo {volume}
  {D3}},\ \bibinfo {pages} {207--223} (\bibinfo {year} {1971})}\BibitemShut
  {NoStop}%
\bibitem [{\citenamefont {Brown}\ \emph {et~al.}(1971)\citenamefont {Brown},
  \citenamefont {Hobbs},\ and\ \citenamefont {Smith}}]{Brown:1971xk}%
  \BibitemOpen
  \bibfield  {author} {\bibinfo {author} {\bibfnamefont {R.~W.}\ \bibnamefont
  {Brown}}, \bibinfo {author} {\bibfnamefont {R.~H.}\ \bibnamefont {Hobbs}}, \
  and\ \bibinfo {author} {\bibfnamefont {J.}~\bibnamefont {Smith}},\ }\bibfield
   {title} {\enquote {\bibinfo {title} {{Intermediate boson. ii. theoretical
  muon spectra in high-energy neutrino experiments}},}\ }\href {\doibase
  10.1103/PhysRevD.4.794} {\bibfield  {journal} {\bibinfo  {journal} {Phys.
  Rev.}\ }\textbf {\bibinfo {volume} {D4}},\ \bibinfo {pages} {794--813}
  (\bibinfo {year} {1971})}\BibitemShut {NoStop}%
\bibitem [{\citenamefont {Arnison}\ \emph {et~al.}(1983)\citenamefont {Arnison}
  \emph {et~al.}}]{Arnison:1983rp}%
  \BibitemOpen
  \bibfield  {author} {\bibinfo {author} {\bibfnamefont {G.}~\bibnamefont
  {Arnison}} \emph {et~al.} (\bibinfo {collaboration} {UA1}),\ }\bibfield
  {title} {\enquote {\bibinfo {title} {{Experimental Observation of Isolated
  Large Transverse Energy Electrons with Associated Missing Energy at s**(1/2)
  = 540-GeV}},}\ }\href {\doibase 10.1016/0370-2693(83)91177-2} {\bibfield
  {journal} {\bibinfo  {journal} {Phys. Lett.}\ }\textbf {\bibinfo {volume}
  {B122}},\ \bibinfo {pages} {103--116} (\bibinfo {year} {1983})}\BibitemShut
  {NoStop}%
\bibitem [{\citenamefont {Ageron}\ \emph {et~al.}(2011)\citenamefont {Ageron}
  \emph {et~al.}}]{Collaboration:2011nsa}%
  \BibitemOpen
  \bibfield  {author} {\bibinfo {author} {\bibfnamefont {M.}~\bibnamefont
  {Ageron}} \emph {et~al.} (\bibinfo {collaboration} {ANTARES}),\ }\bibfield
  {title} {\enquote {\bibinfo {title} {{ANTARES: the first undersea neutrino
  telescope}},}\ }\href {\doibase 10.1016/j.nima.2011.06.103} {\bibfield
  {journal} {\bibinfo  {journal} {Nucl. Instrum. Meth.}\ }\textbf {\bibinfo
  {volume} {A656}},\ \bibinfo {pages} {11--38} (\bibinfo {year} {2011})},\
  \Eprint {http://arxiv.org/abs/1104.1607} {arXiv:1104.1607 [astro-ph.IM]}
  \BibitemShut {NoStop}%
\bibitem [{\citenamefont {Lovseth}\ and\ \citenamefont
  {Radomiski}(1971)}]{Lovseth:1971vv}%
  \BibitemOpen
  \bibfield  {author} {\bibinfo {author} {\bibfnamefont {J.}~\bibnamefont
  {Lovseth}}\ and\ \bibinfo {author} {\bibfnamefont {M.}~\bibnamefont
  {Radomiski}},\ }\bibfield  {title} {\enquote {\bibinfo {title} {{Kinematical
  distributions of neutrino-produced lepton triplets}},}\ }\href {\doibase
  10.1103/PhysRevD.3.2686} {\bibfield  {journal} {\bibinfo  {journal} {Phys.
  Rev.}\ }\textbf {\bibinfo {volume} {D3}},\ \bibinfo {pages} {2686--2706}
  (\bibinfo {year} {1971})}\BibitemShut {NoStop}%
\bibitem [{\citenamefont {Fujikawa}(1971)}]{Fujikawa:1971nx}%
  \BibitemOpen
  \bibfield  {author} {\bibinfo {author} {\bibfnamefont {K.}~\bibnamefont
  {Fujikawa}},\ }\bibfield  {title} {\enquote {\bibinfo {title} {{The
  self-coupling of weak lepton currents in high-energy neutrino and muon
  reactions}},}\ }\href {\doibase 10.1016/0003-4916(71)90244-2} {\bibfield
  {journal} {\bibinfo  {journal} {Annals Phys.}\ }\textbf {\bibinfo {volume}
  {68}},\ \bibinfo {pages} {102--162} (\bibinfo {year} {1971})}\BibitemShut
  {NoStop}%
\bibitem [{\citenamefont {Koike}\ \emph
  {et~al.}(1971{\natexlab{a}})\citenamefont {Koike}, \citenamefont {Konuma},
  \citenamefont {Kurata},\ and\ \citenamefont {Sugano}}]{Koike:1971tu}%
  \BibitemOpen
  \bibfield  {author} {\bibinfo {author} {\bibfnamefont {K.}~\bibnamefont
  {Koike}}, \bibinfo {author} {\bibfnamefont {M.}~\bibnamefont {Konuma}},
  \bibinfo {author} {\bibfnamefont {K.}~\bibnamefont {Kurata}}, \ and\ \bibinfo
  {author} {\bibfnamefont {K.}~\bibnamefont {Sugano}},\ }\bibfield  {title}
  {\enquote {\bibinfo {title} {{Neutrino production of lepton pairs. 1.}}}\
  }\href {\doibase 10.1143/PTP.46.1150} {\bibfield  {journal} {\bibinfo
  {journal} {Prog. Theor. Phys.}\ }\textbf {\bibinfo {volume} {46}},\ \bibinfo
  {pages} {1150--1169} (\bibinfo {year} {1971}{\natexlab{a}})}\BibitemShut
  {NoStop}%
\bibitem [{\citenamefont {Koike}\ \emph
  {et~al.}(1971{\natexlab{b}})\citenamefont {Koike}, \citenamefont {Konuma},
  \citenamefont {Kurata},\ and\ \citenamefont {Sugano}}]{Koike:1971vg}%
  \BibitemOpen
  \bibfield  {author} {\bibinfo {author} {\bibfnamefont {K.}~\bibnamefont
  {Koike}}, \bibinfo {author} {\bibfnamefont {M.}~\bibnamefont {Konuma}},
  \bibinfo {author} {\bibfnamefont {K.}~\bibnamefont {Kurata}}, \ and\ \bibinfo
  {author} {\bibfnamefont {K.}~\bibnamefont {Sugano}},\ }\bibfield  {title}
  {\enquote {\bibinfo {title} {{Neutrino production of lepton pairs. 2.}}}\
  }\href {\doibase 10.1143/PTP.46.1799} {\bibfield  {journal} {\bibinfo
  {journal} {Prog. Theor. Phys.}\ }\textbf {\bibinfo {volume} {46}},\ \bibinfo
  {pages} {1799--1804} (\bibinfo {year} {1971}{\natexlab{b}})}\BibitemShut
  {NoStop}%
\bibitem [{\citenamefont {Brown}\ \emph {et~al.}(1972)\citenamefont {Brown},
  \citenamefont {Hobbs}, \citenamefont {Smith},\ and\ \citenamefont
  {Stanko}}]{Brown:1973ih}%
  \BibitemOpen
  \bibfield  {author} {\bibinfo {author} {\bibfnamefont {R.~W.}\ \bibnamefont
  {Brown}}, \bibinfo {author} {\bibfnamefont {R.~H.}\ \bibnamefont {Hobbs}},
  \bibinfo {author} {\bibfnamefont {J.}~\bibnamefont {Smith}}, \ and\ \bibinfo
  {author} {\bibfnamefont {N.}~\bibnamefont {Stanko}},\ }\bibfield  {title}
  {\enquote {\bibinfo {title} {{Intermediate boson. iii. virtual-boson effects
  in neutrino trident production}},}\ }\href {\doibase 10.1103/PhysRevD.6.3273}
  {\bibfield  {journal} {\bibinfo  {journal} {Phys. Rev.}\ }\textbf {\bibinfo
  {volume} {D6}},\ \bibinfo {pages} {3273--3292} (\bibinfo {year}
  {1972})}\BibitemShut {NoStop}%
\bibitem [{\citenamefont {Belusevic}\ and\ \citenamefont
  {Smith}(1988)}]{Belusevic:1987cw}%
  \BibitemOpen
  \bibfield  {author} {\bibinfo {author} {\bibfnamefont {R.}~\bibnamefont
  {Belusevic}}\ and\ \bibinfo {author} {\bibfnamefont {J.}~\bibnamefont
  {Smith}},\ }\bibfield  {title} {\enquote {\bibinfo {title} {{W - Z
  Interference in Neutrino - Nucleus Scattering}},}\ }\href {\doibase
  10.1103/PhysRevD.37.2419} {\bibfield  {journal} {\bibinfo  {journal} {Phys.
  Rev.}\ }\textbf {\bibinfo {volume} {D37}},\ \bibinfo {pages} {2419} (\bibinfo
  {year} {1988})}\BibitemShut {NoStop}%
\bibitem [{\citenamefont {Geiregat}\ \emph {et~al.}(1990)\citenamefont
  {Geiregat} \emph {et~al.}}]{Geiregat:1990gz}%
  \BibitemOpen
  \bibfield  {author} {\bibinfo {author} {\bibfnamefont {D.}~\bibnamefont
  {Geiregat}} \emph {et~al.} (\bibinfo {collaboration} {CHARM-II}),\ }\bibfield
   {title} {\enquote {\bibinfo {title} {{First observation of neutrino trident
  production}},}\ }\href {\doibase 10.1016/0370-2693(90)90146-W} {\bibfield
  {journal} {\bibinfo  {journal} {Phys. Lett.}\ }\textbf {\bibinfo {volume}
  {B245}},\ \bibinfo {pages} {271--275} (\bibinfo {year} {1990})}\BibitemShut
  {NoStop}%
\bibitem [{\citenamefont {Mishra}\ \emph {et~al.}(1991)\citenamefont {Mishra}
  \emph {et~al.}}]{Mishra:1991bv}%
  \BibitemOpen
  \bibfield  {author} {\bibinfo {author} {\bibfnamefont {S.~R.}\ \bibnamefont
  {Mishra}} \emph {et~al.} (\bibinfo {collaboration} {CCFR}),\ }\bibfield
  {title} {\enquote {\bibinfo {title} {{Neutrino tridents and W Z
  interference}},}\ }\href {\doibase 10.1103/PhysRevLett.66.3117} {\bibfield
  {journal} {\bibinfo  {journal} {Phys. Rev. Lett.}\ }\textbf {\bibinfo
  {volume} {66}},\ \bibinfo {pages} {3117--3120} (\bibinfo {year}
  {1991})}\BibitemShut {NoStop}%
\bibitem [{\citenamefont {Adams}\ \emph {et~al.}(2000)\citenamefont {Adams}
  \emph {et~al.}}]{Adams:1999mn}%
  \BibitemOpen
  \bibfield  {author} {\bibinfo {author} {\bibfnamefont {T.}~\bibnamefont
  {Adams}} \emph {et~al.} (\bibinfo {collaboration} {NuTeV}),\ }\bibfield
  {title} {\enquote {\bibinfo {title} {{Evidence for diffractive charm
  production in muon-neutrino Fe and anti-muon-neutrino Fe scattering at the
  Tevatron}},}\ }\href {\doibase 10.1103/PhysRevD.61.092001} {\bibfield
  {journal} {\bibinfo  {journal} {Phys. Rev.}\ }\textbf {\bibinfo {volume}
  {D61}},\ \bibinfo {pages} {092001} (\bibinfo {year} {2000})},\ \Eprint
  {http://arxiv.org/abs/hep-ex/9909041} {arXiv:hep-ex/9909041 [hep-ex]}
  \BibitemShut {NoStop}%
\bibitem [{\citenamefont {Anelli}\ \emph {et~al.}(2015)\citenamefont {Anelli}
  \emph {et~al.}}]{Anelli:2015pba}%
  \BibitemOpen
  \bibfield  {author} {\bibinfo {author} {\bibfnamefont {M.}~\bibnamefont
  {Anelli}} \emph {et~al.} (\bibinfo {collaboration} {SHiP}),\ }\bibfield
  {title} {\enquote {\bibinfo {title} {{A facility to Search for Hidden
  Particles (SHiP) at the CERN SPS}},}\ }\href@noop {} {\  (\bibinfo {year}
  {2015})},\ \Eprint {http://arxiv.org/abs/1504.04956} {arXiv:1504.04956
  [physics.ins-det]} \BibitemShut {NoStop}%
\bibitem [{\citenamefont {Antonello}\ \emph {et~al.}(2015)\citenamefont
  {Antonello} \emph {et~al.}}]{Antonello:2015lea}%
  \BibitemOpen
  \bibfield  {author} {\bibinfo {author} {\bibfnamefont {M.}~\bibnamefont
  {Antonello}} \emph {et~al.} (\bibinfo {collaboration} {MicroBooNE, LAr1-ND,
  ICARUS-WA104}),\ }\bibfield  {title} {\enquote {\bibinfo {title} {{A Proposal
  for a Three Detector Short-Baseline Neutrino Oscillation Program in the
  Fermilab Booster Neutrino Beam}},}\ }\href@noop {} {\  (\bibinfo {year}
  {2015})},\ \Eprint {http://arxiv.org/abs/1503.01520} {arXiv:1503.01520
  [physics.ins-det]} \BibitemShut {NoStop}%
\bibitem [{\citenamefont {Acciarri}\ \emph {et~al.}(2015)\citenamefont
  {Acciarri} \emph {et~al.}}]{Acciarri:2015uup}%
  \BibitemOpen
  \bibfield  {author} {\bibinfo {author} {\bibfnamefont {R.}~\bibnamefont
  {Acciarri}} \emph {et~al.} (\bibinfo {collaboration} {DUNE}),\ }\bibfield
  {title} {\enquote {\bibinfo {title} {{Long-Baseline Neutrino Facility (LBNF)
  and Deep Underground Neutrino Experiment (DUNE)}},}\ }\href@noop {} {\
  (\bibinfo {year} {2015})},\ \Eprint {http://arxiv.org/abs/1512.06148}
  {arXiv:1512.06148 [physics.ins-det]} \BibitemShut {NoStop}%
\bibitem [{\citenamefont {Soler}(2015)}]{Soler:2015ada}%
  \BibitemOpen
  \bibfield  {author} {\bibinfo {author} {\bibfnamefont {F.~J.~P.}\
  \bibnamefont {Soler}},\ }\bibfield  {title} {\enquote {\bibinfo {title}
  {{nuSTORM: Neutrinos from Stored Muons}},}\ }in\ \href@noop {} {\emph
  {\bibinfo {booktitle} {{Proceedings, Topical Research Meeting on Prospects in
  Neutrino Physics (NuPhys2014): London, UK, December 15-17, 2014}}}}\
  (\bibinfo {year} {2015})\ \Eprint {http://arxiv.org/abs/1507.08836}
  {arXiv:1507.08836 [physics.ins-det]} \BibitemShut {NoStop}%
\bibitem [{\citenamefont {Alwall}\ \emph {et~al.}(2014)\citenamefont {Alwall},
  \citenamefont {Frederix}, \citenamefont {Frixione}, \citenamefont {Hirschi},
  \citenamefont {Maltoni}, \citenamefont {Mattelaer}, \citenamefont {Shao},
  \citenamefont {Stelzer}, \citenamefont {Torrielli},\ and\ \citenamefont
  {Zaro}}]{Alwall:2014hca}%
  \BibitemOpen
  \bibfield  {author} {\bibinfo {author} {\bibfnamefont {J.}~\bibnamefont
  {Alwall}}, \bibinfo {author} {\bibfnamefont {R.}~\bibnamefont {Frederix}},
  \bibinfo {author} {\bibfnamefont {S.}~\bibnamefont {Frixione}}, \bibinfo
  {author} {\bibfnamefont {V.}~\bibnamefont {Hirschi}}, \bibinfo {author}
  {\bibfnamefont {F.}~\bibnamefont {Maltoni}}, \bibinfo {author} {\bibfnamefont
  {O.}~\bibnamefont {Mattelaer}}, \bibinfo {author} {\bibfnamefont {H.~S.}\
  \bibnamefont {Shao}}, \bibinfo {author} {\bibfnamefont {T.}~\bibnamefont
  {Stelzer}}, \bibinfo {author} {\bibfnamefont {P.}~\bibnamefont {Torrielli}},
  \ and\ \bibinfo {author} {\bibfnamefont {M.}~\bibnamefont {Zaro}},\
  }\bibfield  {title} {\enquote {\bibinfo {title} {{The automated computation
  of tree-level and next-to-leading order differential cross sections, and
  their matching to parton shower simulations}},}\ }\href {\doibase
  10.1007/JHEP07(2014)079} {\bibfield  {journal} {\bibinfo  {journal} {JHEP}\
  }\textbf {\bibinfo {volume} {07}},\ \bibinfo {pages} {079} (\bibinfo {year}
  {2014})},\ \Eprint {http://arxiv.org/abs/1405.0301} {arXiv:1405.0301
  [hep-ph]} \BibitemShut {NoStop}%
\bibitem [{\citenamefont {Belyaev}\ \emph {et~al.}(2013)\citenamefont
  {Belyaev}, \citenamefont {Christensen},\ and\ \citenamefont
  {Pukhov}}]{Belyaev:2012qa}%
  \BibitemOpen
  \bibfield  {author} {\bibinfo {author} {\bibfnamefont {Alexander}\
  \bibnamefont {Belyaev}}, \bibinfo {author} {\bibfnamefont {Neil~D.}\
  \bibnamefont {Christensen}}, \ and\ \bibinfo {author} {\bibfnamefont
  {Alexander}\ \bibnamefont {Pukhov}},\ }\bibfield  {title} {\enquote {\bibinfo
  {title} {{CalcHEP 3.4 for collider physics within and beyond the Standard
  Model}},}\ }\href {\doibase 10.1016/j.cpc.2013.01.014} {\bibfield  {journal}
  {\bibinfo  {journal} {Comput. Phys. Commun.}\ }\textbf {\bibinfo {volume}
  {184}},\ \bibinfo {pages} {1729--1769} (\bibinfo {year} {2013})},\ \Eprint
  {http://arxiv.org/abs/1207.6082} {arXiv:1207.6082 [hep-ph]} \BibitemShut
  {NoStop}%
\bibitem [{\citenamefont {Vysotsky}\ \emph {et~al.}(2002)\citenamefont
  {Vysotsky}, \citenamefont {Gaidaenko},\ and\ \citenamefont
  {Novikov}}]{Vysotsky:2002ix}%
  \BibitemOpen
  \bibfield  {author} {\bibinfo {author} {\bibfnamefont {M.~I.}\ \bibnamefont
  {Vysotsky}}, \bibinfo {author} {\bibfnamefont {I.~V.}\ \bibnamefont
  {Gaidaenko}}, \ and\ \bibinfo {author} {\bibfnamefont {V.~A.}\ \bibnamefont
  {Novikov}},\ }\bibfield  {title} {\enquote {\bibinfo {title} {{On lepton pair
  production in neutrino nucleus collisions}},}\ }\href {\doibase
  10.1134/1.1508695} {\bibfield  {journal} {\bibinfo  {journal} {Phys. Atom.
  Nucl.}\ }\textbf {\bibinfo {volume} {65}},\ \bibinfo {pages} {1634--1642}
  (\bibinfo {year} {2002})},\ \bibinfo {note} {[Yad.
  Fiz.65,1676(2002)]}\BibitemShut {NoStop}%
\bibitem [{\citenamefont {Mertig}\ \emph {et~al.}(1991)\citenamefont {Mertig},
  \citenamefont {Bohm},\ and\ \citenamefont {Denner}}]{Mertig:1990an}%
  \BibitemOpen
  \bibfield  {author} {\bibinfo {author} {\bibfnamefont {R.}~\bibnamefont
  {Mertig}}, \bibinfo {author} {\bibfnamefont {M.}~\bibnamefont {Bohm}}, \ and\
  \bibinfo {author} {\bibfnamefont {Ansgar}\ \bibnamefont {Denner}},\
  }\bibfield  {title} {\enquote {\bibinfo {title} {{FEYN CALC: Computer
  algebraic calculation of Feynman amplitudes}},}\ }\href {\doibase
  10.1016/0010-4655(91)90130-D} {\bibfield  {journal} {\bibinfo  {journal}
  {Comput. Phys. Commun.}\ }\textbf {\bibinfo {volume} {64}},\ \bibinfo {pages}
  {345--359} (\bibinfo {year} {1991})}\BibitemShut {NoStop}%
\bibitem [{\citenamefont {Shtabovenko}\ \emph {et~al.}(2016)\citenamefont
  {Shtabovenko}, \citenamefont {Mertig},\ and\ \citenamefont
  {Orellana}}]{Shtabovenko:2016sxi}%
  \BibitemOpen
  \bibfield  {author} {\bibinfo {author} {\bibfnamefont {Vladyslav}\
  \bibnamefont {Shtabovenko}}, \bibinfo {author} {\bibfnamefont {Rolf}\
  \bibnamefont {Mertig}}, \ and\ \bibinfo {author} {\bibfnamefont {Frederik}\
  \bibnamefont {Orellana}},\ }\bibfield  {title} {\enquote {\bibinfo {title}
  {{New Developments in FeynCalc 9.0}},}\ }\href {\doibase
  10.1016/j.cpc.2016.06.008} {\bibfield  {journal} {\bibinfo  {journal}
  {Comput. Phys. Commun.}\ }\textbf {\bibinfo {volume} {207}},\ \bibinfo
  {pages} {432--444} (\bibinfo {year} {2016})},\ \Eprint
  {http://arxiv.org/abs/1601.01167} {arXiv:1601.01167 [hep-ph]} \BibitemShut
  {NoStop}%
\bibitem [{Mur()}]{Murayama_PS}%
  \BibitemOpen
  \href@noop {} {}\bibinfo {howpublished} {\url{
  http://hitoshi.berkeley.edu/233B/phasespace.pdf }}\BibitemShut {NoStop}%
\bibitem [{\citenamefont {Gluck}\ \emph {et~al.}(2002)\citenamefont {Gluck},
  \citenamefont {Pisano},\ and\ \citenamefont {Reya}}]{Gluck:2002fi}%
  \BibitemOpen
  \bibfield  {author} {\bibinfo {author} {\bibfnamefont {M.}~\bibnamefont
  {Gluck}}, \bibinfo {author} {\bibfnamefont {Cristian}\ \bibnamefont
  {Pisano}}, \ and\ \bibinfo {author} {\bibfnamefont {E.}~\bibnamefont
  {Reya}},\ }\bibfield  {title} {\enquote {\bibinfo {title} {{The Polarized and
  unpolarized photon content of the nucleon}},}\ }\href {\doibase
  10.1016/S0370-2693(02)02125-1} {\bibfield  {journal} {\bibinfo  {journal}
  {Phys. Lett.}\ }\textbf {\bibinfo {volume} {B540}},\ \bibinfo {pages}
  {75--80} (\bibinfo {year} {2002})},\ \Eprint
  {http://arxiv.org/abs/hep-ph/0206126} {arXiv:hep-ph/0206126 [hep-ph]}
  \BibitemShut {NoStop}%
\bibitem [{\citenamefont {Martin}\ \emph {et~al.}(2005)\citenamefont {Martin},
  \citenamefont {Roberts}, \citenamefont {Stirling},\ and\ \citenamefont
  {Thorne}}]{Martin:2004dh}%
  \BibitemOpen
  \bibfield  {author} {\bibinfo {author} {\bibfnamefont {A.~D.}\ \bibnamefont
  {Martin}}, \bibinfo {author} {\bibfnamefont {R.~G.}\ \bibnamefont {Roberts}},
  \bibinfo {author} {\bibfnamefont {W.~J.}\ \bibnamefont {Stirling}}, \ and\
  \bibinfo {author} {\bibfnamefont {R.~S.}\ \bibnamefont {Thorne}},\ }\bibfield
   {title} {\enquote {\bibinfo {title} {{Parton distributions incorporating QED
  contributions}},}\ }\href {\doibase 10.1140/epjc/s2004-02088-7} {\bibfield
  {journal} {\bibinfo  {journal} {Eur. Phys. J.}\ }\textbf {\bibinfo {volume}
  {C39}},\ \bibinfo {pages} {155--161} (\bibinfo {year} {2005})},\ \Eprint
  {http://arxiv.org/abs/hep-ph/0411040} {arXiv:hep-ph/0411040 [hep-ph]}
  \BibitemShut {NoStop}%
\bibitem [{\citenamefont {Ball}\ \emph {et~al.}(2013)\citenamefont {Ball},
  \citenamefont {Bertone}, \citenamefont {Carrazza}, \citenamefont
  {Del~Debbio}, \citenamefont {Forte}, \citenamefont {Guffanti}, \citenamefont
  {Hartland},\ and\ \citenamefont {Rojo}}]{Ball:2013hta}%
  \BibitemOpen
  \bibfield  {author} {\bibinfo {author} {\bibfnamefont {Richard~D.}\
  \bibnamefont {Ball}}, \bibinfo {author} {\bibfnamefont {Valerio}\
  \bibnamefont {Bertone}}, \bibinfo {author} {\bibfnamefont {Stefano}\
  \bibnamefont {Carrazza}}, \bibinfo {author} {\bibfnamefont {Luigi}\
  \bibnamefont {Del~Debbio}}, \bibinfo {author} {\bibfnamefont {Stefano}\
  \bibnamefont {Forte}}, \bibinfo {author} {\bibfnamefont {Alberto}\
  \bibnamefont {Guffanti}}, \bibinfo {author} {\bibfnamefont {Nathan~P.}\
  \bibnamefont {Hartland}}, \ and\ \bibinfo {author} {\bibfnamefont {Juan}\
  \bibnamefont {Rojo}} (\bibinfo {collaboration} {NNPDF}),\ }\bibfield  {title}
  {\enquote {\bibinfo {title} {{Parton distributions with QED corrections}},}\
  }\href {\doibase 10.1016/j.nuclphysb.2013.10.010} {\bibfield  {journal}
  {\bibinfo  {journal} {Nucl. Phys.}\ }\textbf {\bibinfo {volume} {B877}},\
  \bibinfo {pages} {290--320} (\bibinfo {year} {2013})},\ \Eprint
  {http://arxiv.org/abs/1308.0598} {arXiv:1308.0598 [hep-ph]} \BibitemShut
  {NoStop}%
\bibitem [{\citenamefont {Martin}\ and\ \citenamefont
  {Ryskin}(2014)}]{Martin:2014nqa}%
  \BibitemOpen
  \bibfield  {author} {\bibinfo {author} {\bibfnamefont {A.~D.}\ \bibnamefont
  {Martin}}\ and\ \bibinfo {author} {\bibfnamefont {M.~G.}\ \bibnamefont
  {Ryskin}},\ }\bibfield  {title} {\enquote {\bibinfo {title} {{The photon PDF
  of the proton}},}\ }\href {\doibase 10.1140/epjc/s10052-014-3040-y}
  {\bibfield  {journal} {\bibinfo  {journal} {Eur. Phys. J.}\ }\textbf
  {\bibinfo {volume} {C74}},\ \bibinfo {pages} {3040} (\bibinfo {year}
  {2014})},\ \Eprint {http://arxiv.org/abs/1406.2118} {arXiv:1406.2118
  [hep-ph]} \BibitemShut {NoStop}%
\bibitem [{\citenamefont {Harland-Lang}\ \emph {et~al.}(2016)\citenamefont
  {Harland-Lang}, \citenamefont {Khoze},\ and\ \citenamefont
  {Ryskin}}]{Harland-Lang:2016kog}%
  \BibitemOpen
  \bibfield  {author} {\bibinfo {author} {\bibfnamefont {L.~A.}\ \bibnamefont
  {Harland-Lang}}, \bibinfo {author} {\bibfnamefont {V.~A.}\ \bibnamefont
  {Khoze}}, \ and\ \bibinfo {author} {\bibfnamefont {M.~G.}\ \bibnamefont
  {Ryskin}},\ }\bibfield  {title} {\enquote {\bibinfo {title}
  {{Photon-initiated processes at high mass}},}\ }\href {\doibase
  10.1103/PhysRevD.94.074008} {\bibfield  {journal} {\bibinfo  {journal} {Phys.
  Rev.}\ }\textbf {\bibinfo {volume} {D94}},\ \bibinfo {pages} {074008}
  (\bibinfo {year} {2016})},\ \Eprint {http://arxiv.org/abs/1607.04635}
  {arXiv:1607.04635 [hep-ph]} \BibitemShut {NoStop}%
\bibitem [{\citenamefont {Manohar}\ \emph {et~al.}(2016)\citenamefont
  {Manohar}, \citenamefont {Nason}, \citenamefont {Salam},\ and\ \citenamefont
  {Zanderighi}}]{Manohar:2016nzj}%
  \BibitemOpen
  \bibfield  {author} {\bibinfo {author} {\bibfnamefont {Aneesh}\ \bibnamefont
  {Manohar}}, \bibinfo {author} {\bibfnamefont {Paolo}\ \bibnamefont {Nason}},
  \bibinfo {author} {\bibfnamefont {Gavin~P.}\ \bibnamefont {Salam}}, \ and\
  \bibinfo {author} {\bibfnamefont {Giulia}\ \bibnamefont {Zanderighi}},\
  }\bibfield  {title} {\enquote {\bibinfo {title} {{How bright is the proton? A
  precise determination of the photon parton distribution function}},}\ }\href
  {\doibase 10.1103/PhysRevLett.117.242002} {\bibfield  {journal} {\bibinfo
  {journal} {Phys. Rev. Lett.}\ }\textbf {\bibinfo {volume} {117}},\ \bibinfo
  {pages} {242002} (\bibinfo {year} {2016})},\ \Eprint
  {http://arxiv.org/abs/1607.04266} {arXiv:1607.04266 [hep-ph]} \BibitemShut
  {NoStop}%
\bibitem [{\citenamefont {Manohar}\ \emph {et~al.}(2017)\citenamefont
  {Manohar}, \citenamefont {Nason}, \citenamefont {Salam},\ and\ \citenamefont
  {Zanderighi}}]{Manohar:2017eqh}%
  \BibitemOpen
  \bibfield  {author} {\bibinfo {author} {\bibfnamefont {Aneesh~V.}\
  \bibnamefont {Manohar}}, \bibinfo {author} {\bibfnamefont {Paolo}\
  \bibnamefont {Nason}}, \bibinfo {author} {\bibfnamefont {Gavin~P.}\
  \bibnamefont {Salam}}, \ and\ \bibinfo {author} {\bibfnamefont {Giulia}\
  \bibnamefont {Zanderighi}},\ }\bibfield  {title} {\enquote {\bibinfo {title}
  {{The Photon Content of the Proton}},}\ }\href {\doibase
  10.1007/JHEP12(2017)046} {\bibfield  {journal} {\bibinfo  {journal} {JHEP}\
  }\textbf {\bibinfo {volume} {12}},\ \bibinfo {pages} {046} (\bibinfo {year}
  {2017})},\ \Eprint {http://arxiv.org/abs/1708.01256} {arXiv:1708.01256
  [hep-ph]} \BibitemShut {NoStop}%
\bibitem [{\citenamefont {Bertone}\ \emph {et~al.}(2018)\citenamefont
  {Bertone}, \citenamefont {Carrazza}, \citenamefont {Hartland},\ and\
  \citenamefont {Rojo}}]{Bertone:2017bme}%
  \BibitemOpen
  \bibfield  {author} {\bibinfo {author} {\bibfnamefont {Valerio}\ \bibnamefont
  {Bertone}}, \bibinfo {author} {\bibfnamefont {Stefano}\ \bibnamefont
  {Carrazza}}, \bibinfo {author} {\bibfnamefont {Nathan~P.}\ \bibnamefont
  {Hartland}}, \ and\ \bibinfo {author} {\bibfnamefont {Juan}\ \bibnamefont
  {Rojo}} (\bibinfo {collaboration} {NNPDF}),\ }\bibfield  {title} {\enquote
  {\bibinfo {title} {{Illuminating the photon content of the proton within a
  global PDF analysis}},}\ }\href {\doibase 10.21468/SciPostPhys.5.1.008}
  {\bibfield  {journal} {\bibinfo  {journal} {SciPost Phys.}\ }\textbf
  {\bibinfo {volume} {5}},\ \bibinfo {pages} {008} (\bibinfo {year} {2018})},\
  \Eprint {http://arxiv.org/abs/1712.07053} {arXiv:1712.07053 [hep-ph]}
  \BibitemShut {NoStop}%
\bibitem [{CT1()}]{CT14web}%
  \BibitemOpen
  \href@noop {} {}\bibinfo {howpublished}
  {\url{https://hep.pa.msu.edu/cteq/public/index.html}}\BibitemShut {NoStop}%
\bibitem [{\citenamefont {Chekanov}\ \emph {et~al.}(2010)\citenamefont
  {Chekanov} \emph {et~al.}}]{Chekanov:2009dq}%
  \BibitemOpen
  \bibfield  {author} {\bibinfo {author} {\bibfnamefont {S.}~\bibnamefont
  {Chekanov}} \emph {et~al.} (\bibinfo {collaboration} {ZEUS}),\ }\bibfield
  {title} {\enquote {\bibinfo {title} {{Measurement of isolated photon
  production in deep inelastic ep scattering}},}\ }\href {\doibase
  10.1016/j.physletb.2010.02.045} {\bibfield  {journal} {\bibinfo  {journal}
  {Phys. Lett.}\ }\textbf {\bibinfo {volume} {B687}},\ \bibinfo {pages}
  {16--25} (\bibinfo {year} {2010})},\ \Eprint {http://arxiv.org/abs/0909.4223}
  {arXiv:0909.4223 [hep-ex]} \BibitemShut {NoStop}%
\bibitem [{\citenamefont {Block}\ \emph {et~al.}(2014)\citenamefont {Block},
  \citenamefont {Durand},\ and\ \citenamefont {Ha}}]{Block:2014kza}%
  \BibitemOpen
  \bibfield  {author} {\bibinfo {author} {\bibfnamefont {Martin~M.}\
  \bibnamefont {Block}}, \bibinfo {author} {\bibfnamefont {Loyal}\ \bibnamefont
  {Durand}}, \ and\ \bibinfo {author} {\bibfnamefont {Phuoc}\ \bibnamefont
  {Ha}},\ }\bibfield  {title} {\enquote {\bibinfo {title} {{Connection of the
  virtual $\gamma^*p$ cross section of ep deep inelastic scattering to real ¿p
  scattering, and the implications for ¿N and ep total cross sections}},}\
  }\href {\doibase 10.1103/PhysRevD.89.094027} {\bibfield  {journal} {\bibinfo
  {journal} {Phys. Rev.}\ }\textbf {\bibinfo {volume} {D89}},\ \bibinfo {pages}
  {094027} (\bibinfo {year} {2014})},\ \Eprint {http://arxiv.org/abs/1404.4530}
  {arXiv:1404.4530 [hep-ph]} \BibitemShut {NoStop}%
\bibitem [{\citenamefont {Argüelles}\ \emph {et~al.}(2015)\citenamefont
  {Argüelles}, \citenamefont {Halzen}, \citenamefont {Wille}, \citenamefont
  {Kroll},\ and\ \citenamefont {Reno}}]{Arguelles:2015wba}%
  \BibitemOpen
  \bibfield  {author} {\bibinfo {author} {\bibfnamefont {Carlos~A.}\
  \bibnamefont {Argüelles}}, \bibinfo {author} {\bibfnamefont {Francis}\
  \bibnamefont {Halzen}}, \bibinfo {author} {\bibfnamefont {Logan}\
  \bibnamefont {Wille}}, \bibinfo {author} {\bibfnamefont {Mike}\ \bibnamefont
  {Kroll}}, \ and\ \bibinfo {author} {\bibfnamefont {Mary~Hall}\ \bibnamefont
  {Reno}},\ }\bibfield  {title} {\enquote {\bibinfo {title} {{High-energy
  behavior of photon, neutrino, and proton cross sections}},}\ }\href {\doibase
  10.1103/PhysRevD.92.074040} {\bibfield  {journal} {\bibinfo  {journal} {Phys.
  Rev.}\ }\textbf {\bibinfo {volume} {D92}},\ \bibinfo {pages} {074040}
  (\bibinfo {year} {2015})},\ \Eprint {http://arxiv.org/abs/1504.06639}
  {arXiv:1504.06639 [hep-ph]} \BibitemShut {NoStop}%
\bibitem [{\citenamefont {Bertone}\ \emph {et~al.}(2019)\citenamefont
  {Bertone}, \citenamefont {Gauld},\ and\ \citenamefont
  {Rojo}}]{Bertone:2018dse}%
  \BibitemOpen
  \bibfield  {author} {\bibinfo {author} {\bibfnamefont {Valerio}\ \bibnamefont
  {Bertone}}, \bibinfo {author} {\bibfnamefont {Rhorry}\ \bibnamefont {Gauld}},
  \ and\ \bibinfo {author} {\bibfnamefont {Juan}\ \bibnamefont {Rojo}},\
  }\bibfield  {title} {\enquote {\bibinfo {title} {{Neutrino Telescopes as QCD
  Microscopes}},}\ }\href {\doibase 10.1007/JHEP01(2019)217} {\bibfield
  {journal} {\bibinfo  {journal} {JHEP}\ }\textbf {\bibinfo {volume} {01}},\
  \bibinfo {pages} {217} (\bibinfo {year} {2019})},\ \Eprint
  {http://arxiv.org/abs/1808.02034} {arXiv:1808.02034 [hep-ph]} \BibitemShut
  {NoStop}%
\bibitem [{\citenamefont {Bauer}\ \emph {et~al.}(2017)\citenamefont {Bauer},
  \citenamefont {Ferland},\ and\ \citenamefont {Webber}}]{Bauer:2017isx}%
  \BibitemOpen
  \bibfield  {author} {\bibinfo {author} {\bibfnamefont {Christian~W.}\
  \bibnamefont {Bauer}}, \bibinfo {author} {\bibfnamefont {Nicolas}\
  \bibnamefont {Ferland}}, \ and\ \bibinfo {author} {\bibfnamefont {Bryan~R.}\
  \bibnamefont {Webber}},\ }\bibfield  {title} {\enquote {\bibinfo {title}
  {{Standard Model Parton Distributions at Very High Energies}},}\ }\href
  {\doibase 10.1007/JHEP08(2017)036} {\bibfield  {journal} {\bibinfo  {journal}
  {JHEP}\ }\textbf {\bibinfo {volume} {08}},\ \bibinfo {pages} {036} (\bibinfo
  {year} {2017})},\ \Eprint {http://arxiv.org/abs/1703.08562} {arXiv:1703.08562
  [hep-ph]} \BibitemShut {NoStop}%
\bibitem [{\citenamefont {Bauer}\ and\ \citenamefont
  {Webber}(2019)}]{Bauer:2018arx}%
  \BibitemOpen
  \bibfield  {author} {\bibinfo {author} {\bibfnamefont {Christian~W.}\
  \bibnamefont {Bauer}}\ and\ \bibinfo {author} {\bibfnamefont {Bryan~R.}\
  \bibnamefont {Webber}},\ }\bibfield  {title} {\enquote {\bibinfo {title}
  {{Polarization Effects in Standard Model Parton Distributions at Very High
  Energies}},}\ }\href {\doibase 10.1007/JHEP03(2019)013} {\bibfield  {journal}
  {\bibinfo  {journal} {JHEP}\ }\textbf {\bibinfo {volume} {03}},\ \bibinfo
  {pages} {013} (\bibinfo {year} {2019})},\ \Eprint
  {http://arxiv.org/abs/1808.08831} {arXiv:1808.08831 [hep-ph]} \BibitemShut
  {NoStop}%
\bibitem [{\citenamefont {Fornal}\ \emph {et~al.}(2018)\citenamefont {Fornal},
  \citenamefont {Manohar},\ and\ \citenamefont {Waalewijn}}]{Fornal:2018znf}%
  \BibitemOpen
  \bibfield  {author} {\bibinfo {author} {\bibfnamefont {Bartosz}\ \bibnamefont
  {Fornal}}, \bibinfo {author} {\bibfnamefont {Aneesh~V.}\ \bibnamefont
  {Manohar}}, \ and\ \bibinfo {author} {\bibfnamefont {Wouter~J.}\ \bibnamefont
  {Waalewijn}},\ }\bibfield  {title} {\enquote {\bibinfo {title} {{Electroweak
  Gauge Boson Parton Distribution Functions}},}\ }\href {\doibase
  10.1007/JHEP05(2018)106} {\bibfield  {journal} {\bibinfo  {journal} {JHEP}\
  }\textbf {\bibinfo {volume} {05}},\ \bibinfo {pages} {106} (\bibinfo {year}
  {2018})},\ \Eprint {http://arxiv.org/abs/1803.06347} {arXiv:1803.06347
  [hep-ph]} \BibitemShut {NoStop}%
\bibitem [{\citenamefont {Kovarik}\ \emph {et~al.}(2016)\citenamefont {Kovarik}
  \emph {et~al.}}]{Kovarik:2015cma}%
  \BibitemOpen
  \bibfield  {author} {\bibinfo {author} {\bibfnamefont {K.}~\bibnamefont
  {Kovarik}} \emph {et~al.},\ }\bibfield  {title} {\enquote {\bibinfo {title}
  {{nCTEQ15 - Global analysis of nuclear parton distributions with
  uncertainties in the CTEQ framework}},}\ }\href {\doibase
  10.1103/PhysRevD.93.085037} {\bibfield  {journal} {\bibinfo  {journal} {Phys.
  Rev.}\ }\textbf {\bibinfo {volume} {D93}},\ \bibinfo {pages} {085037}
  (\bibinfo {year} {2016})},\ \Eprint {http://arxiv.org/abs/1509.00792}
  {arXiv:1509.00792 [hep-ph]} \BibitemShut {NoStop}%
\bibitem [{\citenamefont {Khanpour}\ and\ \citenamefont
  {Atashbar~Tehrani}(2016)}]{Khanpour:2016pph}%
  \BibitemOpen
  \bibfield  {author} {\bibinfo {author} {\bibfnamefont {Hamzeh}\ \bibnamefont
  {Khanpour}}\ and\ \bibinfo {author} {\bibfnamefont {S.}~\bibnamefont
  {Atashbar~Tehrani}},\ }\bibfield  {title} {\enquote {\bibinfo {title}
  {{Global Analysis of Nuclear Parton Distribution Functions and Their
  Uncertainties at Next-to-Next-to-Leading Order}},}\ }\href {\doibase
  10.1103/PhysRevD.93.014026} {\bibfield  {journal} {\bibinfo  {journal} {Phys.
  Rev.}\ }\textbf {\bibinfo {volume} {D93}},\ \bibinfo {pages} {014026}
  (\bibinfo {year} {2016})},\ \Eprint {http://arxiv.org/abs/1601.00939}
  {arXiv:1601.00939 [hep-ph]} \BibitemShut {NoStop}%
\bibitem [{\citenamefont {Wang}\ \emph {et~al.}(2017)\citenamefont {Wang},
  \citenamefont {Chen},\ and\ \citenamefont {Fu}}]{Wang:2016mzo}%
  \BibitemOpen
  \bibfield  {author} {\bibinfo {author} {\bibfnamefont {Rong}\ \bibnamefont
  {Wang}}, \bibinfo {author} {\bibfnamefont {Xurong}\ \bibnamefont {Chen}}, \
  and\ \bibinfo {author} {\bibfnamefont {Qiang}\ \bibnamefont {Fu}},\
  }\bibfield  {title} {\enquote {\bibinfo {title} {{Global study of nuclear
  modifications on parton distribution functions}},}\ }\href {\doibase
  10.1016/j.nuclphysb.2017.04.008} {\bibfield  {journal} {\bibinfo  {journal}
  {Nucl. Phys.}\ }\textbf {\bibinfo {volume} {B920}},\ \bibinfo {pages} {1--19}
  (\bibinfo {year} {2017})},\ \Eprint {http://arxiv.org/abs/1611.03670}
  {arXiv:1611.03670 [hep-ph]} \BibitemShut {NoStop}%
\bibitem [{\citenamefont {Eskola}\ \emph {et~al.}(2017)\citenamefont {Eskola},
  \citenamefont {Paakkinen}, \citenamefont {Paukkunen},\ and\ \citenamefont
  {Salgado}}]{Eskola:2016oht}%
  \BibitemOpen
  \bibfield  {author} {\bibinfo {author} {\bibfnamefont {Kari~J.}\ \bibnamefont
  {Eskola}}, \bibinfo {author} {\bibfnamefont {Petja}\ \bibnamefont
  {Paakkinen}}, \bibinfo {author} {\bibfnamefont {Hannu}\ \bibnamefont
  {Paukkunen}}, \ and\ \bibinfo {author} {\bibfnamefont {Carlos~A.}\
  \bibnamefont {Salgado}},\ }\bibfield  {title} {\enquote {\bibinfo {title}
  {{EPPS16: Nuclear parton distributions with LHC data}},}\ }\href {\doibase
  10.1140/epjc/s10052-017-4725-9} {\bibfield  {journal} {\bibinfo  {journal}
  {Eur. Phys. J.}\ }\textbf {\bibinfo {volume} {C77}},\ \bibinfo {pages} {163}
  (\bibinfo {year} {2017})},\ \Eprint {http://arxiv.org/abs/1612.05741}
  {arXiv:1612.05741 [hep-ph]} \BibitemShut {NoStop}%
\end{thebibliography}%


%

\end{document}